\documentclass[onecolumn,draftcls]{IEEEtran}

\usepackage{epsf,times,amsmath,color}

\usepackage{epsfig,amsfonts,amsmath,graphicx}
\usepackage{subfigure,graphics,amssymb,amsxtra,color}
\usepackage{algorithm}
\usepackage{algorithmic}
\usepackage{flushend}

\usepackage[ansinew]{inputenc}

\usepackage{color,soul}
\usepackage{setspace}
\usepackage[hang,flushmargin,multiple]{footmisc}
\usepackage{epsfig}
\usepackage{wrapfig}
\usepackage{float}
\usepackage{amsfonts}
\usepackage{cite}
 \usepackage{acronym}
\usepackage{mymaths}
\usepackage{tikz}
\usepackage[tikz]{mypics}
\usepackage{kbordermatrix}
\usepackage{hyperref}

\newtheorem{lemma}{Lemma}
\newtheorem{theorem}{Theorem}
\newtheorem{corollary}{Corollary}

\renewcommand{\M}{\mathbf{\Phi}}

\renewcommand{\B}{\mathbf{B}}

\newcommand{\x}{\mathbf{x}}
\newcommand{\y}{\mathbf{y}}

\newcommand{\tr}{\mathrm{tr}}

\newcommand\rank{\mathrm{rank}}
\newcommand\MMSE{\text{MMSE}}
\newcommand\MSE{\text{MSE}}

\acrodef{CP}{cyclic prefix}
\acrodef{ZS}{zero pad suffix}
\acrodef{pdf}{probability density function}
\acrodef{iid}{independent and identically distributed}
\acrodef{BER}{bit error rate}
\acrodef{OFDM}{orthogonal frequency division multiplexing}
\acrodef{GSVD}{generalized singular value decomposition}
\acrodef{SVD}{singular value decomposition}
\acrodef{DMT}{discrete multitone}
\acrodef{ISI}{intersymbol interference}
\acrodef{ICI}{interchannel interference}
\acrodef{LOS}{line of sight}
\acrodef{NLOS}{non line of sight}
\acrodef{SNR}{signal to noise ratio}
\acrodef{SINR}{signal to interference plus noise ratio}
\acrodef{MSE}{mean squared error}
\acrodef{MIMO}{multiple-input multiple-output}
\acrodef{FFT}{fast Fourier transform}
\acrodef{IFFT}{inverse fast Fourier transform}
\acrodef{CDF}{cumulative distribution function}
\acrodef{QAM}{quadrature amplitude modulation}
\acrodef{MMSE}{minimum mean-squared error}
\acrodef{SNR}{signal-to-noise ratio}
\acrodef{i.i.d.}{independent identically distributed}
\acrodef{SVD}{singular value decomposition}
\acrodef{MAP}{maximum \emph{a posteriori}}
\acrodef{MIMO}{multiple input multiple output}
\acrodef{OFDM}{orthogonal frequency division multiplexing}
\acrodef{CSI}{channel state information}
\acrodef{AWGN}{additive white Gaussian noise}
\acrodef{CDF}{cumulative distribution function}
\acrodef{KKT}{Karush-Kuhn-Tucker}
\acrodef{PDP}{power delay profile}
\acrodef{QPSK}{quadrature phase-shift keying}
\acrodef{CS}{Compressive sensing}
\acrodef{GMM}{Gaussian mixture model}
\acrodef{OMP}{orthogonal matching pursuit}
\acrodef{EM}{expectation maximization}
\acrodef{SIC}{classification with side information}
\acrodef{DC}{distributed classification}
\acrodef{pmf}{probability mass function}
\acrodef{SIR}{reconstruction with side information}
\acrodef{DR}{distributed reconstruction}
\acrodef{AMP}{approximate message passing}
\acrodef{RIP}{restricted isometry property}
\acrodef{LDA}{linear discriminant analysis}
\acrodef{PCA}{principal component analysis}
\acrodef{IDA}{information discriminant analysis}
\acrodef{EM}{expectation-maximization}
\acrodef{PSNR}{peak signal-to-noise ratio}
\acrodef{CASSI}{coded aperture snapshot spectral imager}

\newcommand{\rev}[1]{{\color{black}#1}}

\IEEEoverridecommandlockouts

\title{\color{black}Classification and Reconstruction of High-Dimensional Signals from Low-Dimensional Features in the Presence of Side Information}

\author{\normalsize Francesco~Renna, Liming~Wang, Xin~Yuan, Jianbo~Yang, Galen~Reeves, Robert~Calderbank, Lawrence~Carin, and~Miguel~R.~D.~Rodrigues

\thanks{This paper was presented in part at the 2015 IEEE International Symposium on Information Theory.}
   
\thanks{F. Renna is with the Instituto de Telecomunica\c{c}\~{o}es and the Departamento de Ci\^{encia} de Computadores, Faculdade de Ci\^{e}ncias da Universidade do Porto, Porto, Portugal (e-mail: frarenna@dcc.fc.up.pt) and with the Department of E\&EE, University College London, London, UK (email: f.renna@ee.ucl.ac.uk).}
\thanks{L. Wang, X. Yuan, J. Yang, G. Reeves, R. Calderbank and L. Carin are with the Department of Electrical and Computer Engineering, Duke University, Durham NC, USA (e-mail: \{liming.w, xin.yuan, jianbo.yang, galen.reeves, robert.calderbank, lcarin\}@duke.edu).}
\thanks{M. R. D. Rodrigues is with the Department of E\&EE, University College London, London, UK (email: m.rodrigues@ucl.ac.uk).}
}

\begin{document}

\maketitle

\vspace{-15mm}

\begin{abstract}
This paper offers a characterization of fundamental limits on the classification and reconstruction of high-dimensional signals from low-dimensional features, in the presence of side information. We consider a scenario where a decoder has access both to linear features of the signal of interest and to linear features of the side information signal; while the side information may be in a compressed form, the objective is recovery or classification of the primary signal, not the side information. \rev{The signal of interest and the side information are each assumed to have (distinct) latent discrete labels; conditioned on these two labels, the signal of interest and side information are drawn from a multivariate Gaussian distribution, that correlates the two. With joint probabilities on the latent labels, the overall signal-(side information) representation is defined by a Gaussian mixture model.}

By considering bounds to the misclassification probability associated with the recovery of the underlying signal label, and bounds to the reconstruction error associated with recovery of the signal of interest itself, we then provide sharp sufficient and/or necessary conditions \rev{for these quantities to approach zero when the covariance matrices of the Gaussians are nearly low-rank.} These conditions, which are reminiscent of the well-known Slepian-Wolf and Wyner-Ziv conditions, are a function of the number of linear features extracted from the signal of interest, the number of linear features extracted from the side information signal, and the geometry of these signals and their interplay. \rev{Moreover, on assuming that the signal of interest and the side information obey such an approximately low-rank model, we derive expansions of the reconstruction error as a function of the deviation from an exactly low-rank model; such expansions also allow identification of operational regimes where the impact of side information on signal reconstruction is most relevant.}

Our framework, which offers a principled mechanism to integrate side information in high-dimensional data problems, is also tested in the context of imaging applications. In particular, we report state-of-the-art results in compressive hyperspectral imaging applications, where the accompanying side information is a conventional digital photograph.
\end{abstract}


\section{Introduction}
\label{introduction}

A significant focus of recent research concerns approaches to represent and extract the salient information of a high-dimensional signal from low-dimensional signal features. Methods such as feature extraction, supervised dimensionality reduction and unsupervised dimensionality reduction have thus been \rev{studied in} various disciplines~\cite{Duda00,Jain00,Han01,Guyon03}.

Linear dimensionality reduction methods based on the second-order statistics of the source have been developed, such as \ac{LDA} \cite{Duda00} or \ac{PCA} \cite{Duda00}. Linear dimensionality reduction methods based on higher-order statistics of the data have also been developed~\cite{Carson12,Chen12,Erdogmus04,Hild06,Kaski03,Liu12,Nenadic07,Tao09,Torkkola01,Torkkola03,Wright09,Adcock13,Bogdan14}. In particular, an information-theoretic supervised approach, which uses the mutual information \cite{Chen12,Carson12} or approximations of the mutual information, such as quadratic mutual information (with quadratic R\'enyi entropy) \cite{Torkkola01,Torkkola03,Hild06} as a criterion to linearly reduce dimensionality, have been shown to lead to state-of-the-art classification and reconstruction results. A generalization of Bregman divergence has also been used to express in a unified way the gradient of mutual information for Gaussian and Poisson channels, thus enabling efficient projection design for both signal classification and reconstruction~\cite{WangNIPS13,WangIT14}. In addition, nonlinear (supervised) dimensionality reduction methods have also become popular recently~\cite{Tenenbaum00,WangICML14}.

\ac{CS} -- a signal acquisition paradigm that offers the means to simultaneously sense and compress a signal without any (or minimal) loss of information ~\cite{Candes06,CandesTao06,Donoho06,Candes05,Candes06IT,Baraniuk08}  -- also seeks to extract a set of low-dimensional features from a high-dimensional signal. In particular, this emerging paradigm shows that it is possible to perfectly reconstruct an $n$-dimensional $s$-sparse signal (sparse in some orthonormal dictionary or frame) with overwhelming probability with only $\mathcal{O} (s \log (n/s))$ linear random measurements or projections \cite{Baraniuk08,Candes06,Donoho06} using tractable $\ell_1$ minimization methods~\cite{Candes06IT} or iterative methods, like greedy matching pursuit \cite{Mallat93,Chen98,Tropp10}. Generalizations of the compressive sensing paradigm to settings where one wishes to perform other signal processing operations in the compressive domain, such as detection and classification, have also become popular recently \cite{Davenport10}. 

These dimensionality-reduction methods often attempt to explore structure in the signal, to aid in the dimensionality reduction process. Some prominent models that are used to capture the structure of a high-dimensional signal include union-of-subspaces~\cite{Blumensath09,Stojnic09,Eldar09,Eldar10}, wavelet trees~\cite{Blumensath09,Baraniuk10} and manifolds~\cite{Baraniuk09,Chen10}. A signal drawn from a union-of-subspaces is assumed to lie in one out of a collection of $K$ linear subspaces with dimension less than or equal to $s$. By leveraging such structure, reliable reconstruction can be performed with a number of projections of the order $\mathcal{O}(s + \log(2 K))$~\cite{Blumensath09} by using mixed $\ell_2/\ell_1$-norm approaches~\cite{Eldar09}. Tree models are usually adopted in conjunction with wavelet dictionaries, as they leverage the property that non-zero coefficients of wavelet transforms of smooth signals or images are usually organized in a rooted, connected tree~\cite{Crouse98}. In this case, the number of features needed for reliable reconstruction can be reduced to $\mathcal{O}(s)$~\cite{Baraniuk10}. Finally, manifold structures are shown to provide perfect recovery with a number of projections that grows linearly with the dimension of the manifold $s$, logarithmically with the product of signal size $n$ and parameters that characterize the volume and the regularity of the manifold~\cite{Baraniuk09}.

However, it is often the case that one is also presented at the encoder, at the decoder, or at both with additional information -- known as \emph{side information} -- beyond signal structure, in the form of another signal that exhibits some correlation with the signal of interest. The key question concerns how to leverage side information to enhance the classification and reconstruction of high-dimensional signals from low-dimensional features. This paper proposes to study this aspect by using models that capture key attributes of high-dimensional signals, namely the fact that such signals often live on a union of low-dimensional subspaces or affine spaces, or on a union of approximately low-dimensional spaces. The high-dimensional signal to be measured and the side information are assumed to have distinct low-dimensional representations of this type, with shared or correlated latent structure.

\subsection{Related Work}

Our problem connects to source coding with side information and distributed source coding, as the number of features extracted from high-dimensional signals can be related to the compression rate, whereas performance metrics for classification and reconstruction can be related to distortion. The foundations of distributed source coding theory were laid by Slepian and Wolf~\cite{Slepian73}, whereas those of source coding with side information by Ahlswede and K\"orner \cite{Ahlswede75}, and by Wyner and Ziv~\cite{Wyner76}. Namely, \cite{Slepian73} characterized the rates at which two discrete input sources can be compressed independently by guaranteeing lossless reconstruction at the decoder side. Perhaps surprisingly, the rates associated with independent compression at the two sources are shown to be identical to those associated with joint compression at the encoders. On the other hand, \cite{Ahlswede75} determined the rate at which a discrete source input can be compressed without losses in the presence of coded side information. In the lossy compression case, Wyner and Ziv \cite{Wyner76} proposed an encoding scheme to achieve the optimum tradeoff between compression rate and distortion when side information is available at the decoder. In contrast with the result in~\cite{Slepian73}, they proved that lossy compression without side information at the encoder suffers in general a rate loss compared to lossy compression with side information both at the encoder and the decoder~\cite{Cover91}. However, such loss was shown to be vanishingly small for the case of memoryless Gaussian sources and squared-error distortion metrics~\cite{Wyner76}. 

Our problem also relates to the problems of compressive sensing with side information/prior information \cite{Vaswani10TSP,Herzet13IT,WangLiang13,Chen08,Weizman14,Mota14,Mota15icassp,Mota15tsp}, distributed compressive sensing \cite{Baron09,Duarte13,Yang10,Haghighatshoar14,Donoho13IT,Eftekhari14,Hormati10} and multi-task compressive sensing \cite{Ji09}. The problem of compressive sensing with side information or prior information entails the reconstruction of a sparse signal in the presence of partial information about the desired signal, using reconstruction algorithms akin to those from \ac{CS}. For example, \cite{Vaswani10TSP,Herzet13IT} consider the reconstruction of a signal by leveraging partial information about the support of the signal at the decoder side; \cite{WangLiang13} considers the reconstruction of the signal by using an additional noisy version of the signal at the decoder side. \cite{Chen08} takes the side information to be associated with the previous scans of a certain subject in dynamic tomographic imaging. In this case, $\ell_1$-norm based minimization is used for recovery, by adding an additional term that accounts for the distance between the recovered image and the side information snapshot. A similar approach has been adopted recently in \cite{Weizman14}, that is shown to require a smaller number of measurements than traditional \ac{CS} in recovering magnetic resonance images. A theoretical analysis of the number of measurements sufficient for reliable recovery with high probability in the presence of side information for both $\ell_1/\ell_1$ and mixed $\ell_1/\ell_2$ reconstruction strategies is provided in \cite{Mota14}. \rev{The application of such approaches to compressive video foreground extraction is presented in \cite{Mota15icassp,Mota15tsp}.}

The problem of distributed compressive sensing, which has been considered by \cite{Baron09,Yang10,Duarte13,Haghighatshoar14,Donoho13IT,Eftekhari14,Hormati10}, involves the joint reconstruction of multiple correlated sparse signals. In\cite{Baron09,Duarte13} necessary and sufficient conditions on the minimum number of  measurements needed for perfect recovery (via $\ell_0$-norm minimization) are derived. Multiple signals are described there via joint sparsity models that involve a common component for all signals and innovation components specific to each signal. \cite{Haghighatshoar14} also provides conditions on the number of measurements for approximately zero-distortion recovery using an inversion procedure based on a generalized, multi-terminal \ac{AMP} algorithm. Reconstruction via \ac{AMP} methods for distributed \ac{CS} was also considered in \cite{Donoho13IT}, where the minimum number of measurements needed for successful signal recovery was derived assuming that measurements extracted from different signals are spatially coupled. Reconstruction obtained via $\ell_1$-norm minimization methods is considered in \cite{Eftekhari14}, where \ac{RIP} conditions for block-diagonal, random linear projection matrices are discussed. Namely, such matrices are shown to verify the \ac{RIP} if the total number of rows scales linearly with the signal sparsity $s$ and poly-logarithmically with the signal ambient dimension $n$. \cite{Hormati10} considers the problem of distributed recovery of two signals that are related through a sparse time-domain filtering operation, and it derives sufficient conditions on the number of samples needed for reliable recovery as well as a computationally-efficient reconstruction algorithm.
 
Multi-task compressive sensing \cite{Ji09} involves the description of multiple signals through a hierarchical Bayesian framework, where a prior is imposed on the wavelet coefficients for the different signals. Such a prior is inferred statistically from features extracted from the data and then used in the recovery process, thus demonstrating reconstruction reliability and robustness with various types of experimental data. 

\subsection{Contributions}

This paper studies the impact of side information on the \emph{classification} and \emph{reconstruction} of a high-dimensional signal from low-dimensional, linear and random features, by assuming that both the signal of interest and the side information are drawn from a joint \ac{GMM}. Unlike distributed and multi-task \ac{CS}, here we are generally only interested in recovering or classifying the primary signal, and not necessarily interested in recovering the underlying side information that is represented compressively.

There are multiple reasons for adopting a \ac{GMM} representation, which is often used in conjunction with the Bayesian \ac{CS} formalism~\cite{Ji08}:
\begin{itemize}
\item A \ac{GMM} model represents the Bayesian counterpart of well-known high-dimensional signal models in the literature~\cite{Blumensath09,Stojnic09,Eldar09,Eldar10,Chen10}. In particular, signals drawn from a \ac{GMM} can be seen to lie in a union of (linear or affine) subspaces, where each subspace is associated with the translation of the image of the (possibly low-rank) covariance matrix of each Gaussian component within the \ac{GMM}. Moreover, low-rank \ac{GMM} priors have been shown to approximate signals in compact manifolds~\cite{Chen10}. Also, a \ac{GMM} can represent complex distributions subject to mild regularity conditions~\cite{Sorenson71}. 
\item A \ac{GMM} model has also been shown to provide state-of-the-art results in practical problems in image processing~\cite{Yu11,YuSapiro12,DuarteC13}, dictionary learning~\cite{Chen10}, image classification \cite{Chen12} and video compression \cite{Yang14TIP}.
\item Optimal inversion of \ac{GMM} sources from linear features can be performed via a closed-form classifier or estimator, which has computational complexity proportional to the number of Gaussian classes within the \ac{GMM}. Moreover, moderate numbers of classes have been shown to model reliably real-world data as, for example, patches extracted from natural images or video frames \cite{Carson12,RecJournal,Yang14TIP}.
\end{itemize}
Of particular relevance, the adoption of \ac{GMM} priors also offers an opportunity to analyze \rev{conditions for reliable} classification or reconstruction: in particular, and in line with the contributions in
 \cite{Reboredo13,Reboredo13G,Reboredo14,RecJournal}, it is possible to adopt wireless communications-inspired metrics, akin to the diversity gain or the measurement gain \cite{Tarokh98,Tarokh99}, in order to characterize performance more finely in certain asymptotic regimes. 

Our main contributions, which generalize the analysis carried out in \cite{Reboredo14,RecJournal} to the scenario where the decoder has access to side information, include:
\begin{itemize}
\item The definition of a joint \ac{GMM} model both for the signal of interest and the side information, that generalizes the joint sparsity models in \cite{Baron09,Duarte13}.
\item Sufficient conditions for perfect signal classification in the asymptotic limit of low-rank that are a function of the geometry of the signal of interest, the geometry of the side information, their interaction, and the number of features.
\item Sufficient and necessary conditions for perfect signal reconstruction in the asymptotic limit of low-rank that are also a function of the geometries of the signal of interest, the side information, as well as the number of features.
{\color{black}
\item Expansions of the classification error and reconstruction error for the case when signals are described via approximately low-rank models, which are expressed as a function of the deviation from exactly low-rank models, that illuminate the impact of side information on performance.
}
\item A range of results that illustrate not only how theory aligns with practice, but also how to use the ideas in real-world applications, such as compressive hyperspectral imaging in the presence of side information (here a traditional photograph constitutes the side information).
\end{itemize}

These contributions differ from other contributions in the literature in various aspects. Unlike previous works on the characterization of the minimum number of measurements needed for reliable reconstruction in distributed compressive sensing \cite{Baron09,Duarte13}, our Bayesian framework allows consideration of signals with different sizes that are sparse over different bases; our model also allows characterization of conditions for reliable classification and reconstruction error. In addition, and unlike previous studies in the literature associated with $\ell_1$-norm minimization or \ac{AMP} algorithms for reconstruction, the analysis carried out in this work is also valid in the finite signal length regime, providing a sharp characterization of signal processing performance as a function of the number of features extracted from both the input and the side information. To the best of our knowledge, this work represents the first contribution in the context of structured or model-based \ac{CS} to consider both classification and reconstruction of signals in the presence of side information \rev{for approximately low-rank models.}

\subsection{Organization}

The remainder of the paper is organized as follows: Section~\ref{par:2} defines the signal and the system model used throughout the article. Section~\ref{par:PerrAnalysis} provides results for classification with side information, containing an analysis of an upper bound to the misclassification probability, that also leads to a characterization of sufficient conditions for \rev{perfect classification in the low-rank regime}. Section~\ref{par:reconstruction} provides results for reconstruction with side information, most notably sufficient and necessary conditions for \rev{perfect reconstruction} in the asymptotic limit of low-rank models; the sufficient and necessary conditions differ within a single measurement. {\color{black}Moreover, it contains expansions of the reconstruction error for the case when signals are described via to approximately low-rank models.} 
Numerical examples both with synthetic and real data are presented in Section~\ref{par:NumRes}. Finally, conclusions are drawn in Section~\ref{par:conclusions}. The Appendices contain the proofs of the main theorems.

\subsection{Notation}
In the remainder of the paper, we adopt the following notation: 
boldface upper-case letters denote matrices (${\bf X}$) and boldface lower-case letters denote column vectors (${\bf x}$); the context defines whether the quantities are deterministic or random. The symbols ${\bf I}_n$ and $\mathbf{0}_{m \times n}$ represent the identity matrix of dimension $n \times n$ and the all-zero-entries matrix of dimension $m \times n$, respectively (subscripts will be dropped when the dimensions are clear from the context). $\left(\cdot\right)\tra$, $\tr(\cdot)$,  $\rank(\cdot)$ represent the transpose, trace and the rank operators, respectively. $(\cdot)^{\dag}$ represents the Moore-Penrose pseudoinverse of a matrix. $\mathrm{Im}(\cdot)$ and $\mathrm{Null}(\cdot)$ denote the (column) image and null space of a matrix, respectively, and $\dim (\cdot )$ denotes the dimension of a linear subspace. $\E{\cdot}$ represents the expectation operator. The Gaussian distribution with mean $\boldsymbol{\mu}$ and covariance matrix ${\mathbf{\Sigma}}$ is denoted by $\mathcal{N}(\boldsymbol{\mu},{\mathbf{\Sigma}})$. The symbol $\mathrm{Cov}(\cdot)$ denotes the covariance matrix of a given random vector.

\section{Model}
\label{par:2}

We consider both the classification and reconstruction of a high-dimensional signal from linear features in the presence of side information, as shown in Fig.~\ref{fig:Model1}. In particular, we assume that the decoder has access to a set of linear features $\mathbf{y}_1 \in \mathbb{R}^{m_1}$ associated with the desired signal $\mathbf{x}_1  \in \mathbb{R}^{n_1}$ given by:
{\color{black}
\begin{equation}
\mathbf{y}_1 = \mathbf{\Phi}_1 \, \mathbf{x}_1,
\label{eq:sys1}
\end{equation}}
where $\mathbf{\Phi}_1 \in \mathbb{R}^{m_1 \times n_1}$ is the projection kernel.\footnote{In the remainder of the paper, we will use interchangeably the terms projection/measurement/sensing kernel or matrix.} 
We also assume that the decoder has access to another set of features $\mathbf{y}_2 \in \mathbb{R}^{m_2}$ -- called \emph{side information} -- associated with another signal $\mathbf{x}_2 \in \mathbb{R}^{n_2}$ given by:
{\color{black}
\begin{equation}
\mathbf{y}_2 = \mathbf{\Phi}_2 \, \mathbf{x}_2,
\label{eq:sys2}
\end{equation}}
where $\mathbf{\Phi}_2 \in \mathbb{R}^{m_2 \times n_2}$ is the projection kernel associated with the side information. 
For the sake of compact notation, we re-write the models in (\ref{eq:sys1}) and (\ref{eq:sys2}) as:
\begin{equation}
\mathbf{y} = \mathbf{\Phi} \,  \mathbf{x},
\label{eq:sys_model_synthetic}
\end{equation}
where
\begin{equation}
\mathbf{x} = \left[
\begin{array}{cc}
\mathbf{x}_1\\
\mathbf{x}_2
\end{array}
\right]
\qv
\mathbf{y} = \left[
\begin{array}{cc}
\mathbf{y}_1\\
\mathbf{y}_2
\end{array}
\right]
\end{equation} 
and
\begin{equation}
\mathbf{\Phi} = \left[
\begin{array}{ccc}
\mathbf{\Phi}_1 & \mathbf{0} \\
\mathbf{0}  & \mathbf{\Phi}_2
\end{array}
\right].
\label{eq:Phi}
\end{equation}
We focus on random projection kernels, where both matrices $\mathbf{\Phi}_1$ and $\mathbf{\Phi}_2$ are assumed to be drawn from left rotation-invariant distributions\footnote{A random matrix $\mathbf{A}\in \mathbb{R}^{m \times n}$ is said to be (left or right) rotation-invariant if the joint \ac{pdf} of its entries $p(\mathbf{A})$ satisfies $p(\mathbf{\Theta}\mathbf{A})=p(\mathbf{A})$, or $p(\mathbf{A} \mathbf{\Psi})=p(\mathbf{A})$, respectively, for any orthogonal matrix $\mathbf{\Theta}$ or $\mathbf{\Psi}$. A special case of (left and right)  rotation-invariant random matrices is represented by matrices with \ac{i.i.d.}, zero-mean Gaussian entries with fixed variance.}. {\color{black} We also assume that the rotation kernels are modified so that their rows are orthonormal, i.e., so that it holds $\mathbf{\Phi}_1 \mathbf{\Phi}_1\tra = \mathbf{I}_{m_1}$ and $\mathbf{\Phi}_2 \mathbf{\Phi}_2\tra = \mathbf{I}_{m_2}$.
}

{\color{black}
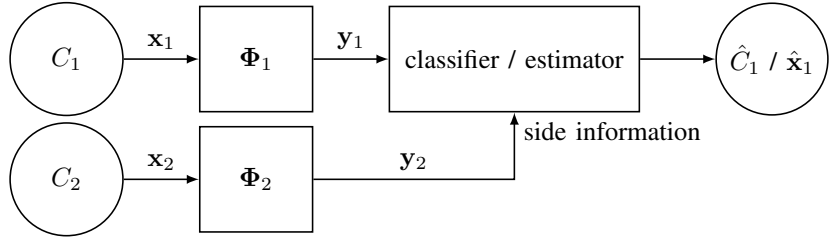
\begin{figure}
\begin{center}
\begin{tikzpicture}[x=1cm,y=1cm,semithick]
\draw (0,0) node(sou)[draw,circle,minimum size=15mm]{$C_1$};
\draw[-latex] (sou.east) --++(1,0) node[midway,above]{$\mathbf{x}_1$}
node(mod)[draw,inner sep=6pt,minimum height=1.4cm,minimum width=1.5cm,anchor=west]{$\mathbf{\Phi}_1$};
\draw[-latex] (mod.east) --++(1,0) node[midway,above]{$\mathbf{y}_1$}
node(dem)[draw,inner sep=6pt,minimum height=1.4cm,minimum width=1.5cm,anchor=west]{classifier / estimator};
\draw[-latex] (dem.east) --++(1,0) node[midway,above]{}
node(use)[draw,circle,minimum size=14mm,anchor=west]{$\hat{C}_1$ / $\hat{\mathbf{x}}_1$};
\draw (0,-1.6) node(sou2)[draw,circle,minimum size=15mm]{$C_2$};
\draw[-latex] (sou2.east) --++(1,0) node[midway,above]{$\mathbf{x}_2$}
node(hJ)[draw,inner sep=6pt,minimum height=1.4cm,minimum width=1.5cm,anchor=west]{$\mathbf{\Phi}_2$};
\draw[-latex] (hJ.east) -| (dem.south) node(x)[near start,above]{$\mathbf{y}_2$};
\node(SI)[anchor=north west] at (dem.south){side information};
\end{tikzpicture}
\end{center}
\caption{Classification and reconstruction with side information. The user attempts to generate an estimate $\hat{C}_1$ of the index of the component from  which the input signal $\mathbf{x}_1$ was drawn (classification) or it aims to generate an estimate $\hat{\mathbf{x}}_1$ of the input signal itself (reconstruction) on the basis of the observation of both feature vectors $\mathbf{y}_1$ and $\mathbf{y}_2$.}
\label{fig:Model1}
\end{figure}
}


We consider underlying class labels $C_1 \in \{ 1, \ldots, K_1 \}$ and $C_2 \in \{ 1, \ldots, K_2 \}$, where $C_1$ is associated with the signal of interest $\mathbf{x}_1$ and $C_2$ is associated with the side information signal $\mathbf{x}_2$. We assume that  $\mathbf{x}_1$ and $\mathbf{x}_2$, conditioned on the underlying class labels $C_1=i$ and $C_2=k$, are drawn from a joint distribution $p(\mathbf{x}_1,\mathbf{x}_2|C_1=i,C_2=k)$, with the class labels drawn from probability $p_{C_1,C_2}(i,k)$. 
We assume that the decoder, for both classification and reconstruction purposes, knows perfectly the joint  \ac{pmf} $p_{C_1,C_2}(i,k)$ of the discrete random variables corresponding to the class labels of $\mathbf{x}_1$ and $\mathbf{x}_2$, and the conditional distributions $p(\mathbf{x}_1,\mathbf{x}_2|C_1=i,C_2=k)$. 
For the problem of classification with side information, the objective is to estimate the value of the index $C_1$ that identifies the distribution/component from which $\mathbf{x}_1$ was drawn, on the basis of the observation of both vectors $\mathbf{y}_1$ and $\mathbf{y}_2$. The minimum average error probability in classifying $C_1$ from $\mathbf{y}_1$ and $\mathbf{y}_2$ is achieved by the \ac{MAP} classifier~\cite{Duda00}, given by
\begin{IEEEeqnarray}{rCl}
\hat{C}_1 & = & \arg \max_{i  \in \{ 1,\ldots,K_1 \}}  p(C_1=i | \mathbf{y}_1,\mathbf{y}_2)  \\
 &=& \arg \max_{i \in \{ 1,\ldots,K_1 \} }  \sum_{k=1}^{K_2} p_{C_1,C_2}(i,k) p(\mathbf{y}_1, \mathbf{y}_2 | C_1=i, C_2=k), \IEEEeqnarraynumspace
\end{IEEEeqnarray}
where $p(C_1=i | \mathbf{y}_1, \mathbf{y}_2) $ is the \emph{a posteriori} probability of class $C_1=i$ conditioned on $\mathbf{y}_1$ and $\mathbf{y}_2$. 

For the problem of reconstruction with side information, the objective of the decoder is to estimate the signal $\mathbf{x}_1$ from the observation of $\mathbf{y}_1$ and $\mathbf{y}_2$. In particular, we consider reconstruction obtained via the conditional mean estimator
\begin{equation}
\hat{\mathbf{x}}_1(\mathbf{y}_1,\mathbf{y}_2) = \E{\mathbf{x}_1 | \mathbf{y}_1, \mathbf{y}_2} =\int \mathbf{x}_1  p(\mathbf{x}_1|\mathbf{y}_1,\mathbf{y}_2) d\mathbf{x}_1,
\label{eq:CondMeanEst}
\end{equation}
where $p(\mathbf{x}_1|\mathbf{y}_1,\mathbf{y}_2)$ is the \emph{posterior} \ac{pdf} of $\mathbf{x}_1$ given the observations $\mathbf{y}_1$ and $\mathbf{y}_2$, which minimizes the reconstruction error.

We emphasize the key distinction between the previously studied problems of distributed \cite{Baron09,Duarte13} or multi-task compressive sensing \cite{Ji09}: our goal is to recover $\mathbf{x}_1$ or its label $C_1$, based upon compressive $\mathbf{y}_1$ and $\mathbf{y}_2$, while previous work considered \emph{joint} recovery of $\mathbf{x}_1$ and $\mathbf{x}_2$ (or joint estimation of $C_1$ and $C_2$). Note that our theory allows the special case for which $\mathbf{\Phi}_2$ is the identity matrix, in which case $\mathbf{y}_2=\mathbf{x}_2$ and the side information is not measured compressively. 

\subsection{Signal, Side Information and Correlation Models}
\label{par:SigModel}
The key aspect now relates to the definition of the signal, side information, and the respective correlation models. In particular, we adopt a multivariate Gaussian model for the distribution of $\mathbf{x}_1$ and $\mathbf{x}_2$, conditioned on $(C_1,C_2) = (i,k)$, i.e.
%
\begin{equation}
p(\mathbf{x}_1,\mathbf{x}_2 | C_1=i, C_2=k) = \mathcal{N} (\boldsymbol{\mu}_{\mathbf{x}}^{(ik)}, {\mathbf{\Sigma}}_{\mathbf{x}}^{(ik)}),
\label{eq:JointG}
\end{equation}
where
\begin{equation}
\boldsymbol{\mu}_{\mathbf{x}}^{(ik)} = \left[
\begin{array}{cc}
\boldsymbol{\mu}_{\mathbf{x}_1}^{(ik)} \\
\boldsymbol{\mu}_{\mathbf{x}_2}^{(ik)}
\end{array}
\right]
\qv
{\mathbf{\Sigma}}_{\mathbf{x}}^{(ik)} = \left[
\begin{array}{ccc}
{\mathbf{\Sigma}}_{\mathbf{x}_1}^{(ik)} & {\mathbf{\Sigma}}_{\mathbf{x}_{12}}^{(ik)}    \\
  {\mathbf{\Sigma}}_{\mathbf{x}_{21}}^{(ik)}   & {\mathbf{\Sigma}}_{\mathbf{x}_2}^{(ik)}
\end{array}
\right],
\label{eq:GaussDistr}
\end{equation}
so that $p(\mathbf{x}_1|C_1=i,C_2=k)= \mathcal{N}(\boldsymbol{\mu}_{\mathbf{x}_1}^{(ik)} , {\mathbf{\Sigma}}_{\mathbf{x}_1}^{(ik)})$ and  $p(\mathbf{x}_2|C_1=i,C_2=k)= \mathcal{N}(\boldsymbol{\mu}_{\mathbf{x}_2}^{(ik)} , {\mathbf{\Sigma}}_{\mathbf{x}_2}^{(ik)})$, where $\boldsymbol{\mu}_{\mathbf{x}_1}^{(ik)}$ and ${\mathbf{\Sigma}}_{\mathbf{x}_1}^{(ik)}$ are the mean and covariance matrix of $\mathbf{x}_1$ conditioned on the pair of classes $(i,k)$, respectively, $\boldsymbol{\mu}_{\mathbf{x}_2}^{(ik)}$ and $ {\mathbf{\Sigma}}_{\mathbf{x}_2}^{(ik)}$ are the mean and covariance matrix of $\mathbf{x}_2$ conditioned on the pair of classes $(i,k)$, respectively, and ${\mathbf{\Sigma}}_{\mathbf{x}_{12}}^{(ik)} $ is the cross-covariance matrix between $\mathbf{x}_1$ and $\mathbf{x}_2$ conditioned on the pair of classes $(i,k)$.

The motivation for this choice is associated by the fact that this apparently simple model can accommodate a wide range of signal distributions. In fact, note that the joint pdf of $\mathbf{x}_1$ and $\mathbf{x}_2$ follows a \ac{GMM} model:
\begin{equation}
p(\mathbf{x}_1,\mathbf{x}_2) =\sum_{i=1}^{K_1}\sum_{k=1}^{K_2} p_{C_1,C_2}(i,k)p(\mathbf{x}_1,\mathbf{x}_2 | C_1=i, C_2=k),
\end{equation}
so that we can in principle approximate very complex distributions by incorporating additional terms in the decomposition~\cite{Sorenson71}. Note also that the conditional marginal \ac{pdf}s of $\mathbf{x}_1$ and $\mathbf{x}_2$ also follow \ac{GMM} models:
\begin{IEEEeqnarray}{rCl}
p(\mathbf{x}_1| C_1=i) &=& \sum_{k=1}^{K_2} p_{C_2|C_1}(k|i) \int d\mathbf{x}_2  p(\mathbf{x}_1,\mathbf{x}_2  |  C_1=i, C_2=k) \\
 & = & \sum_{k=1}^{K_2} p_{C_2|C_1}(k|i) \  \mathcal{N}(\boldsymbol{\mu}_{\mathbf{x}_1}^{(ik)}, {\mathbf{\Sigma}}_{\mathbf{x}_1}^{(ik)})
\label{eq:x1c1}
\end{IEEEeqnarray}
and
\begin{IEEEeqnarray}{rCl}
p(\mathbf{x}_2| C_2=k)& = & \sum_{i=1}^{K_1} p_{C_1|C_2}(i|k) \int d\mathbf{x}_1  p(\mathbf{x}_1,\mathbf{x}_2  |  C_1=i, C_2=k)\\
 & = & \sum_{i=1}^{K_1} p_{C_1|C_2}(i|k) \  \mathcal{N}(\boldsymbol{\mu}_{\mathbf{x}_2}^{(ik)}, {\mathbf{\Sigma}}_{\mathbf{x}_2}^{(ik)}),
 \label{eq:x2c2}
\end{IEEEeqnarray}
where $p_{C_2|C_1}(k|i)=\frac{p_{C_1,C_2}(i,k)}{p_{C_1}(i)}$ and $p_{C_1|C_2}(i|k)=\frac{p_{C_1,C_2}(i,k)}{p_{C_2}(k)}$ are the conditional \ac{pmf}s of $C_2$ and $C_1$. Therefore, our model naturally subsumes the standard \ac{GMM} models used in the literature to deliver state-of-the-art results in reconstruction and classification problems, hyperspectral imaging and digit recognition applications \cite{Chen12}.

{\color{black}
In this work, we consider a framework in which the signal of interest and the side information are described via approximately low-rank models. In particular, conditioned on class labels $(C_1,C_2)=(i,k)$, the signals $\mathbf{x}_1$ and $\mathbf{x}_2$ can be expressed as 
\begin{IEEEeqnarray}{rCl}
\mathbf{x}_1  & = & \bar{\mathbf{x}}_1 + \mathbf{w}_1 \\
\mathbf{x}_2  & = & \bar{\mathbf{x}}_2 + \mathbf{w}_2, 
\end{IEEEeqnarray}
where
\begin{equation}
p(\bar{\mathbf{x}}_1, \bar{\mathbf{x}}_2 | C_1=i,C_2=k)  = \mathcal{N}({\boldsymbol{\mu}}_{\mathbf{x}}^{(ik)},\bar{\mathbf{\Sigma}}_{\mathbf{x}}^{(ik)}),
\end{equation}
and
\begin{equation}
\bar{\mathbf{\Sigma}}_{\mathbf{x}}^{(ik)} = \left[
\begin{array}{ccc}
\bar{\mathbf{\Sigma}}_{\mathbf{x}_1}^{(ik)} & \bar{\mathbf{\Sigma}}_{\mathbf{x}_{12}}^{(ik)}    \\
 \bar{\mathbf{\Sigma}}_{\mathbf{x}_{21}}^{(ik)}   & \bar{\mathbf{\Sigma}}_{\mathbf{x}_2}^{(ik)}
\end{array}
\right],
\label{eq:GaussDistr_LowRank}
\end{equation}
and where $\mathbf{w}_1 \sim \mathcal{N}(\mathbf{0}, \sigma_1^2 \mathbf{I}_{n_1})$, $\mathbf{w}_2 \sim \mathcal{N}(\mathbf{0}, \sigma_2^2 \mathbf{I}_{n_2})$ are independent. We assume that $\bar{\mathbf{\Sigma}}_{\mathbf{x}_1}^{(ik)}, \bar{\mathbf{\Sigma}}_{\mathbf{x}_2}^{(ik)} ,\bar{\mathbf{\Sigma}}_{\mathbf{x}}^{(ik)}$ are low-rank, so that the vectors $\bar{\mathbf{x}}_1,\bar{\mathbf{x}}_2$ represent the components of $\mathbf{x}_1, \mathbf{x}_2$ that are contained in a low-dimensional affine subspace, whereas the vectors $\mathbf{w}_1, \mathbf{w}_2$ accounts for small deviations of the signals $\mathbf{x}_1, \mathbf{x}_2$ from an exactly low-rank model.

%
%
%
%
%
}

We also adopt a framework that allows common and innovative components in the representation of $\mathbf{x}_1$ and $\mathbf{x}_2$ conditioned on $(C_1,C_2) = (i,k)$, generalizing the one in \cite{Baron09,Duarte13}. In particular, note that
{\color{black}
\begin{equation}
p(\bar{\mathbf{x}}_1, \bar{\mathbf{x}}_2 | C_1=i, C_2=k) =  \mathcal{N} (\boldsymbol{\mu}_{\mathbf{x}}^{(ik)}, \bar{\mathbf{\Sigma}}_{\mathbf{x}}^{(ik)})
\end{equation}
is equivalent to expressing $\bar{\mathbf{x}}_1$ and $\bar{\mathbf{x}}_2$ conditioned on the pair of classes $(i,k)$ as
\begin{IEEEeqnarray}{rCl}
\label{eq:x1comp}
\bar{\mathbf{x}}_1 & = &  \mathbf{x}\sub{c_1} + \mathbf{x}_1'  + \boldsymbol{\mu}_{\mathbf{x}_1}^{(ik)}   = \mathbf{P}\sub{c_1}^{(ik)} \mathbf{z}\sub c + \mathbf{P}_1^{(ik)} \mathbf{z}_1 + \boldsymbol{\mu}_{\mathbf{x}_1}^{(ik)}  \\
\label{eq:x2comp}
\bar{\mathbf{x}}_2 & = & \mathbf{x}\sub{c_2} + \mathbf{x}_2' + \boldsymbol{\mu}_{\mathbf{x}_2}^{(ik)}  =  \mathbf{P}\sub{c_2}^{(ik)} \mathbf{z}\sub c + \mathbf{P}_2^{(ik)} \mathbf{z}_2 + \boldsymbol{\mu}_{\mathbf{x}_2}^{(ik)},
\end{IEEEeqnarray}}
for an appropriate choice of the matrices $\mathbf{P}^{(ik)}\sub{c_1} \in \mathbb{R}^{n_1 \times s\sub c^{(ik)}}, \mathbf{P}^{(ik)}\sub{c_2}  \in \mathbb{R}^{n_2 \times s\sub c^{(ik)}}, \mathbf{P}^{(ik)}_1  \in \mathbb{R}^{n_1 \times s_1^{(ik)}}, \mathbf{P}^{(ik)}_2 \in \mathbb{R}^{n_2 \times s_2^{(ik)}}$ and where the vectors $\mathbf{z}\sub c \sim \mathcal{N}(\boldsymbol{0}, \mathbf{I}_{s\sub c^{(ik)}}), \mathbf{z}_1\sim \mathcal{N}(\boldsymbol{0}, \mathbf{I}_{s_1^{(ik)}})$ and $\mathbf{z}_2 \sim \mathcal{N}(\boldsymbol{0}, \mathbf{I}_{s_2^{(ik)}})$ are independent. 
In our scenario, the {\color{black}low-rank component of the} covariance matrix of $\mathbf{x}_1$ and $\mathbf{x}_2$ conditioned on the pair of classes $(i,k)$ can be also written as $\bar{\mathbf{\Sigma}}_{\mathbf{x}}^{(ik)}  = \mathbf{P}^{(ik)}(\mathbf{P}^{(ik)})\tra $, with
\begin{equation}
\mathbf{P}^{(ik)}  =\left[
\begin{array}{ccc}
\mathbf{P}\sub{c_1}^{(ik)} & \mathbf{P}_1^{(ik)} & \mathbf{0} \\
\mathbf{P}\sub{c_2}^{(ik)} & \mathbf{0} & \mathbf{P}_2^{(ik)}
\end{array}
\right],
\label{eq:Pik}
\end{equation}
where $\mathbf{P}\sub{c_1}^{(ik)},\mathbf{P}\sub{c_2}^{(ik)},\mathbf{P}_1^{(ik)}$ and $\mathbf{P}_2^{(ik)}$ are such that
\footnote{Note that the common and innovation component representation proposed here is redundant, i.e., there are various choices of matrices $\mathbf{P}^{(ik)}\sub{c_1}, \mathbf{P}^{(ik)}\sub{c_2}, \mathbf{P}^{(ik)}_1, \mathbf{P}^{(ik)}_2 $ that satisfy (\ref{eq:conditionsP}). 
We also emphasize that the results obtained in the following analysis hold irrespective of the particular choice of the matrices $\mathbf{P}^{(ik)}\sub{c_1}, \mathbf{P}^{(ik)}\sub{c_2}, \mathbf{P}^{(ik)}_1, \mathbf{P}^{(ik)}_2 $ that satisfy (\ref{eq:conditionsP}). Then, although the adoption of the common and innovation component representation is not required to prove the results contained in this work, we leverage such representation in order to give a clear interpretation of the interaction between $\mathbf{x}_1$ and $\mathbf{x}_2$ and to underline the connection of our work with previous results in the literature.
  }
\begin{equation}
\bar{\mathbf{\Sigma}}_{\mathbf{x}_1}^{(ik)}  = \mathbf{P}\sub{c_1}^{(ik)}(\mathbf{P}\sub{c_1}^{(ik)})\tra +  \mathbf{P}_1^{(ik)}(\mathbf{P}_1^{(ik)})\tra
\qv
\bar{\mathbf{\Sigma}}_{\mathbf{x}_2}^{(ik)}  = \mathbf{P}\sub{c_2}^{(ik)}(\mathbf{P}\sub{c_2}^{(ik)})\tra +  \mathbf{P}_2^{(ik)}(\mathbf{P}_2^{(ik)})\tra
\qv
\bar{\mathbf{\Sigma}}_{\mathbf{x}_{12}}^{(ik)}  = \mathbf{P}\sub{c_1}^{(ik)}(\mathbf{P}\sub{c_2}^{(ik)})\tra .
\label{eq:conditionsP}
\end{equation}

Note that (\ref{eq:x1comp}) and (\ref{eq:x2comp}) correspond to a factor or union-of-subspace model; the vector $\mathbf{z}\sub{c}$ characterizes a shared latent process, and $\mathbf{P}\sub{c_1}^{(ik)}$ and $\mathbf{P}\sub{c_2}^{(ik)}$ are linear subspaces (dictionaries) that are a function of the properties of the signal and side information, respectively. The vectors $\mathbf{z}_1$ and $\mathbf{z}_2$ are distinct latent processes, associated with respective linear subspaces $\mathbf{P}_1^{(ik)}$ and $\mathbf{P}_2^{(ik)}$. So the model may be viewed from the perspective of generalizing previous union-of-subspaces models~\cite{Blumensath09,Stojnic09,Eldar09,Eldar10}.



We refer to the vectors $\mathbf{x}\sub{c_1}\sim \mathcal{N}(\mathbf{0}, \mathbf{P}\sub{c_1} (\mathbf{P}^{(ik)}\sub{c_1})\tra)$ and $\mathbf{x}\sub{c_2} \sim \mathcal{N}(\mathbf{0}, \mathbf{P}\sub{c_2} (\mathbf{P}^{(ik)}\sub{c_2})\tra)$ as the \emph{common components}: these components of $\mathbf{x}_1$ and $\mathbf{x}_2$ are correlated, as they are obtained as linear combinations of atoms in two different dictionaries (the columns of $\mathbf{P}\sub{c_1}^{(ik)}$ and $\mathbf{P}\sub{c_2}^{(ik)}$, respectively) but with the same weights, that are contained in the vector $\mathbf{z}\sub{c}$, and therefore can be seen to model some underlying phenomena common to both $\mathbf{x}_1$ and $\mathbf{x}_2$ (conditioned on the classes). On the other hand, we refer to $\mathbf{x}_1' \sim \mathcal{N}(\mathbf{0}, \mathbf{P}_1^{(ik)} (\mathbf{P}_1^{(ik)})\tra)$ and $\mathbf{x}_2' \sim \mathcal{N}(\mathbf{0}, \mathbf{P}_2^{(ik)} (\mathbf{P}_2^{(ik)})\tra)$ as \emph{innovation components}: these components are statistically independent and thus can be seen to model phenomena specific to $\mathbf{x}_1$ and $\mathbf{x}_2$ (conditioned on the classes).\footnote{
The representation in (\ref{eq:x1comp}) and (\ref{eq:x2comp}) is reminiscent of the joint sparsity models JSM-1 and JSM-3 in~\cite{Baron09}, where signals sensed by multiple sensors were also described in terms of the sum of a common component plus innovation components. However, fundamental differences characterize our formulation with respect to such models: i) we consider a Bayesian framework in which the input signal and side information signal are picked from a mixture of components, where each component is described by a \ac{GMM} distribution, whereas in JSM-1 and JSM-3 all the components are deterministic; ii) in our model, the common components are correlated, but they are not exactly the same for $\mathbf{x}_1$ and $\mathbf{x}_2$, as it is instead for signals in JSM-1 and JSM-3; iii) in our case, the common and innovation components can be sparse over four different bases, corresponding to the ranges of the matrices $\mathbf{P}\sub{c_1}^{(ik)},\mathbf{P}\sub{c_2}^{(ik)},\mathbf{P}_1^{(ik)}$ and $\mathbf{P}_2^{(ik)}$; on the other hand, all signals in JSM-1 and JSM-3 are assumed to be sparse over the same basis.
} 

Therefore, we can now express the ranks of the matrices appearing in (\ref{eq:GaussDistr_LowRank}) as a function of ranks of the matrices appearing in the models in (\ref{eq:x1comp}) and (\ref{eq:x2comp}) as follows:
\begin{equation}
r_{\mathbf{x}_1}^{(ik)} = \rank(\bar{\mathbf{\Sigma}}_{\mathbf{x}_1}^{(ik)})=\rank [\mathbf{P}\sub{c_1}^{(ik)}  \   \mathbf{P}_1^{(ik)} ]
\label{eq:firstrank}
\end{equation}
which represents the dimension of the subspace spanned by input signals $\bar{\mathbf{x}}_1$ drawn from the Gaussian distribution corresponding to the indices $C_1=i, C_2=k$;
\begin{equation}
r_{\mathbf{x}_2}^{(ik)} = \rank(\bar{\mathbf{\Sigma}}_{\mathbf{x}_2}^{(ik)})=\rank [\mathbf{P}\sub{c_2}^{(ik)}  \   \mathbf{P}_2^{(ik)} ]
\end{equation}
which represents the dimension of the subspace spanned by side information signals $\bar{\mathbf{x}}_2$ drawn from the Gaussian distribution corresponding to the indices $C_1=i, C_2=k$;
\begin{equation}
r_{\mathbf{x}_1}^{(ik,j\ell)} = \rank(\bar{\mathbf{\Sigma}}_{\mathbf{x}_1}^{(ik)}+ \bar{\mathbf{\Sigma}}_{\mathbf{x}_1}^{(j\ell)})=\rank [\mathbf{P}\sub{c_1}^{(ik)} \ \mathbf{P}\sub{c_1}^{(j\ell)}  \   \mathbf{P}_1^{(ik)} \  \mathbf{P}_1^{(j\ell)} ]
\end{equation}
which represents the dimension of the sum of the subspaces spanned by input signals drawn from the Gaussian distribution corresponding to the indices $C_1=i, C_2=k$ and those from the Gaussian distribution corresponding to the indices $C_1=j, C_2=\ell$;
\begin{equation}
r_{\mathbf{x}_2}^{(ik,j\ell)} = \rank(\bar{\mathbf{\Sigma}}_{\mathbf{x}_2}^{(ik)}+ \bar{\mathbf{\Sigma}}_{\mathbf{x}_2}^{(j\ell)})=\rank [\mathbf{P}\sub{c_2}^{(ik)} \ \mathbf{P}\sub{c_2}^{(j\ell)}  \   \mathbf{P}_2^{(ik)} \  \mathbf{P}_2^{(j\ell)} ]
\end{equation}
which represents the dimension of the sum of the subspaces spanned by side information signals drawn from the Gaussian distribution corresponding to the indices $C_1=i, C_2=k$ and those from the Gaussian distribution corresponding to the indices $C_1=j, C_2=\ell$; and finally, the corresponding dimensions spanned collectively by input and side information signals are given by
\begin{IEEEeqnarray}{rCl}
r_{\mathbf{x}}^{(ik)} & = & \rank(\bar{\mathbf{\Sigma}}_{\mathbf{x}}^{(ik)}) =\rank
\left[
\begin{array}{ccc}
\mathbf{P}\sub{c_1}^{(ik)} & \mathbf{P}_1^{(ik)} & \mathbf{0} \\
\mathbf{P}\sub{c_2}^{(ik)} & \mathbf{0} & \mathbf{P}_2^{(ik)}
\end{array}
\right] \\
r_{\mathbf{x}}^{(ik,j\ell)} & = & \rank(\bar{\mathbf{\Sigma}}_{\mathbf{x}}^{(ik)}+\bar{\mathbf{\Sigma}}_{\mathbf{x}}^{(j\ell)})  =\rank
\left[
\begin{array}{ccc}
\mathbf{P}\sub{c_1}^{(ik,j\ell)} & \mathbf{P}_1^{(ik,j\ell)} & \mathbf{0} \\
\mathbf{P}\sub{c_2}^{(ik,j\ell)} & \mathbf{0} & \mathbf{P}_2^{(ik,j\ell)}
\end{array}
\right],
\end{IEEEeqnarray}
where we have introduced the compact notation $\mathbf{P}\sub{c_1}^{(ik,j\ell)}  = [\mathbf{P}\sub{c_1}^{(ik)}  \ \mathbf{P}\sub{c_1}^{(j\ell)}  ]$, $\mathbf{P}\sub{c_2}^{(ik,j\ell)}  = [\mathbf{P}\sub{c_2}^{(ik)}  \ \mathbf{P}\sub{c_2}^{(j\ell)}  ]$, $\mathbf{P}_1^{(ik,j\ell)}  = [\mathbf{P}_1^{(ik)}  \ \mathbf{P}_1^{(j\ell)}  ]$ and $\mathbf{P}_2^{(ik,j\ell)}  = [\mathbf{P}_2^{(ik)}  \ \mathbf{P}_2^{(j\ell)}  ]$.

We also define the rank:
\begin{equation}
r^{(ik)}=\rank\left(   \mathbf{\Phi}  \bar{\mathbf{\Sigma}}_{\mathbf{x}}^{(ik)} \mathbf{\Phi}\tra \right),
\end{equation}
that represents the dimension of the subspace in $\mathbb{R}^{m_1 + m_2}$ spanned collectively by the projections of input signals and the projections of side information signals drawn from the Gaussian distribution identified by the component indices $C_1=i,C_2=k$, and
\begin{equation}
r^{(ik,j\ell)}=\rank\left(   \mathbf{\Phi} ( \bar{\mathbf{\Sigma}}_{\mathbf{x}}^{(ik)} + \bar{\mathbf{\Sigma}}_{\mathbf{x}}^{(j\ell)}) \mathbf{\Phi}\tra  \right),
\label{eq:lastrank}
\end{equation}
that represents the dimension of the subspace obtained by summing the subspace in $\mathbb{R}^{m_1 + m_2}$ spanned collectively by the projections of input signals and the projections of side information signals drawn from the Gaussian distribution identified by the component indices $C_1=i,C_2=k$ with the subspace spanned by the projections of input signals and the projections of side information signals drawn from the Gaussian distribution identified by the component indices $C_1=j,C_2=\ell$.

The quantities in (\ref{eq:firstrank})--(\ref{eq:lastrank}), which provide a concise description of the geometry of the input source, the side information source, and the geometry of the interaction of such sources with the projections kernels, will be fundamental to determining the performance of the classification and reconstruction of high-dimensional signals from low-dimensional features in the presence of side information. {\color{black}In particular, they will allow the expression of necessary/sufficient conditions for reliable classification and reconstruction in the asymptotic low-rank regime, i.e., when $\sigma_1^2, \sigma_2^2 \to 0$, and of expansions of the reconstruction error as a function of $\sigma_1^2$ and $\sigma_2^2$.}

\section{Classification with Side Information}
\label{par:PerrAnalysis}

\rev{We first consider signal classification in the presence of side information, which will be instrumental in order to understand reconstruction.} 
The basis of the analysis is an asymptotic characterization -- in the limit of {\color{black}$\sigma_1^2,\sigma_2^2 \to 0$} -- of the behavior of an upper bound to the misclassification probability associated with the optimal \ac{MAP} classifier (rather than the exact misclassification probability which is not tractable). In particular, for a two class problem\footnote{The number of classes corresponding to the side information signal, $K_2$, can be arbitrary.}, i.e., when $K_1 =2$, via the Bhattacharyya bound~\cite{Duda00}, the misclassification probability can be upper bounded as follows:
\begin{IEEEeqnarray}{rCl}
\bar{P}\sub{err} & = & \sqrt{p_{C_1}(1) p_{C_1} (2)} \int \sqrt{p(\mathbf{y} | C_1=1)p(\mathbf{y} | C_1=2)}d\mathbf{y} \\
&=& \sqrt{p_{C_1}(1) p_{C_1} (2)} \int  \sqrt{    \sum_{k.\ell=1}^{K_2}  p_{C_2|C_1}(k|1) p_{C_2|C_1}(\ell |2) p(\mathbf{y} | C_1=1, C_2=k)  p(\mathbf{y} | C_1=2, C_2=\ell)  }d\mathbf{y}. \IEEEeqnarraynumspace  
\label{eq:BhattM1}
\end{IEEEeqnarray}
For a multiple class problem, via the Bhattacharyya bound in conjunction with the union bound, the misclassification probability can be upper bounded as follows:
\begin{equation}
\bar{P}\sub{err} = \sum_{i=1}^{K_1}\sum_{\substack{j=1\\ j\neq i}}^{K_1}  {\color{black}\sqrt{p_{C_1}(i) p_{C_1} (j)}}  \int   \sqrt{    \sum_{k.\ell=1}^{K_2}  p_{C_2|C_1}(k|i) p_{C_2|C_1}(\ell |j) p(\mathbf{y} |C_1= i, C_2=k)  p(\mathbf{y} | C_1=j, C_2=\ell)  }d\mathbf{y}.
\label{eq:BhattM1union}
\end{equation}
{\color{black}We assume $\sigma_1^2=\sigma_2^2 = \sigma^2$ and we provide an} asymptotic characterization -- akin to that in \cite{Reboredo14} -- that is based on two key metrics. The first one identifies the presence or absence of an error floor in the upper bound to the misclassification probability as $\sigma^2 \to 0$, 
leading to conditions on the number of features that guarantee perfect classification in the low-rank regime, i.e.,
\begin{equation}
\lim_{\sigma^2 \to 0} \bar{P}\sub{err}(\sigma^2) =0.
\label{eq:phasetrans}
\end{equation}
\rev{Note that conditions on the number of features $m_1$ and $m_2$ required for $\lim_{\sigma^2 \to 0} \bar{P}\sub{err}(\sigma^2) = 0$ represent also sufficient conditions for the true error probability to approach zero when $\sigma^2 \to 0$.}


The second metric offers a more refined description of the behavior of the upper bound to the misclassification probability by considering the slope at which $\log \bar{P}\sub{err}$ decays (in a $\log \sigma^2$ scale) in the low-rank regime. This value is named the \emph{diversity-order} and is given by
\begin{equation}
d = \lim_{\sigma^2 \to 0} \frac{\log  \bar{P}\sub{err}(\sigma^2)}{\log \sigma^2}.
\label{diversity}	
\end{equation}
Note also that the diversity-order associated with the upper bound of the error probability represents a lower bound on (the absolute value of) the slope of the true error probability in the low-rank regime.

We next characterize these quantities as a function of the number of features/measurements $m_1$ and $m_2$ and as a function of the underlying geometry of the signal and the side information, both for zero-mean classes (signal lives in a union of linear subspaces) and nonzero-mean ones (signal lives in a union of affine spaces). We also characterize the quantities in (\ref{eq:phasetrans}) and (\ref{diversity}) in terms of the diversity-order associated with the classification of two Gaussian distributions $\mathcal{N}(\boldsymbol{\mu}_{\mathbf{x}}^{(ik)},{\mathbf{\Sigma}}_{\mathbf{x}}^{(ik)})$ and $\mathcal{N}(\boldsymbol{\mu}_{\mathbf{x}}^{(j\ell)},{\mathbf{\Sigma}}_{\mathbf{x}}^{(j\ell)})$ from the observation of the noisy linear features $\mathbf{y}$ in (\ref{eq:sys_model_synthetic}),
\begin{equation}
d(ik,j\ell) = \lim_{\sigma^2 \to 0} \frac{1}{\log \sigma^2} \log \left( \sqrt{p_{C_1,C_2}(i,k) p_{C_1,C_2} (j,\ell)} \int \sqrt{p(\mathbf{y} | C_1=i,C_2=k)p(\mathbf{y} | C_1=j,C_2=\ell)}d\mathbf{y} \right).
\label{eq:defdikjl}
\end{equation}
Moreover, all the pairs of indices $(i,k)$ such that $p_{C_1,C_2}(i,k)=0$ clearly do not affect the diversity-order associated to classification with side information. Therefore, we can define the set of index pairs of interest as
\begin{equation}
\mathcal{S}= \left\{ (i,k) \in \{1,\ldots,K_1\} \times \{1,\ldots,K_2\} :  p_{C_1,C_2}(i,k)>0   \right\}.
\end{equation}
We also define the sets of index quadruples
\begin{equation}
\mathcal{S}\sub{SIC}=\{(i,k,j,\ell): (i,k), (j,\ell)\in \mathcal{S} ,  i\neq j\},
\label{eq:defSIC}
\end{equation}
{\color{black}
and
\begin{equation}
\mathcal{S}\sub{DC}=\{(i,k,j,\ell): (i,k), (j,\ell)\in \mathcal{S} ,  (i,k) \neq (j,\ell)\}.
\end{equation}
}

\subsection{Zero-Mean Classes}
We now provide a low-rank expansion of the upper bound to the misclassification probability associated with the system with side information in (\ref{eq:sys1}) and (\ref{eq:sys2}), when assuming that the signals involved are all zero-mean, i.e., $\boldsymbol{\mu}_{\mathbf{x}}^{(ik)}=\mathbf{0}, \forall (i,k)$. 
 
\begin{theorem}
\label{theo:zeromean}
Consider the model in (\ref{eq:sys1}) and (\ref{eq:sys2}), where the input signal $\mathbf{x}_1$ is drawn according to the class-conditioned distribution (\ref{eq:x1c1}), the side information $\mathbf{x}_2$ is drawn according to the class-conditioned distribution (\ref{eq:x2c2}), and the class-conditioned joint distribution of $\mathbf{x}_1$ and $\mathbf{x}_2$ is given by (\ref{eq:JointG}) with $\boldsymbol{\mu}_{\mathbf{x}}^{(ik)}=\mathbf{0}, \forall (i,k)$. Then, with probability 1, in the low-rank regime, i.e., when $\sigma^2 \to 0$, the upper bound to the misclassification probability (\ref{eq:BhattM1union}) can be expanded as
\begin{equation}
\bar{P}\sub{err}(\sigma^2) = A \cdot (\sigma^2)^{d} + o\left((\sigma^2)^{d}\right),
\label{eq:expzeromean}
\end{equation}
for a fixed constant $A>0$, where
\begin{equation}
d =  \min_{(i,k,j,\ell) \in \mathcal{S}\sub{SIC}} d(ik, j\ell),
\label{eq:diversitySIC}
\end{equation}
with
\begin{equation}
d(ik, j\ell) = \frac{1}{2} \left( r^{(ik,j\ell)} -  \frac{r^{(ik)} +r^{(j\ell)}}{2}   \right),
\label{eq:dpairwise}
\end{equation}
and
\begin{IEEEeqnarray}{rCl}
r^{(ik,j\ell)}&=&\rank\left(   \mathbf{\Phi} ( \bar{\mathbf{\Sigma}}_{\mathbf{x}}^{(ik)} + \bar{\mathbf{\Sigma}}_{\mathbf{x}}^{(j\ell)}) \mathbf{\Phi}\tra  \right)\\
 \label{eq:rikjell}
& = & \min \{ r_{\mathbf{x}}^{(ik,j\ell)},  \min \{m_1,r_{\mathbf{x}_1}^{(ik,j\ell)}\} +  \min \{ m_2,  r_{\mathbf{x}_2}^{(ik,j\ell)} \}  \}, \\
r^{(ik)}&=&\rank\left(   \mathbf{\Phi}  \bar{\mathbf{\Sigma}}_{\mathbf{x}}^{(ik)} \mathbf{\Phi}\tra \right) \\
 & = &  \min \{ r_{\mathbf{x}}^{(ik)}, \min \{m_1,r_{\mathbf{x}_1}^{(ik)}\} +  \min \{ m_2,  r_{\mathbf{x}_2}^{(ik)} \}  \}
 \label{eq:rik}
\end{IEEEeqnarray}
and $r^{(j\ell)}$ is obtained as $r^{(ik)}$.
\end{theorem}
\begin{IEEEproof}
See Appendix~\ref{app:B}.
\end{IEEEproof}

Theorem~\ref{theo:zeromean} provides a complete characterization of the slope of the upper bound to the misclassification probability for the case of zero-mean classes, in terms of the number of features and the geometrical description of the sources. 
In particular, observe that:
\begin{itemize}
\item The diversity-order $d$ associated with the estimation of the component index $C_1$ from noisy linear features with side information is given by the worst-case diversity-order term $d(ik,j\ell)$ associated with pair-wise classification problems for which the indices corresponding to $C_1$ are not the same ($i\neq j$). 
\item The diversity-order in (\ref{eq:diversitySIC}), which depends on the pairwise diversity-order in (\ref{eq:dpairwise}), can also be seen to depend on the difference between the dimension of the sum of the linear spaces collectively spanned by signals $\mathbf{\Phi}_1 \bar{\mathbf{x}}_1$ and $\mathbf{\Phi}_2 \bar{\mathbf{x}}_2$ drawn from the Gaussian distributions with indices $(i,k)$ and $(j,\ell)$ and the dimension of those spaces taken individually. This dependence in the presence of side information is akin to that in the absence of side information: the additional information, however, provides subspaces with increased dimensions over which it is possible to discriminate among signals belonging to different classes.
\item The effect of the correlation between $\mathbf{x}_1$ and $\mathbf{x}_2$ is embodied in the rank expressions (\ref{eq:rikjell}) and (\ref{eq:rik}). In particular, we note that, in case $\mathbf{x}_1$ and $\mathbf{x}_2$ are conditionally independent given any pairs of classes $(C_1,C_2)$, i.e., $p(\mathbf{x}_1,\mathbf{x}_2|C_1=i,C_2=k)=p(\mathbf{x}_1|C_1=i,C_2=k)p(\mathbf{x}_2|C_1=i,C_2=k)$, then $r_{\mathbf{x}}^{(ik)}=r_{\mathbf{x}_1}^{(ik)}+r_{\mathbf{x}_2}^{(ik)}$, $r_{\mathbf{x}}^{(j\ell)}=r_{\mathbf{x}_1}^{(j\ell)}+r_{\mathbf{x}_2}^{(j\ell)}$ and $r_{\mathbf{x}}^{(ik,j\ell)}=r_{\mathbf{x}_1}^{(ik,j\ell)}+r_{\mathbf{x}_2}^{(ik,j\ell)}$. Then, the diversity-order is given by the sum of the diversity-order values corresponding to the classification of $\mathbf{x}_1$ from $\mathbf{y}_1$ and that corresponding to the classification of $\mathbf{x}_2$ from $\mathbf{y}_2$. From a geometrical point of view, when $\mathbf{x}_1$ and $\mathbf{x}_2$ are conditionally independent, the linear spaces spanned by the side information offer new dimensions over which the decoder can discriminate among classes, which are completely decoupled from the dimensions corresponding to linear spaces spanned by the realizations of $\mathbf{x}_1$. Otherwise, when $\mathbf{x}_1$ and $\mathbf{x}_2$ are not conditionally independent, the diversity-order can be in general larger than, smaller than, or equal to the sum of the diversity-order values corresponding to the classification of $\mathbf{x}_1$ from $\mathbf{y}_1$ and that corresponding to the classification of $\mathbf{x}_2$ from $\mathbf{y}_2$.
\end{itemize}

A direct consequence of the asymptotic characterization of the upper bound to the misclassification probability in (\ref{eq:BhattM1union}) is access to conditions on the number of features $m_1$ and $m_2$ that are both necessary and sufficient to drive the upper bound to the misclassification probability to zero when $\sigma^2 \to 0$, and hence a condition on the number of features $m_1$ and $m_2$ that is sufficient to drive the true misclassification probability to zero when $\sigma^2 \to 0$.





\begin{corollary}

\label{cor:phasetrans}
Consider the model in (\ref{eq:sys1}) and (\ref{eq:sys2}), where the input signal $\mathbf{x}_1$ is drawn according to the class-conditioned distribution (\ref{eq:x1c1}), the side information $\mathbf{x}_2$ is drawn according to the class-conditioned distribution (\ref{eq:x2c2}), and the class-conditioned joint distribution of $\mathbf{x}_1$ and $\mathbf{x}_2$ is given by (\ref{eq:JointG}) with $\boldsymbol{\mu}_{\mathbf{x}}^{(ik)}=\mathbf{0}, \forall (i,k)$. 

If there exists an index quadruple $(i,k,j,\ell) \in \mathcal{S}\sub{SIC}$ such that $r_{\mathbf{x}}^{(ik,j\ell)} =r_{\mathbf{x}}^{(ik)}=r_{\mathbf{x}}^{(j\ell)}$, then, $d=0$ and the upper bound to the misclassification probability (\ref{eq:BhattM1union}) exhibits an error floor in the low-rank regime. Otherwise, if $r_{\mathbf{x}}^{(ik,j\ell)} > r_{\mathbf{x}}^{(ik)},r_{\mathbf{x}}^{(j\ell)}$, $\forall (i,k,j,\ell) \in \mathcal{S}\sub{SIC}$, then, with probability 1, the upper bound to the misclassification probability (\ref{eq:BhattM1union}) approaches zero when $\sigma^2 \to 0$ if and only if the following conditions hold $\forall (i,k,j,\ell) \in \mathcal{S}\sub{SIC}$:
\begin{enumerate}
\item if $r_{\mathbf{x}_1}^{(ik,j\ell)}>r_{\mathbf{x}_1}^{(ik)},r_{\mathbf{x}_1}^{(j\ell)}$ and $r_{\mathbf{x}_2}^{(ik,j\ell)}>r_{\mathbf{x}_2}^{(ik)},r_{\mathbf{x}_2}^{(j\ell)}$:
\begin{equation}
m_1 > \min \{r_{\mathbf{x}_1}^{(ik)},r_{\mathbf{x}_1}^{(j\ell)}\} \quad \mathrm{or} \quad m_2 > \min \{r_{\mathbf{x}_2}^{(ik)},r_{\mathbf{x}_2}^{(j\ell)}\} \quad \mathrm{or} \quad  
m_1+m_2 > \min \{r_{\mathbf{x}}^{(ik)},r_{\mathbf{x}}^{(j\ell)}\};
\label{eq:cor1}
\end{equation}
\item if $r_{\mathbf{x}_1}^{(ik,j\ell)}=r_{\mathbf{x}_1}^{(ik)}=r_{\mathbf{x}_1}^{(j\ell)}$ and $r_{\mathbf{x}_2}^{(ik,j\ell)}=r_{\mathbf{x}_2}^{(ik)}=r_{\mathbf{x}_2}^{(j\ell)}$:
\begin{equation}
\left\{
\begin{array}{lll}
m_1> \min \{  r_{\mathbf{x}}^{(ik)} -r_{\mathbf{x}_2}^{(ik)},  r_{\mathbf{x}}^{(j\ell)} -r_{\mathbf{x}_2}^{(j\ell)}   \} \\
m_2> \min \{  r_{\mathbf{x}}^{(ik)} -r_{\mathbf{x}_1}^{(ik)},  r_{\mathbf{x}}^{(j\ell)} -r_{\mathbf{x}_1}^{(j\ell)}   \} \\
m_1+ m_2> \min \{  r_{\mathbf{x}}^{(ik)} ,  r_{\mathbf{x}}^{(j\ell)}  \} \\
\end{array}
\right. ;
\label{eq:cor4}
\end{equation}
\item if $r_{\mathbf{x}_1}^{(ik,j\ell)}>r_{\mathbf{x}_1}^{(ik)},r_{\mathbf{x}_1}^{(j\ell)}$ and $r_{\mathbf{x}_2}^{(ik,j\ell)}=r_{\mathbf{x}_2}^{(ik)}=r_{\mathbf{x}_2}^{(j\ell)}$:
\begin{equation}
m_1 > \min \{ r_{\mathbf{x}_1}^{(ik)},r_{\mathbf{x}_1}^{(j\ell)}  \}  \quad \mathrm{or}  \quad 
\left\{
\begin{array}{lll}
m_1> \min \{  r_{\mathbf{x}}^{(ik)} -r_{\mathbf{x}_2}^{(ik)},  r_{\mathbf{x}}^{(j\ell)} -r_{\mathbf{x}_2}^{(j\ell)}   \} \\
m_1+ m_2> \min \{  r_{\mathbf{x}}^{(ik)} ,  r_{\mathbf{x}}^{(j\ell)}  \} \\
\end{array}
\right. ;
\label{eq:cor2}
\end{equation}
\item if $r_{\mathbf{x}_1}^{(ik,j\ell)}=r_{\mathbf{x}_1}^{(ik)}=r_{\mathbf{x}_1}^{(j\ell)}$ and $r_{\mathbf{x}_2}^{(ik,j\ell)}>r_{\mathbf{x}_2}^{(ik)},r_{\mathbf{x}_2}^{(j\ell)}$:
\begin{equation}
m_2 > \min \{ r_{\mathbf{x}_2}^{(ik)},r_{\mathbf{x}_2}^{(j\ell)}  \}  \quad \mathrm{or}  \quad 
\left\{
\begin{array}{lll}
m_2> \min \{  r_{\mathbf{x}}^{(ik)} -r_{\mathbf{x}_1}^{(ik)},  r_{\mathbf{x}}^{(j\ell)} -r_{\mathbf{x}_1}^{(j\ell)}   \} \\
m_1+ m_2> \min \{  r_{\mathbf{x}}^{(ik)} ,  r_{\mathbf{x}}^{(j\ell)}  \} \\
\end{array}
\right. .
\label{eq:cor3}
\end{equation}
\end{enumerate}


\end{corollary}
\begin{IEEEproof}
See Appendix~\ref{app:C}.
\end{IEEEproof}

\begin{figure}
\begin{center}
\subfigure[]{\input{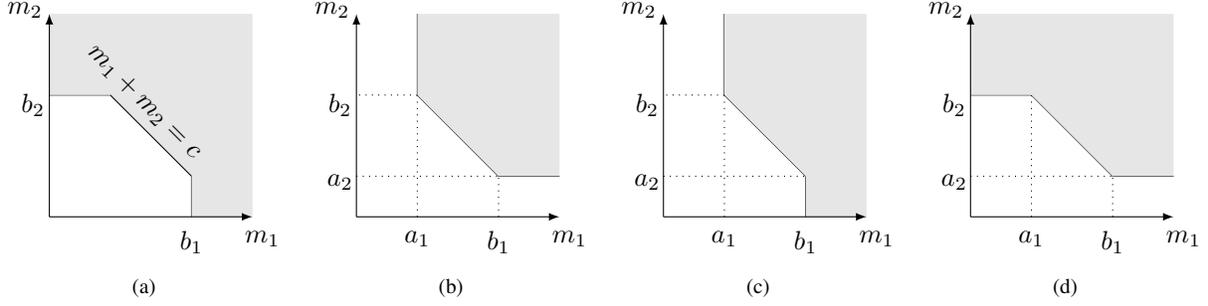}}
\subfigure[]{
\Assi[b;2.7cm,2.7cm](0,0)(1,1){
\wt(-0.12,1.1;t){$m_2$}
\wt(1.05,-0.03;t){$m_1$}
\wt(0.7,-0.03;t){$b_1$}
\wt(0.3,-0.03;t){$a_1$}
\wt(-0.08,0.25;t){$a_2$}
\wt(-0.08,0.65;t){$b_2$}
\draw (0.3,0.6) -- (0.7,0.2);
\draw [dotted] (0.3,0.6) -- (0,0.6);
\draw [dotted](0.7,0) -- (0.7,0.2);
\draw[dotted] (0.3,0)-- (0.3,0.6);
\draw[dotted] (0,0.2)-- (0.7,0.2);
\draw (0.3,0.6) -- (0.3,1);
\draw (0.7,0.2) -- (1,0.2);
\fill[fill=gray!20] (0.3,1) -- (1,1) -- (1,0.2) -- (0.7,0.2) -- (0.3,0.6) -- cycle;
} 
}
\subfigure[]{
\Assi[b;2.7cm,2.7cm](0,0)(1,1){
\wt(-0.12,1.1;t){$m_2$}
\wt(1.05,-0.03;t){$m_1$}
\wt(0.7,-0.03;t){$b_1$}
\wt(0.3,-0.03;t){$a_1$}
\wt(-0.08,0.25;t){$a_2$}
\wt(-0.08,0.65;t){$b_2$}
\draw (0.3,0.6) -- (0.7,0.2);
\draw [dotted] (0.3,0.6) -- (0,0.6);
\draw (0.7,0) -- (0.7,0.2);
\draw[dotted] (0.3,0)-- (0.3,0.6);
\draw[dotted] (0,0.2)-- (0.7,0.2);
\draw (0.3,0.6) -- (0.3,1);
\fill[fill=gray!20] (0.3,1) -- (1,1) -- (1,0) -- (0.7,0) -- (0.7,0.2) -- (0.3,0.6) -- cycle;
} 
}
\subfigure[]{
\Assi[b;2.7cm,2.7cm](0,0)(1,1){
\wt(-0.12,1.1;t){$m_2$}
\wt(1.05,-0.03;t){$m_1$}
\wt(0.7,-0.03;t){$b_1$}
\wt(0.3,-0.03;t){$a_1$}
\wt(-0.08,0.25;t){$a_2$}
\wt(-0.08,0.65;t){$b_2$}
\draw (0.3,0.6) -- (0.7,0.2);
\draw (0.3,0.6) -- (0,0.6);
\draw [dotted](0.7,0) -- (0.7,0.2);
\draw[dotted] (0.3,0)-- (0.3,0.6);
\draw[dotted] (0,0.2)-- (0.7,0.2);
\draw (0.7,0.2) -- (1,0.2);
\fill[fill=gray!20]  (1,1) -- (1,0.2) -- (0.7,0.2) -- (0.3,0.6) -- (0,0.6) -- (0,1) -- cycle;
} 
}
\caption{Representation of the conditions on $m_1$ and $m_2$ for \rev{$\lim_{\sigma^2\to 0}\bar{P}\sub e(\sigma^2)=0$}, for the 4 different cases encapsulated in Corollary \ref{cor:phasetrans}. In all cases $a_1=\min \{  r_{\mathbf{x}}^{(ik)} -r_{\mathbf{x}_2}^{(ik)},  r_{\mathbf{x}}^{(j\ell)} -r_{\mathbf{x}_2}^{(j\ell)}   \}+1, b_1=\min \{ r_{\mathbf{x}_1}^{(ik)},r_{\mathbf{x}_1}^{(j\ell)}  \}+1, a_2=\min \{  r_{\mathbf{x}}^{(ik)} -r_{\mathbf{x}_1}^{(ik)},  r_{\mathbf{x}}^{(j\ell)} -r_{\mathbf{x}_1}^{(j\ell)}   \}+1$, $b_2= \min \{ r_{\mathbf{x}_2}^{(ik)},r_{\mathbf{x}_2}^{(j\ell)}  \}+1$ and $c=\min \{  r_{\mathbf{x}}^{(ik)} ,  r_{\mathbf{x}}^{(j\ell)}  \} +1$. The shaded regions represent values of $m_1$ and $m_2$ that satisfy the conditions (\ref{eq:cor1})--(\ref{eq:cor3}).}
\label{fig:CorRegions}
\end{center}
\end{figure}

The characterization of the numbers of features $m_1$ and $m_2$ that are both necessary and sufficient \rev{to drive the upper bound to the misclassification probability to zero in the the low-rank regime} is divided in 4 cases, depending on whether the range spaces $\mathrm{Im}(\bar{\mathbf{\Sigma}}_{\mathbf{x}_1}^{(ik)})$ and $\mathrm{Im}(\bar{\mathbf{\Sigma}}_{\mathbf{x}_1}^{(j\ell)})$, or the range spaces $\mathrm{Im}(\bar{\mathbf{\Sigma}}_{\mathbf{x}_2}^{(ik)})$ and $\mathrm{Im}(\bar{\mathbf{\Sigma}}_{\mathbf{x}_2}^{(j\ell)})$,  are distinct or not\footnote{We recall that, given two positive semidefinite matrices $\mathbf{A}$ and $\mathbf{B}$ with ranks $r_{\mathbf{A}}= \rank (\mathbf{A}), r_{\mathbf{B}}= \rank (\mathbf{B}), r_{\mathbf{AB}}= \rank (\mathbf{A} + \mathbf{B})$, $\mathrm{Im}(\mathbf{A})=\mathrm{Im}(\mathbf{B})$ if and only if $r_{\mathbf{A}\mathbf{B}}= \frac{r_{\mathbf{A}}+r_{\mathbf{B}}}{2}$~\cite[Lemma 2]{RecJournal} and then, if and only if $r_{\mathbf{A}\mathbf{B}}=r_{\mathbf{A}}=r_{\mathbf{B}}$.}. Fig.~\ref{fig:CorRegions} depicts the tradeoff between the values of $m_1$ and $m_2$ associated with these different cases. Note also that the values of $m_1$ and $m_2$ \rev{needed for the upper bound of the misclassification probability to approach zero when $\sigma^2\to 0$} lie in the intersection of the regions corresponding to index quadruples $(i,k,j,\ell) \in \mathcal{S}\sub{SIC}$. 

In case 1), the range spaces associated to the input covariance matrices are all distinct, and by observing (\ref{eq:cor1}) we can clearly determine the beneficial effect of the correlation between $\mathbf{x}_1$ and $\mathbf{x}_2$ in guaranteeing \rev{reliable classification.} Namely, we note that \rev{the upper bound to the misclassification probability reaches zero in the low-rank regime} either when error-free classification is possible from the observation of $\mathbf{y}_1$ alone ($m_1 > \min \{r_{\mathbf{x}_1}^{(ik)},r_{\mathbf{x}_1}^{(j\ell)}\}$) or from the observation of $\mathbf{y}_2$ alone ($m_2 > \min \{r_{\mathbf{x}_2}^{(ik)},r_{\mathbf{x}_2}^{(j\ell)}\}$) cf.~\cite{Reboredo14}, but, more importantly, the condition $m_1+m_2 > \min \{r_{\mathbf{x}}^{(ik)},r_{\mathbf{x}}^{(j\ell)}\}$ shows the benefit of side information in order to obtain the \rev{reliable classification} with a lower number of features. In fact, when $r_{\mathbf{x}}^{(ik)} < r_{\mathbf{x}_1}^{(ik)}+ r_{\mathbf{x}_2}^{(ik)}$, joint classification of $\mathbf{y}_1$ and $\mathbf{y}_2$ leads to a clear advantage in the number of features needed to achieve \rev{zero error probability in the low-rank regime} with respect to the case in which classification is carried independently from $\mathbf{y}_1$ and $\mathbf{y}_2$, despite the fact that linear features are extracted independently from $\mathbf{x}_1$ and $\mathbf{x}_2$.

In case 2), the range spaces associated to the input covariance matrices are such that  $\mathrm{Im}(\bar{\mathbf{\Sigma}}_{\mathbf{x}_1}^{(ik)})=\mathrm{Im}(\bar{\mathbf{\Sigma}}_{\mathbf{x}_1}^{(j\ell)})$ and $\mathrm{Im}(\bar{\mathbf{\Sigma}}_{\mathbf{x}_2}^{(ik)})=\mathrm{Im}(\bar{\mathbf{\Sigma}}_{\mathbf{x}_2}^{(j\ell)})$ so that classification based on the observation of $\mathbf{y}_1$ or $\mathbf{y}_2$ alone yields an error floor in the upper bound of the misclassification probability~\cite{Reboredo14}. In other terms, input signals and side information signals from classes $(i,k)$ and $(j,\ell)$ are never perfectly distinguishable. In this case, the impact of correlation between the input signal and the side information signal is clear when observing (\ref{eq:cor4}). In fact, when combining features extracted independently from the vectors $\mathbf{x}_1$ and $\mathbf{x}_2$, it is possible to drive to zero the misclassification probability, in the low-rank regime, provided that the number of features extracted $m_1$ and $m_2$ verify the conditions in (\ref{eq:cor4}). 

Finally, cases 3) and 4) represent intermediate scenarios in which range spaces associated to $\mathbf{x}_1$ are distinct, but those related to $\mathbf{x}_2$ are completely overlapping, and vice versa. We note then how the necessary and sufficient conditions \rev{to drive to zero the upper bound of the misclassification probability} in (\ref{eq:cor2}) and (\ref{eq:cor3}) are given by combinations of the conditions in (\ref{eq:cor1}) and (\ref{eq:cor4}).

We further note in passing that the conditions in (\ref{eq:cor4}) are reminiscent of the conditions on compression rates for lossless joint source coding in \cite{Slepian73}.


\subsection{Nonzero-Mean Classes}

We now provide a low-rank expansion of the upper bound to the misclassification probability associated with the feature extraction system with side information in (\ref{eq:sys1}) and (\ref{eq:sys2}),  for the case of nonzero-mean classes, i.e., $\boldsymbol{\mu}_{\mathbf{x}}^{(ik)} \neq\mathbf{0}$. 
The presence of non-zero mean classes -- as already noted in \cite[Theorem 3]{Reboredo14}, for compressive classification without side information -- offers a unique characteristic, that is, the misclassification probability can decay exponentially with $1/\sigma^2$ (i.e., the diversity-order tends to infinity) under certain conditions on the number of linear features extracted and the geometrical description of the source. 
\begin{theorem}
\label{theo:nonzero_SI}
Consider the model in (\ref{eq:sys1}) and (\ref{eq:sys2}), where the input signal $\mathbf{x}_1$ is drawn according to the class-conditioned distribution (\ref{eq:x1c1}), the side information $\mathbf{x}_2$ is drawn according to the class-conditioned distribution (\ref{eq:x2c2}), and the class-conditioned joint distribution of $\mathbf{x}_1$ and $\mathbf{x}_2$ is given by (\ref{eq:JointG}).

If, for all the index quadruples $(i,k,j,\ell) \in \mathcal{S}\sub{SIC}$ it holds, $\boldsymbol{\mu}_{\mathbf{x}}^{(ik)}-\boldsymbol{\mu}_{\mathbf{x}}^{(j\ell)}  \notin \mathrm{Im} (\bar{\mathbf{\Sigma}}_{\mathbf{x}}^{(ik)}+\bar{\mathbf{\Sigma}}_{\mathbf{x}}^{(j\ell)})$, then, with probability 1, in the low-rank regime, i.e., when $\sigma^2 \to 0$, the upper bound to the misclassification probability for classification with side information (\ref{eq:BhattM1union}) can be expanded as
\begin{equation}
\bar{P}\sub{err}(\sigma^2) = B \cdot e^{- C / \sigma^2} + o\left( e^{- C / \sigma^2}   \right),
\end{equation}
for fixed constants $B,C >0$, if and only if the following conditions hold $\forall (i,k,j,\ell) \in \mathcal{S}\sub{SIC}$:
\begin{enumerate}
\item if $\boldsymbol{\mu}_{\mathbf{x}_1}^{(ik)}-\boldsymbol{\mu}_{\mathbf{x}_1}^{(j\ell)}  \notin \mathrm{Im} (\bar{\mathbf{\Sigma}}_{\mathbf{x}_1}^{(ik)}+\bar{\mathbf{\Sigma}}_{\mathbf{x}_1}^{(j\ell)})$ and $\boldsymbol{\mu}_{\mathbf{x}_2}^{(ik)}-\boldsymbol{\mu}_{\mathbf{x}_2}^{(j\ell)}  \notin \mathrm{Im} (\bar{\mathbf{\Sigma}}_{\mathbf{x}_2}^{(ik)}+\bar{\mathbf{\Sigma}}_{\mathbf{x}_2}^{(j\ell)})$: 
\begin{equation}
m_1 > r_{\mathbf{x}_1}^{(ik,j\ell)} \quad \mathrm{or} \quad m_2 > r_{\mathbf{x}_2}^{(ik,j\ell)} \quad \mathrm{or} \quad  
m_1+m_2 > r_{\mathbf{x}}^{(ik,j\ell)};
\label{eq:nonzero1}
\end{equation}
\item if $\boldsymbol{\mu}_{\mathbf{x}_1}^{(ik)}-\boldsymbol{\mu}_{\mathbf{x}_1}^{(j\ell)}  \in \mathrm{Im} (\bar{\mathbf{\Sigma}}_{\mathbf{x}_1}^{(ik)}+\bar{\mathbf{\Sigma}}_{\mathbf{x}_1}^{(j\ell)})$ and $\boldsymbol{\mu}_{\mathbf{x}_2}^{(ik)}-\boldsymbol{\mu}_{\mathbf{x}_2}^{(j\ell)}  \in \mathrm{Im} (\bar{\mathbf{\Sigma}}_{\mathbf{x}_2}^{(ik)}+\bar{\mathbf{\Sigma}}_{\mathbf{x}_2}^{(j\ell)})$:
\begin{equation}
\left\{
\begin{array}{lll}
m_1>   r_{\mathbf{x}}^{(ik,j\ell)} -r_{\mathbf{x}_2}^{(ik,j\ell)} \\
m_2>  r_{\mathbf{x}}^{(ik,j\ell)} -r_{\mathbf{x}_1}^{(ik,j\ell)} \\
m_1+ m_2> r_{\mathbf{x}}^{(ik,j\ell)} \\
\end{array}
\right. ;
\end{equation}
\item if $\boldsymbol{\mu}_{\mathbf{x}_1}^{(ik)}-\boldsymbol{\mu}_{\mathbf{x}_1}^{(j\ell)}  \notin \mathrm{Im} (\bar{\mathbf{\Sigma}}_{\mathbf{x}_1}^{(ik)}+\bar{\mathbf{\Sigma}}_{\mathbf{x}_1}^{(j\ell)})$ and $\boldsymbol{\mu}_{\mathbf{x}_2}^{(ik)}-\boldsymbol{\mu}_{\mathbf{x}_2}^{(j\ell)}  \in \mathrm{Im} (\bar{\mathbf{\Sigma}}_{\mathbf{x}_2}^{(ik)}+\bar{\mathbf{\Sigma}}_{\mathbf{x}_2}^{(j\ell)})$:
\begin{equation}
m_1 > r_{\mathbf{x}_1}^{(ik,j\ell)}  \quad \mathrm{or}  \quad 
\left\{
\begin{array}{lll}
m_1> r_{\mathbf{x}}^{(ik,j\ell)} -r_{\mathbf{x}_2}^{(ik,j\ell)} \\
m_1+ m_2>   r_{\mathbf{x}}^{(ik,j\ell)}  \\
\end{array}
\right. ;
\end{equation}
\item if $\boldsymbol{\mu}_{\mathbf{x}_1}^{(ik)}-\boldsymbol{\mu}_{\mathbf{x}_1}^{(j\ell)}  \in \mathrm{Im} (\bar{\mathbf{\Sigma}}_{\mathbf{x}_1}^{(ik)}+\bar{\mathbf{\Sigma}}_{\mathbf{x}_1}^{(j\ell)})$ and $\boldsymbol{\mu}_{\mathbf{x}_2}^{(ik)}-\boldsymbol{\mu}_{\mathbf{x}_2}^{(j\ell)}  \notin \mathrm{Im} (\bar{\mathbf{\Sigma}}_{\mathbf{x}_2}^{(ik)}+\bar{\mathbf{\Sigma}}_{\mathbf{x}_2}^{(j\ell)})$:
\begin{equation}
m_2 > r_{\mathbf{x}_2}^{(ik,j\ell)} \quad \mathrm{or}  \quad 
\left\{
\begin{array}{lll}
m_2>  r_{\mathbf{x}}^{(ik,j\ell)} -r_{\mathbf{x}_1}^{(ik,j\ell)}    \\
m_1+ m_2>   r_{\mathbf{x}}^{(ik,j\ell)}\\
\end{array}
\right. .
\label{eq:nonzero4}
\end{equation}
\end{enumerate}
Otherwise, denote by $\mathcal{S}'$ the set of quadruples $(i,k,j,\ell) \in \mathcal{S}\sub{SIC}$ for which either $\boldsymbol{\mu}_{\mathbf{x}}^{(ik)}-\boldsymbol{\mu}_{\mathbf{x}}^{(j\ell)}  \in \mathrm{Im} (\bar{\mathbf{\Sigma}}_{\mathbf{x}}^{(ik)}+\bar{\mathbf{\Sigma}}_{\mathbf{x}}^{(j\ell)})$ or conditions (\ref{eq:nonzero1})--(\ref{eq:nonzero4})  do not hold. Then, with probability 1, in the low-rank regime, i.e., when $\sigma^2 \to 0$, the upper bound to the misclassification probability for classification with side information (\ref{eq:BhattM1union}) can be expanded as
\begin{equation}
\bar{P}\sub{err}(\sigma^2) = A \cdot (\sigma^2)^{d} + o\left((\sigma^2)^{d}\right),
\end{equation}
for a fixed constant $A>0$, and
\begin{equation}
d =  \min_{(i,k,j,\ell) \in \mathcal{S}' }d(ik, j\ell),
\end{equation}
where $d(ik,j\ell)$ is obtained as in Theorem \ref{theo:zeromean}.
\end{theorem}
\begin{IEEEproof}
See Appendix~\ref{app:D}.
\end{IEEEproof}

Note that classification based on the joint observation of $\mathbf{y}_1$ and $\mathbf{y}_2$ can guarantee infinite diversity-order even when classification based on $\mathbf{y}_1$ or $\mathbf{y}_2$ alone cannot. In particular, if there exists an index quadruple for which both $\boldsymbol{\mu}_{\mathbf{x}_1}^{(ik)}-\boldsymbol{\mu}_{\mathbf{x}_1}^{(j\ell)}  \in \mathrm{Im} (\bar{\mathbf{\Sigma}}_{\mathbf{x}_1}^{(ik)}+\bar{\mathbf{\Sigma}}_{\mathbf{x}_1}^{(j\ell)})$ and $\boldsymbol{\mu}_{\mathbf{x}_2}^{(ik)}-\boldsymbol{\mu}_{\mathbf{x}_2}^{(j\ell)}  \in \mathrm{Im} (\bar{\mathbf{\Sigma}}_{\mathbf{x}_2}^{(ik)}+\bar{\mathbf{\Sigma}}_{\mathbf{x}_2}^{(j\ell)})$, then, irrespective of the number of features $m_1$ and $m_2$ and of the specific values of the projection kernels $\mathbf{\Phi}_1$ and $\mathbf{\Phi}_2$, we have
\begin{equation}
\mathbf{\Phi}_1( \boldsymbol{\mu}_{\mathbf{x}_1}^{(ik)}-\boldsymbol{\mu}_{\mathbf{x}_1}^{(j\ell)} ) \in \mathrm{Im} (\mathbf{\Phi}_1 (\bar{\mathbf{\Sigma}}_{\mathbf{x}_1}^{(ik)}+\bar{\mathbf{\Sigma}}_{\mathbf{x}_1}^{(j\ell)}) \mathbf{\Phi}_1\tra)
\qv
\mathbf{\Phi}_2 (\boldsymbol{\mu}_{\mathbf{x}_2}^{(ik)}-\boldsymbol{\mu}_{\mathbf{x}_2}^{(j\ell)} ) \in \mathrm{Im} (\mathbf{\Phi}_2 (\bar{\mathbf{\Sigma}}_{\mathbf{x}_2}^{(ik)}+\bar{\mathbf{\Sigma}}_{\mathbf{x}_2}^{(j\ell)}) \mathbf{\Phi}_2\tra)
\end{equation}
and, therefore, the conditions in \cite[Theorem 3]{Reboredo14} are not verified, thus implying that both the upper bounds to the error probability associated to  classification based on $\mathbf{y}_1$ or $\mathbf{y}_2$ do not decay exponentially with $1/\sigma^2$ when $\sigma^2 \to 0$. On the other hand, if $\boldsymbol{\mu}_{\mathbf{x}}^{(ik)}-\boldsymbol{\mu}_{\mathbf{x}}^{(j\ell)}  \notin \mathrm{Im} (\bar{\mathbf{\Sigma}}_{\mathbf{x}}^{(ik)}+\bar{\mathbf{\Sigma}}_{\mathbf{x}}^{(j\ell)})$ for all index quadruples $(i,k,j,\ell) \in \mathcal{S}\sub{SIC}$, then classification based on both $\mathbf{y}_1$ and $\mathbf{y}_2$ is characterized by an exponential decay of the upper bound to the misclassification probability, provided that  conditions (\ref{eq:nonzero1})--(\ref{eq:nonzero4}) on the numbers of features extracted from $\mathbf{x}_1$ and $\mathbf{x}_2$ are verified.

Moreover, the conditions on the number of features needed to achieve an exponential decay in $1/\sigma^2$ of the upper bound to the misclassification probability depend on whether the affine spaces spanned by signal and side information realization in the Gaussian classes $(i,k)$ and $(j,\ell)$ do intersect or not, for all index quadruples $(i,k,j,\ell) \in \mathcal{S}\sub{SIC}$. From a geometrical point of view, if the affine spaces spanned by the overall signal $\mathbf{x}$ obtained by the concatenation of input signal and side information do not intersect, then equations (\ref{eq:nonzero1})--(\ref{eq:nonzero4}) determine conditions on the number of extracted features $m_1$ and $m_2$ such that the affine spaces spanned by the projected signals $\mathbf{\Phi}\mathbf{x}$ do not intersect as well, thus guaranteeing enhanced discrimination among classes.

\section{Reconstruction with Side Information}
\label{par:reconstruction}

\rev{We now consider signal reconstruction in the presence of side information. In particular, by leveraging the classification results, we will address two scenarios: i) the case where the signals obey asymptotically a low-rank model; and ii) the case where the signals obey an approximately low-rank model that is often used in practice \cite{Chen10,RecJournal}.
%
%
%
}

{\color{black}
\subsection{Low-Rank Model}
}
\rev{We focus first on the analysis of the asymptotic regime when $\sigma_1^2, \sigma_2^2 \to 0$. In this case, without loss of generality, we assume $\sigma_1^2=\sigma_2^2 =\sigma^2$.} 
We are interested in the asymptotic characterization of the \ac{MMSE} incurred in reconstructing $\mathbf{x}_1$ from the observation of the signal features $\mathbf{y}_1$ and the side information features $\mathbf{y}_2$, given by\footnote{We emphasize that $\MMSE_{1|1,2} (\sigma^2)$ is a function of $\sigma^2$.}
\begin{equation}
\MMSE_{1|1,2}(\sigma^2)  = \E{\|  \mathbf{x}_1 - \hat{\mathbf{x}}_1(\mathbf{y}_1,\mathbf{y}_2)  \|^{2}},
\label{eq:MMSE1}
\end{equation}
where $\hat{\mathbf{x}}_1(\mathbf{y}_1,\mathbf{y}_2)$ is the conditional mean estimator in (\ref{eq:CondMeanEst}). 
In particular, we are interested in determining conditions on the number of linear features $m_1$ and $m_2$ that guarantee perfect reconstruction in the low-rank regime, i.e., when $\sigma^2 \to 0$, that is
\begin{equation}
\lim_{\sigma^2 \to 0} \MMSE_{1|1,2} (\sigma^2) =0,
\end{equation}
thus generalizing the results in \cite{RecJournal} to the case when side information is available at the decoder; the misclassification results will be key to address this problem.

\subsubsection{Gaussian Sources}
\label{par:recG}

We first consider the simplified case in which $K_1=K_2=1$, i.e., when the signals $\mathbf{x}_1$ and $\mathbf{x}_2$ obey the joint Gaussian distribution $\mathcal{N}(\boldsymbol{\mu}_{\mathbf{x}}, {\mathbf{\Sigma}}_{\mathbf{x}})$, where
\rev{
\begin{equation}
\boldsymbol{\mu}_{\mathbf{x}} = \left[
\begin{array}{cc}
\boldsymbol{\mu}_{\mathbf{x}_1} \\
\boldsymbol{\mu}_{\mathbf{x}_2}
\end{array}
\right]
\qv
\mathbf{\Sigma}_{\mathbf{x}}=\bar{\mathbf{\Sigma}}_{\mathbf{x}} +\sigma^2 \mathbf{I}= \left[
\begin{array}{ccc}
\bar{\mathbf{\Sigma}}_{\mathbf{x}_1} & \bar{\mathbf{\Sigma}}_{\mathbf{x}_{12}}    \\
  \bar{\mathbf{\Sigma}}_{\mathbf{x}_{21}}   & \bar{\mathbf{\Sigma}}_{\mathbf{x}_2}
\end{array}
\right]
+
\left[
\begin{array}{ccc}
\sigma^2 \mathbf{I}& \mathbf{0}    \\
  \mathbf{0}    & \sigma^2 \mathbf{I}
  \end{array}
\right],
\label{eq:Gdist}
\end{equation}}
and with ranks $r_{\mathbf{x}_1 } = \rank (\bar{\mathbf{\Sigma}}_{\mathbf{x}_1}),r_{\mathbf{x}_2 } = \rank (\bar{\mathbf{\Sigma}}_{\mathbf{x}_2}) $ and $r_{\mathbf{x}} = \rank (\bar{\mathbf{\Sigma}}_{\mathbf{x}})$.

For this case, the conditional mean estimator is given by~\cite{Hassibi}
\begin{equation}
\hat{\mathbf{x}}_1(\mathbf{y}) = \boldsymbol{\mu}_{\mathbf{x}_1} + \mathbf{W}_{\mathbf{x}_1}\left(
\mathbf{y}-\mathbf{\Phi} 
\boldsymbol{\mu}_{\mathbf{x}}
\right),
\end{equation}
where
\rev{
\begin{equation}
\mathbf{W}_{\mathbf{x}_1} = \left[  (\bar{\mathbf{\Sigma}}_{\mathbf{x}_1} + \sigma^2 \mathbf{I} )\  \bar{\mathbf{\Sigma}}_{\mathbf{x}_{12}} \right] \mathbf{\Phi}\tra \left( \sigma^2 \mathbf{I}  + \mathbf{\Phi} \bar{\mathbf{\Sigma}}_{\mathbf{x}} \mathbf{\Phi}\tra   \right)^{-1}.
\end{equation}}
Moreover, the \ac{MMSE} in this case can be expressed as
\rev{
\begin{equation}
\MMSE_{1|1,2}^{\sf G} (\sigma^2) = \tr \left(  \bar{\mathbf{\Sigma}}_{\mathbf{x}_1} + \sigma^2 \mathbf{I} -  \left[  (\bar{\mathbf{\Sigma}}_{\mathbf{x}_1} ++ \sigma^2 \mathbf{I})  \  \bar{\mathbf{\Sigma}}_{\mathbf{x}_{12}}  \right] \mathbf{\Phi}\tra   \left( \sigma^2 \mathbf{I}  + \mathbf{\Phi} \bar{\mathbf{\Sigma}}_{\mathbf{x}} \mathbf{\Phi}\tra   \right)^{-1}
\mathbf{\Phi}
 \left[
\begin{array}{cc}
(\bar{\mathbf{\Sigma}}_{\mathbf{x}_1}  + \sigma^2 \mathbf{I})\tra \\
\bar{\mathbf{\Sigma}}_{\mathbf{x}_2}\tra
\end{array}
\right]
   \right).
   \label{eq:MMSEGsi}
\end{equation}}

In the following, we provide necessary and sufficient conditions on the number of features $m_1,m_2$ that guarantee that, in the low-rank regime, the reconstruction \ac{MMSE} for Gaussian sources approaches zero. Sufficient conditions are based on the analysis of two different upper bounds to $\MMSE_{1|1,2}^{\sf G}(\sigma^2)$. The first upper bound is obtained by considering the \ac{MMSE} associated with the reconstruction of the signal $\mathbf{x}_1$ from the observation of $\mathbf{y}_1$ alone, i.e., without side information, which is denoted by
\begin{equation}
\MMSE_{1|1}^{\sf G}(\sigma^2) = \E{  \|  \mathbf{x}_1 - \hat{\mathbf{x}}_1(\mathbf{y}_1)  \|^2},
\end{equation}
where $\hat{\mathbf{x}}_1(\mathbf{y}_1) = \E{\mathbf{x}_1 | \mathbf{y}_1}$ and whose behavior in the low-rank regime has been analyzed in \cite{RecJournal}.\rev{\footnote{\rev{In fact, the analysis in \cite{RecJournal} is based on a slightly different framework, where signals $\mathbf{x}_1$ are described by exactly low-rank models, and the features $\mathbf{y}_1$ are affected by additive Gaussian noise. Nevertheless, it will be shown in the following that the results presented in \cite{RecJournal} on necessary and sufficient conditions for reliable reconstruction generalize to the framework considered in this paper.}} }

The second upper bound is obtained by considering the \ac{MMSE} associated with the distributed reconstruction problem, i.e., the joint recovery of $\mathbf{x}_1$ and $\mathbf{x}_2$ from the observation of both $\mathbf{y}_1$ and $\mathbf{y}_2$ (i.e., the reconstruction of $\mathbf{x}$ from $\mathbf{y}$), which is denoted by
\begin{equation}
\MMSE_{1,2|1,2}^{\sf G}(\sigma^2) = \E{  \|  \mathbf{x} - \hat{\mathbf{x}}(\mathbf{y})  \|^2},
\end{equation}
where
\begin{equation}
\hat{\mathbf{x}}(\mathbf{y}) = \E{\mathbf{x}  |  \mathbf{y}} = \int \mathbf{x}  \, p(\mathbf{x} | \mathbf{y})  d\mathbf{x}.
\label{eq:cond_mean_distributed}
\end{equation}
Note that the analysis of the second upper bound cannot be directly performed on the basis of the results in \cite{RecJournal}, due to the particular block diagonal structure of $\mathbf{\Phi}$. 

Based on the properties of the \ac{MMSE} \cite{Hassibi}, it is straightforward to show that $\MMSE_{1|1,2}^{\sf G}(\sigma^2) \leq \MMSE_{1|1}^{\sf G}(\sigma^2)$  and $\MMSE_{1|1,2}^{\sf G}(\sigma^2) \leq \MMSE_{1,2|1,2}^{\sf G}(\sigma^2)$. 

On the other hand, necessary conditions are derived from the analysis of the lower bound to the \ac{MMSE} obtained by feeding the decoder not only with the noisy features $\mathbf{y}_1$ and $\mathbf{y}_2$, but also with the values of the realizations of the vectors $\mathbf{w}_1$ and $\mathbf{w}_2$, \rev{that represent the deviation from an exactly low-rank model (see Section \ref{par:SigModel} for details).} The following theorem stems from the fact that the necessary and sufficient conditions for \rev{error free reconstruction in the low-rank regime} coincide.

\begin{theorem}
\label{theo:recG}
Consider the model in (\ref{eq:sys1}) and (\ref{eq:sys2}). Assume that the vectors $\mathbf{x}_1, \mathbf{x}_2$ are jointly Gaussian, with distribution $\mathcal{N}(\boldsymbol{\mu}_{\mathbf{x}}, {\mathbf{\Sigma}}_{\mathbf{x}})$, with mean and covariance matrix specified in (\ref{eq:Gdist}), and with $r_{\mathbf{x}_1 } = \rank (\bar{\mathbf{\Sigma}}_{\mathbf{x}_1}),r_{\mathbf{x}_2 } = \rank (\bar{\mathbf{\Sigma}}_{\mathbf{x}_2}) $ and $r_{\mathbf{x}} = \rank (\bar{\mathbf{\Sigma}}_{\mathbf{x}})$. 
Then, with probability 1, we have
\begin{equation}
\lim_{\sigma^2 \to 0} \MMSE_{1|1,2}^{\sf G}(\sigma^2)= 0
\Leftrightarrow 
m_1 \geq r_{\mathbf{x}_1} \quad \mathrm{or}  \quad 
\left\{
\begin{array}{lll}
m_1 \geq  r_{\mathbf{x}} -r_{\mathbf{x}_2}    \\
m_1+ m_2 \geq   r_{\mathbf{x}}\\
\end{array}
\right.  . 
\label{eq:recG}
\end{equation}
\end{theorem}
\begin{IEEEproof}
See Appendix~\ref{app:E}.
\end{IEEEproof}


Without side information, it is known that $m_1 \geq r_{\mathbf{x}_1}$ represents a necessary and sufficient condition on the number of features needed to drive the \ac{MMSE} to zero in the low-rank regime~\cite{RecJournal}. With side information, it is possible to reliably recover the input signal $\mathbf{x}_1$ with a lower number of features, as described by the conditions in (\ref{eq:recG}). In fact, whenever $r_{\mathbf{x}} < r_{\mathbf{x}_1} + r_{\mathbf{x}_2}$, it is possible to perfectly reconstruct $\mathbf{x}_1$ when $\sigma^2 \to 0$ even with less than $r_{\mathbf{x}_1}$ features, provided that $m_1 + m_2 \geq r_{\mathbf{x}}$. This happens when the dimension of the overall space spanned by the projected signals obtained by concatenating the input signal and the side information signal, i.e., $\mathbf{\Phi} \mathbf{x}$, is greater than or equal to the dimension of the space spanned by $\mathbf{x}$ in the signal domain. Moreover, the $m_1$ features extracted from $\mathbf{x}_1$ need to be enough to span a space with dimension equal to the difference between the dimension of the space spanned by $\mathbf{x}$ and that spanned by $\mathbf{x}_2$ alone. In this sense, linear projections extracted from the input signal must be enough to capture signal features that are characteristic of $\mathbf{x}_1$ and are not ``shared'' with $\mathbf{x}_2$, meaning that they are not correlated.

\begin{figure}
\begin{center}
\input{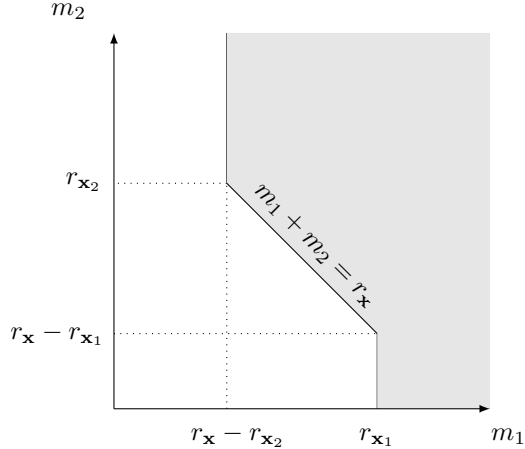}
\caption{Representation of the conditions on $m_1$ and $m_2$ for \rev{reliable reconstruction} for Gaussian sources. The shaded region represents values of $m_1$ and $m_2$ that satisfy the conditions (\ref{eq:recG}).}
\label{fig:FigrecG}
\end{center}
\end{figure}

The values of $m_1$ and $m_2$ that satisfy the necessary and sufficient conditions (\ref{eq:recG}) are reported in Fig.~\ref{fig:FigrecG}. 

\subsubsection{GMM Sources}
\label{par:recGMM}

We now consider the more general case where the signals $\mathbf{x}_1$ and $\mathbf{x}_2$ follow the models in Section~\ref{par:SigModel}. 
It is possible to express the conditional mean estimator in closed form, but not the \ac{MMSE}, which we denote by $\MMSE_{1|1,2}^{\sf GM}(\sigma^2)$. Therefore, we will determine necessary and sufficient conditions on the numbers of features $m_1$ and $m_2$ that guarantee $\MMSE_{1|1,2}^{\sf GM}(\sigma^2) \to 0$ in the low-rank regime, by leveraging the result in Theorem~\ref{theo:recG} together with steps akin to those in \cite[Section IV]{RecJournal}.


In order to provide sufficient conditions for the \ac{MMSE} to approach zero in the low-rank regime,  we analyze the upper bound to the \ac{MMSE} corresponding to the \ac{MSE} associated with a (sub-optimal) \emph{classify and reconstruct} decoder, which we denote by $\MSE^{\mathrm{CR}}(\sigma^2)$. This decoder operates in two steps as follows:
\begin{itemize}
\item First, the decoder estimates the pair of class indices associated to the input signal and the side information signal via the \ac{MAP} classifier\footnote{This \ac{MAP} classifier is associated with the distributed classification problem, \rev{which consists in estimating the labels $C_1$ and $C_2$ from the observation of the feature vectors $\mathbf{y}_1$ and $\mathbf{y}_2$.}} 
\begin{equation}
(\hat{C}_1,\hat{C}_2) = \arg \max_{(i,k)} p(  \mathbf{y}| C_1=i, C_2=k ) p_{C_1,C_2}(i,k);
\end{equation}
\item Second, in view of the fact that, conditioned on $(C_1,C_2)=(i,k)$, the vectors $\mathbf{x}_1 $ and $\mathbf{x}_2$ are jointly Gaussian distributed with mean $\boldsymbol{\mu}_{\mathbf{x}}^{(ik)}$ and covariance $\bar{\mathbf{\Sigma}}_{\mathbf{x}}^{(ik)}$, the decoder reconstructs the input signal $\mathbf{x}_1$ by using the conditional mean estimator corresponding to the estimated classes $\hat{C}_1, \hat{C}_2$
\begin{equation}
\hat{\mathbf{x}}_1(\mathbf{y}; C_1=\hat{C}_1,C_2=\hat{C}_2)  =  \boldsymbol{\mu}_{\mathbf{x}_1}^{(\hat{C}_1 \hat{C}_2)} + \mathbf{W}_{\mathbf{x}_1}^{(\hat{C}_1 \hat{C}_2)}\left(
\mathbf{y}
-\mathbf{\Phi} 
\boldsymbol{\mu}_{\mathbf{x}}^{(\hat{C}_1 \hat{C}_2)}
\right),
\end{equation}
where
\rev{
\begin{equation}
\mathbf{W}_{\mathbf{x}_1}^{(\hat{C}_1 \hat{C}_2)} = \left[ ( \bar{\mathbf{\Sigma}}_{\mathbf{x}_1}^{(\hat{C}_1 \hat{C}_2)} + \sigma^2 \mathbf{I}  ) \  \bar{\mathbf{\Sigma}}_{\mathbf{x}_{12}}^{(\hat{C}_1 \hat{C}_2)} \right] \mathbf{\Phi}\tra \left( \sigma^2 \mathbf{I}  + \mathbf{\Phi} \bar{\mathbf{\Sigma}}_{\mathbf{x}}^{(\hat{C}_1 \hat{C}_2)} \mathbf{\Phi}\tra   \right)^{-1}.
\label{eq:Wiener_indexed}
\end{equation}}
\end{itemize}

The optimality of the \ac{MMSE} estimator immediately implies that $\MMSE_{1|1,2}^{\sf GM}(\sigma^2) \leq \MSE^{\mathrm{CR}}(\sigma^2)$. Therefore, we can immediately leverage the analysis of the misclassification probability carried out in Section~\ref{par:PerrAnalysis} and the result in Theorem~\ref{theo:recG} in order to characterize the behavior of $\MSE^{\mathrm{CR}}(\sigma^2)$ in the low-rank regime, in order to determine sufficient condition for \rev{$\lim_{\sigma^2\to 0}\MMSE_{1|1,2}^{\sf GM}(\sigma^2)=0$.}

\begin{theorem}
\label{theo:recGMM}
Consider the model in (\ref{eq:sys1}) and (\ref{eq:sys2}). Assume that the input signal $\mathbf{x}_1$ is drawn according to the class-conditioned distribution (\ref{eq:x1c1}), $\mathbf{x}_2$ is drawn according to the class-conditioned distribution (\ref{eq:x2c2}) and the class-conditioned joint distribution of $\mathbf{x}_1$ and $\mathbf{x}_2$ is given by (\ref{eq:JointG}). Then, with probability 1, we have
\begin{equation}
m_1 > r_{\mathbf{x}_1}^{(ik)} \quad \mathrm{or}  \quad 
\left\{
\begin{array}{lll}
m_1 >  r_{\mathbf{x}}^{(ik)} -r_{\mathbf{x}_2}^{(ik)}    \\
m_1+ m_2 >   r_{\mathbf{x}}^{(ik)}\\
\end{array}
\right. , \forall (i,k) \in \mathcal{S}
\Rightarrow
\lim_{\sigma^2 \to 0} \MMSE_{1|1,2}^{\sf GM}(\sigma^2) =0.
\label{eq:recGMM}
\end{equation}
\end{theorem}
\begin{IEEEproof}
See Appendix~\ref{app:F}.
\end{IEEEproof}


The sufficient conditions in (\ref{eq:recGMM}) show that -- akin to the Gaussian case -- the numbers of features extracted from $\mathbf{x}_1$ and $\mathbf{x}_2$ have to be collectively greater than the largest among the dimensions of the spaces spanned by signals $\mathbf{x}=[\mathbf{x}_1\tra \ \mathbf{x}_2\tra]\tra$ in the Gaussian components corresponding to indices $(C_1,C_2)=(i,k)$, for $i=1,\ldots,K_1$, $k=1,\ldots,K_2$. Moreover, the features extracted from $\mathbf{x}_1$ need to be enough to capture signal components which are not correlated with the side information, for all Gaussian components. Finally, the condition $m_1 > r_{\mathbf{x}_1}^{(ik)}$ is obtained trivially by considering reconstruction of $\mathbf{x}_1$ from the features collected in the vector $\mathbf{y}_1$, thus disregarding side information.

Note that the values of $m_1$ and $m_2$ that are sufficient \rev{to drive the \ac{MMSE} to zero} are obtained by considering the intersection of regions akin to that in Fig.~\ref{fig:FigrecG} for all the pairs of classes $(i,k) \in \mathcal{S}$.

Appendix \ref{app:F} shows that the conditions in (\ref{eq:recGMM}) guarantee that the decoder can reliably estimate the class indices $(C_1,C_2)$ and hence reliably reconstruct the signal $\mathbf{x}_1$ in the low-rank regime. 


We now derive necessary conditions for \rev{reliable reconstruction} of \ac{GMM} signals with side information. We obtain such conditions from the analysis of a lower bound to the \ac{MMSE} that is obtained by observing that
\begin{IEEEeqnarray}{rCl}
\MMSE_{1|1,2}^{\sf GM}(\sigma^2) & = & \E{\| \mathbf{x}_1 - \hat{\mathbf{x}}_1(\mathbf{y}_1,\mathbf{y}_2)  \|^2} \\
\label{eq:total_prob}
&= & \sum_{(i,k) \in \mathcal{S}}  p_{C_1,C_2}(i,k)   \E{\| \mathbf{x}_1 - \hat{\mathbf{x}}_1(\mathbf{y}_1,\mathbf{y}_2)  \|^2   |   C_1=i, C_2=k}    \\
\label{eq:optMMSE}
& \geq & \sum_{(i,k) \in \mathcal{S}} p_{C_1,C_2}(i,k)    \MMSE_{1|1,2}^{{\sf G} (i,k)}(\sigma^2) = \MSE_{1|1,2}^{\mathrm{LB}}(\sigma^2),
\end{IEEEeqnarray}
where $ \MMSE_{1|1,2}^{{\sf G} (i,k)}(\sigma^2)$ denotes the \ac{MMSE} associated with the reconstruction of the Gaussian signal $\mathbf{x}_1$ corresponding to class indexes $(i,k)$ from the observation of the vector $\mathbf{y}_1$ and the side information $\mathbf{y}_2$. Note that the equality in (\ref{eq:total_prob}) is obtained via the total probability formula and the inequality in (\ref{eq:optMMSE}) is a consequence of the optimality of the \ac{MMSE} estimator for joint Gaussian input and side information signals.

The analysis of $\MSE_{1|1,2}^{\mathrm{LB}}(\sigma^2)$ leads to the derivation of the following necessary conditions on the number of features $m_1$
 and $m_2$ needed to drive $\MMSE_{1|1,2}^{\sf GM}(\sigma^2)$ to zero when $\sigma^2 \to 0$.
 
 \begin{theorem}
\label{theo:recGMM_nec}
Consider the model in (\ref{eq:sys1}) and (\ref{eq:sys2}). Assume that the input signal $\mathbf{x}_1$ is drawn according to the class-conditioned distribution (\ref{eq:x1c1}), $\mathbf{x}_2$ is drawn according to the class-conditioned distribution (\ref{eq:x2c2}) and the class-conditioned joint distribution of $\mathbf{x}_1$ and $\mathbf{x}_2$ is given by (\ref{eq:JointG}). Then, with probability 1, we have
\begin{equation}
\lim_{\sigma^2 \to 0} \MMSE_{1|1,2}^{\sf GM}(\sigma^2) =0
\Rightarrow
m_1 \geq r_{\mathbf{x}_1}^{(ik)} \quad \mathrm{or}  \quad 
\left\{
\begin{array}{lll}
m_1 \geq  r_{\mathbf{x}}^{(ik)} -r_{\mathbf{x}_2}^{(ik)}    \\
m_1+ m_2 \geq   r_{\mathbf{x}}^{(ik)}\\
\end{array}
\right. , \forall (i,k) \in \mathcal{S}.
\label{eq:recGMM_nec}
\end{equation}
\end{theorem}
\begin{IEEEproof}
The proof is based on the result in Theorem \ref{theo:recG}, which implies that, if $ \MMSE_{1|1,2}^{{\sf G} (i,k)}(\sigma^2)\to 0$ when $\sigma^2 \to 0$, $\forall (i,k) \in \mathcal{S}$, then, with probability 1, the conditions on the numbers of features $m_1$ and $m_2$ in (\ref{eq:recGMM_nec}) must be satisfied for all $(i,k) \in \mathcal{S}$.
\end{IEEEproof}

It is interesting to note that  the necessary conditions for \rev{reliable reconstruction} of \ac{GMM} inputs are one feature away from the corresponding sufficient conditions, akin to our previous results for the case without side information \cite{RecJournal}. In this way, Theorems \ref{theo:recGMM} and \ref{theo:recGMM_nec} provide a sharp characterization of the region associated to \rev{vanishing} \ac{MMSE} of \ac{GMM} inputs with side information \rev{in the low-rank regime}.

{\color{black}

\subsection{Approximately Low-Rank Model}
\label{par:GMM_approx}

We now consider the case when the signal of interest and the side information obey an approximately low-rank model, that is when both quantities $\sigma_1^2>0$ and $\sigma_2^2>0$. We are interested in determining the merit of side information in this case, therefore we consider expansions of the \ac{MMSE} as a function of $\sigma_1^2, \sigma_2^2$ for both cases when side information features $\mathbf{y}_2$ are available to the decoder or not.

We study first the behavior of the \ac{MMSE} without side information, i.e., $\MMSE_{1|1}(\sigma_1^2)$. The following lemma offers a characterization of the lower bound to $\MMSE_{1|1}(\sigma_1^2)$ obtained by noting that 
\begin{IEEEeqnarray}{rCl}
\MMSE_{1|1}(\sigma_1^2)  & = & \E{\| \mathbf{x}_1 - \hat{\mathbf{x}}_1(\mathbf{y}_1)  \|^2} \\
& = & \sum_{(i,k) \in \mathcal{S}}  p_{C_1,C_2}(i,k)   \E{\| \mathbf{x}_1 - \hat{\mathbf{x}}_1(\mathbf{y}_1)  \|^2   |   C_1=i, C_2=k}    \\
& \geq & \sum_{(i,k) \in \mathcal{S}} p_{C_1,C_2}(i,k)    \MMSE_{1|1}^{{\sf G} (i,k)}(\sigma_1^2) = \MSE_{1|1}^{\mathrm{LB}}(\sigma_1^2),
\end{IEEEeqnarray}
where $ \MMSE_{1|1}^{{\sf G} (i,k)}(\sigma_1^2)$ denotes the \ac{MMSE} associated with the reconstruction of the Gaussian signal $\mathbf{x}_1$ corresponding to class indexes $(i,k)$ from the observation of the vector $\mathbf{y}_1$. 
\begin{lemma}
\label{lem:lower_bound_nosi}
Consider the model in (\ref{eq:sys1}). Assume that the input signal $\mathbf{x}_1$ is drawn according to the class-conditioned distribution (\ref{eq:x1c1}). Then, when $\sigma_1^2 \to 0$, the MMSE lower bound $\MSE_{1|1}^{\mathrm{LB}}(\sigma_1^2)$ can be expanded as
\begin{equation}
\MSE^{\mathrm{LB}}_{1|1}(\sigma_1^2) =  \mathcal{M}_{1|1} + \mathcal{D}_{1|1} \cdot \sigma_1^2  + o(\sigma_1^2)
\label{eq:lower_bound_exp_nosi}
\end{equation}
where 
\begin{equation}
\mathcal{M}_{1|1} = \sum_{(i,k) \in \mathcal{S}} p_{C_1,C_2}(i,k)  \mathcal{M}_{1|1}^{(i,k)} \qv \mathcal{D}_{1|1} = \sum_{(i,k) \in \mathcal{S}} p_{C_1,C_2}(i,k)  \mathcal{D}_{1|1}^{(i,k)}.
\label{eq:MD_1_nosi}
\end{equation}
The terms $\mathcal{M}_{1|1}^{(i,k)}$ and $\mathcal{D}_{1|1}^{(i,k)}$ are obtained by considering the following eigenvalue decomposition:
 \begin{IEEEeqnarray}{rCl}
 \mathbf{\Xi}^{(ik)} = (\bar{\mathbf{\Sigma}}_{\mathbf{x}_1}^{(ik)})^{\frac{1}{2}} \mathbf{\Phi}_1\tra \mathbf{\Phi}_1 (\bar{\mathbf{\Sigma}}_{\mathbf{x}_1}^{(ik)})^{\frac{1}{2}} = \mathbf{U}_{\mathbf{\Xi}}^{(ik)} \mathbf{\Lambda}_{\mathbf{\Xi}}^{(ik)} (\mathbf{U}_{\mathbf{\Xi}}^{(ik)} )\tra.
 \label{eq:def_Xi}
 \end{IEEEeqnarray}
In particular, on writing $\mathbf{\Lambda}_{\mathbf{\Xi}}^{(ik)}  = \diag (  \lambda_{\mathbf{\Xi},1}^{(ik)}, \ldots, \lambda_{\mathbf{\Xi},r_{\mathbf{\Xi}}^{(ik)} }^{(ik)}, 0, \ldots, 0    )$, where $r_{\mathbf{\Xi}}^{(ik)} = \rank (\mathbf{\Xi}^{(ik)})$, and on denoting by $\mathbf{u}_{\mathbf{\Xi},t}^{(ik)}$ the $t$-th column of $\mathbf{U}_{\mathbf{\Xi}}^{(ik)} $, we have
\begin{IEEEeqnarray}{rCl}
\mathcal{M}_{1|1}^{(i,k)}  & = & \sum_{t= r_{\mathbf{\Xi}}^{(ik)}  +1}^{r_{\mathbf{x}_1}^{(ik)}}  (\mathbf{u}_{\mathbf{\Xi},t}^{(ik)})\tra \bar{\mathbf{\Sigma}}_{\mathbf{x}_1}^{(ik)} \mathbf{u}_{\mathbf{\Xi},t}^{(ik)} \\
\mathcal{D}_{1|1}^{(i,k)} & = & n_1 - m_1 -r_{\mathbf{\Xi}}^{(ik)}  + \sum_{t=1}^{r_{\mathbf{\Xi}}^{(ik)} }    \frac{1}{\lambda_{\mathbf{\Xi},t}^{(ik)}}   (\mathbf{u}_{\mathbf{\Xi},t}^{(ik)})\tra \bar{\mathbf{\Sigma}}_{\mathbf{x}_1}^{(ik)} \mathbf{u}_{\mathbf{\Xi},t}^{(ik)}.
\label{eq:MD_2_nosi}
\end{IEEEeqnarray}
\end{lemma}

\begin{IEEEproof}
See Appendix \ref{app:lower_bound_nosi}.
\end{IEEEproof}

The expansion of the lower bound $\MSE^{\mathrm{LB}}_{1|1}(\sigma_1^2) $, which is based on the results in \cite{RecJournal}, allows one to quantify the effect of small deviations from an exactly low-rank model on the reconstruction \ac{MMSE} of the signal of interest when side information is not available at the decoder. 

We can note that $\mathcal{M}_{1|1}>0$ if there exist indexes $(i,k) \in \mathcal{S}$ such that $m_1<r_{\mathbf{x}_1}^{(ik)} $. In this case, the zeroth-order term $\mathcal{M}_{1|1}$ represents the error floor of the lower bound of the \ac{MMSE}, which is achieved asymptotically when $\sigma_1^2 \to 0$. On the other hand, if $m_1 \geq r_{\mathbf{x}_1}^{(ik)}$ for all $(i,k) \in \mathcal{S}$, then $\mathcal{M}_{1|1}=0$, the lower bound of the \ac{MMSE} decays to zero as $1/\sigma_1^2$ when $\sigma_1^2 \to 0$, and the value $\mathcal{D}_{1|1}$ determines the horizontal offset of $\log \MSE^{\mathrm{LB}}_{1|1}(\sigma_1^2) $ (in a $\log \sigma_1^2$ scale).

The following lemma provides conditions that guarantee that the the expansion of the \ac{MMSE} lower bound in Lemma \ref{lem:lower_bound_nosi} is tight, thus it captures the behavior of the true \ac{MMSE} with respect to the deviation from an exactly low-rank model, expressed via the parameter $\sigma_1^2$. 

\begin{lemma}
\label{lem:upper_bound_nosi}
Consider the model in (\ref{eq:sys1}). Assume that the input signal $\mathbf{x}_1$ is drawn according to the class-conditioned distribution (\ref{eq:x1c1}). If $m_1$ is such that $d^{\mathrm{NOSI}}(ik,j\ell)>1, \forall (i,k,j,\ell) \in \mathcal{S}\sub{DC}$, where
\begin{equation}
d^{\mathrm{NOSI}}(ik,j\ell) = \frac{1}{2} \left(  \min \{m_1,r_{\mathbf{x}_1}^{(ik,j\ell)}\}    - \frac{  \min \{m_1,r_{\mathbf{x}_1}^{(ik)}\}+ \min \{m_1,r_{\mathbf{x}_1}^{(j\ell)}\}      }{2}     \right),
\label{eq:d_nosi_ikjl}
\end{equation}
then, when $\sigma_1^2 \to 0$, the \ac{MMSE} can be expanded as
\begin{equation}
\MMSE_{1|1}(\sigma_1^2) =  \mathcal{M}_{1|1} + \mathcal{D}_{1|1} \cdot \sigma_1^2  + o(\sigma_1^2),
\label{eq:upper_bound_exp_nosi}
\end{equation}
where $\mathcal{M}_{1|1} $ and $\mathcal{D}_{1|1}$ are given by \eqref{eq:MD_1_nosi}-\eqref{eq:MD_2_nosi}.
\end{lemma}
\begin{IEEEproof}
See Appendix \ref{app:upper_bound_nosi}.
\end{IEEEproof}

Note that the conditions stem from the analysis of a classify and reconstruct upper bound akin to to that described in Section~\ref{par:recGMM}, which leverages the characterization of the upper bound to the misclassification probability developed in Section \ref{par:PerrAnalysis}. In particular, such conditions guarantee that the error probability decays as $o(\sigma_1^2)$ when $\sigma_1^2 \to 0$. 
In fact, as will be confirmed by the numerical results presented in Section \ref{par:NumRes}, in certain regimes, the decay rate of the \ac{MMSE} function is dictated by the corresponding decay of the misclassification probability as a function of $\sigma_1^2$.

Consider now the case when side information is available at the decoder. The following lemma provides an expansion of the \ac{MMSE} lower bound in \eqref{eq:optMMSE}\footnote{In fact, the lower bound in \eqref{eq:optMMSE} was expressed as a function of $\sigma^2$, whereas in this case we express the lower bound in terms of $\sigma_1^2$ and $\sigma_2^2$, which can be different in general.} akin to the expansion \eqref{eq:lower_bound_exp_nosi} obtained for the case of reconstruction without side information.
\begin{lemma}
\label{lem:lower_bound}
Consider the model in (\ref{eq:sys1}) and (\ref{eq:sys2}). Assume that the input signal $\mathbf{x}_1$ is drawn according to the class-conditioned distribution (\ref{eq:x1c1}), $\mathbf{x}_2$ is drawn according to the class-conditioned distribution (\ref{eq:x2c2}) and the class-conditioned joint distribution of $\mathbf{x}_1$ and $\mathbf{x}_2$ is given by (\ref{eq:JointG}). Then, when $\sigma_1^2 \to 0$, the lower bound $\MSE_{1|1,2}^{\mathrm{LB}}(\sigma_1^2)$ can be expanded as
\begin{equation}
\MSE_{1|1,2}^{\mathrm{LB}}(\sigma_1^2)=  \mathcal{M}_{1|1,2} + \mathcal{D}_{1|1,2} \cdot \sigma_1^2  + o(\sigma_1^2)
\label{eq:lower_bound_exp}
\end{equation}
where 
\begin{equation}
\mathcal{M}_{1|1,2} = \sum_{(i,k) \in \mathcal{S}} p_{C_1,C_2}(i,k)  \mathcal{M}_{1|1,2}^{(i,k)} \qv \mathcal{D}_{1|1,2} = \sum_{(i,k) \in \mathcal{S}} p_{C_1,C_2}(i,k)  \mathcal{D}_{1|1,2}^{(i,k)}.
\label{eq:MD_1}
\end{equation}
The terms $\mathcal{M}_{1|1,2}^{(i,k)}$ and $\mathcal{D}_{1|1,2}^{(i,k)}$ are obtained by defining 
\begin{equation}
\bar{\mathbf{\Sigma}}_{\mathbf{z}}^{(i,k)}  =   \bar{\mathbf{\Sigma}}_{\mathbf{x}_1}^{(ik)} - \bar{\mathbf{\Sigma}}_{\mathbf{x}_{12}}^{(ik)}  \mathbf{\Phi}_2\tra (  \mathbf{\Phi}_2\tra {\bar{\mathbf{\Sigma}}_{\mathbf{x}_2}^{(ik)}}   \mathbf{\Phi}_2\tra + \mathbf{I}\sigma_2^2   )^{-1}  \mathbf{\Phi}_2 \bar{\mathbf{\Sigma}}_{\mathbf{x}_{21}}^{(ik)},
\end{equation}
and by considering the following eigenvalue decomposition:
 \begin{IEEEeqnarray}{rCl}
 \mathbf{\Theta}^{(ik)} = (\bar{\mathbf{\Sigma}}_{\mathbf{z}}^{(ik)})^{\frac{1}{2}} \mathbf{\Phi}_1\tra \mathbf{\Phi}_1 (\bar{\mathbf{\Sigma}}_{\mathbf{z}}^{(ik)})^{\frac{1}{2}} = \mathbf{U}_{\mathbf{\Theta}}^{(ik)} \mathbf{\Lambda}_{\mathbf{\Theta}}^{(ik)} (\mathbf{U}_{\mathbf{\Theta}}^{(ik)} )\tra.
 \end{IEEEeqnarray}
In particular, on introducing the symbols $r_{\mathbf{z}}^{(ik)} = \rank (\bar{\mathbf{\Sigma}}_{\mathbf{z}}^{(ik)})$ and  $r_{\mathbf{\Theta}}^{(ik)} = \rank (\mathbf{\Theta}^{(ik)})$, on writing $\mathbf{\Lambda}_{\mathbf{\Theta}}^{(ik)}  = \diag (  \lambda_{\mathbf{\Theta},1}^{(ik)}, \ldots, \lambda_{\mathbf{\Theta},r_{\mathbf{\Theta}}^{(ik)}}^{(ik)}, 0, \ldots, 0    )$, and on denoting by $\mathbf{u}_{\mathbf{\Theta},t}^{(ik)}$ the $t$-th column of $\mathbf{U}_{\mathbf{\Theta}}^{(ik)} $, we have
\begin{IEEEeqnarray}{rCl}
\mathcal{M}_{1|1,2}^{(i,k)}  & = & \sum_{t= r_{\mathbf{\Theta}}^{(ik)} +1}^{r_{\mathbf{z}}^{(ik)}}  (\mathbf{u}_{\mathbf{\Theta},t}^{(ik)})\tra \bar{\mathbf{\Sigma}}_{\mathbf{z}}^{(ik)} \mathbf{u}_{\mathbf{\Theta},t}^{(ik)} \\
\mathcal{D}_{1|1,2}^{(i,k)} & = &  n_1- m_1- r_{\mathbf{\Theta}}^{(ik)}+   \sum_{t=1}^{r_{\mathbf{\Theta}}^{(ik)}}    \frac{1}{\lambda_{\mathbf{\Theta},t}^{(ik)}}   (\mathbf{u}_{\mathbf{\Theta},t}^{(ik)})\tra \bar{\mathbf{\Sigma}}_{\mathbf{z}}^{(ik)} \mathbf{u}_{\mathbf{\Theta},t}^{(ik)}.
\label{eq:MD_2}
\end{IEEEeqnarray}
\end{lemma}
\begin{IEEEproof}
See Appendix \ref{app:lower_bound}.
\end{IEEEproof}

Note that we have also expressed the expansion of the lower bound to the \ac{MMSE} for the case with side information as a function of the deviation of the signal of interest with respect to an exactly low-rank model, $\sigma_1^2 \to 0$. However, the expansion terms  $\mathcal{M}_{1|1,2}$ and $\mathcal{D}_{1|1,2}$ are functions of the number of features extracted from the side information signal $m_2$, the corresponding projection kernel $\mathbf{\Phi}_2$, the correlation between $\mathbf{x}_1$ and $\mathbf{x}_2$, and the deviation from an exactly low-rank model associated to the side information signal, since they are defined via the matrices $\bar{\mathbf{\Sigma}}_{\mathbf{z}}^{(ik)}$.

The following lemma now provides conditions that guarantee that the lower bound expansion in \eqref{eq:lower_bound_exp} is tight. Also this result is obtained by leveraging the analysis of an upper bound to the \ac{MMSE} based on a classify and reconstruct approach akin to that described in Section~\ref{par:recGMM}.

\begin{lemma}
\label{lem:upper_bound}
Consider the model in (\ref{eq:sys1}) and (\ref{eq:sys2}). Assume that the input signal $\mathbf{x}_1$ is drawn according to the class-conditioned distribution (\ref{eq:x1c1}), $\mathbf{x}_2$ is drawn according to the class-conditioned distribution (\ref{eq:x2c2}) and the class-conditioned joint distribution of $\mathbf{x}_1$ and $\mathbf{x}_2$ is given by (\ref{eq:JointG}). If $m_1$ is such that $d^{\mathrm{NOSI}}(ik,j\ell)>1, \forall (i,k,j,\ell) \in \mathcal{S}\sub{DC}$, where $d^{\mathrm{NOSI}}(ik,j\ell)$ is as in \eqref{eq:d_nosi_ikjl}, then, when $\sigma_1^2 \to 0$, the \ac{MMSE} can be expanded as
\begin{equation}
\MMSE_{1|1,2}(\sigma_1^2) =  \mathcal{M}_{1|1,2} + \mathcal{D}_{1|1,2} \cdot \sigma_1^2  + o(\sigma_1^2),
\label{eq:upper_bound_exp}
\end{equation}
where $\mathcal{M}_{1|1,2} $ and $\mathcal{D}_{1|1,2}$ are given by \eqref{eq:MD_1}-\eqref{eq:MD_2}.
\end{lemma}
\begin{IEEEproof}
See Appendix \ref{app:upper_bound}.
\end{IEEEproof}

It is interesting to note that the conditions guaranteeing the tightness of the lower bound expansion in \eqref{eq:lower_bound_exp} for the case of reconstruction with side information are exactly the same as obtained for the case without side information. 

Finally, the following theorem provides a characterization of the impact of side information on the reconstruction of signals drawn from approximately low-rank models, which is based on the analysis of the expansions provided in Lemmas \ref{lem:lower_bound_nosi}-\ref{lem:upper_bound}.

\begin{theorem}
\label{theo:impact_si}
Consider the model in (\ref{eq:sys1}) and (\ref{eq:sys2}). Assume that the input signal $\mathbf{x}_1$ is drawn according to the class-conditioned distribution (\ref{eq:x1c1}), $\mathbf{x}_2$ is drawn according to the class-conditioned distribution (\ref{eq:x2c2}) and the class-conditioned joint distribution of $\mathbf{x}_1$ and $\mathbf{x}_2$ is given by (\ref{eq:JointG}). 
Consider the expansion for the \ac{MMSE} without and with side information in \eqref{eq:lower_bound_exp_nosi}, \eqref{eq:upper_bound_exp_nosi} and \eqref{eq:lower_bound_exp}, \eqref{eq:upper_bound_exp}. If $\exists (i,k) \in \mathcal{S}$ such that $m_1 < r_{\mathbf{x}_1}^{(ik)}$, then, $\mathcal{M}_{1|1}>0 ,\mathcal{M}_{1|1,2}  >0$, and
\begin{equation}
\mathcal{M}_{1|1,2} \leq \mathcal{M}_{1|1}. 
\end{equation}
On the other hand, if $m_1 \geq r_{\mathbf{x}_1}^{(ik)}, \forall (i,k)  \in \mathcal{S}$, then $\mathcal{M}_{1|1} = \mathcal{M}_{1|1,2}   =0$, and
\begin{equation}
\mathcal{D}_{1|1,2} = \mathcal{D}_{1|1}.
\end{equation}
\end{theorem}

\begin{IEEEproof}
See Appendix \ref{app:impact_si}.
\end{IEEEproof}

This theorem -- which capitalizes on the analysis of the MMSE expansions presented in Lemmas \ref{lem:lower_bound_nosi}-\ref{lem:upper_bound} -- offers an important insight about the impact of side information in the reconstruction of signals described via the presence of two different regimes (in terms of number of features $m_1$ that are extracted from the signal of interest) in which side information has a substantially different impact to reconstruction, when assuming that signals are described via approximately low-rank models. 
In particular, when the number of features $m_1$ is less than the maximum dimension spanned by signals $\bar{\mathbf{x}}_1$ in the different Gaussian components, the MMSE is dominated by the zeroth-order expansion value when $\sigma_1^2 \to 0$. In this case, side information can lower the reconstruction error (as will be confirmed by the numerical results in Section \ref{par:NumRes}). On the other hand, if the number of features $m_1$ exceeds the maximum dimension spanned by signals $\bar{\mathbf{x}}_1$ in the different Gaussian components, then the \ac{MMSE} decays to zero with $1/\sigma_1^2$. Moreover, the first order expansions of the MMSE with and without side information coincide. In this case, collecting features from the side information signal has no significant value (with respect to a first order approximation).
}

\section{Numerical Results}
\label{par:NumRes}

We now report a series of numerical results, both with synthetic and real data, that cast further light on the role of side information to aid signal classification or reconstruction. Results with synthetic data aim to showcase how theory is able to predict \rev{the number of features needed to achieve reliable classification}, and the diversity-order of the true misclassification probability for classification problems. \rev{They also show how theory approximates well the number of features needed to guarantee reliable reconstruction and the behavior of the true reconstruction error as a function of the deviation from exactly low-rank models}.


\subsection{Synthetic Data: Classification}
\label{par:NumResClass}

We first present numerical results that showcase how the predictions on the diversity-order characterization based on the upper bound (from Theorem~\ref{theo:zeromean}) match well the behavior of the experimental misclassification probability.

We consider $\mathbf{x}_1$ and $\mathbf{x}_2$ with dimensions respectively $n_1=20$ and $n_2=12$, with $K_1 = K_2 =2$, so that the marginal \ac{pdf}s for both signals are given by the mixture of two \ac{GMM}s, each of them consisting of two Gaussian classes. All Gaussian classes are assumed to be zero-mean, i.e., $\boldsymbol{\mu}_{\mathbf{x}}^{(ik)}=\mathbf{0}$, $\forall i,k\in \{  1,2\}$. The columns of the matrices $\mathbf{P}\sub{c_1}^{(ik)},\mathbf{P}\sub{c_2}^{(ik)},\mathbf{P}_1^{(ik)}$ and $\mathbf{P}_2^{(ik)}$ are generated with \ac{i.i.d.}, zero-mean, unit-variance Gaussian entries. The dimensions of the linear spaces spanned by signals in the different classes are such that $r_{\mathbf{x}_1}^{(ik)}=7,r_{\mathbf{x}_2}^{(ik)}=6$ and $r_{\mathbf{x}}^{(ik)}=9$, $\forall i,k\in \{  1,2\}$. Moreover, the matrices $\mathbf{P}\sub{c_1}^{(ik)},\mathbf{P}\sub{c_2}^{(ik)},\mathbf{P}_1^{(ik)}$ and $\mathbf{P}_2^{(ik)}$ associated with different classes share some of their columns, so that the dimensions of the sums of spaces spanned by signals in different classes are such that the corresponding ranks associated to pairs of class-conditioned input covariance matrices are given in Table~\ref{table:ranks}. 
The projection kernels $\mathbf{\Phi}_1, \mathbf{\Phi}_2$ are generated with \ac{i.i.d.}, zero-mean, Gaussian entries, with fixed variance. \rev{After that, the projection kernel are modified in order to verify $\mathbf{\Phi}_1\mathbf{\Phi}_1\tra=\mathbf{I}$ and $\mathbf{\Phi}_2\mathbf{\Phi}_2\tra=\mathbf{I}$.}

\begin{table}
\begin{center}
\caption{Ranks associated to pairs of class-conditioned input covariance matrices in the numerical example of classification with side information.}
\begin{tabular}{|r||c|c|c|c|c|c|c|}
\hline
$(ik,j\ell)$ & (11,12) & (21,22) & (11,21) & (11,22) & (12,21) & (12,22)\\ 
\hline
\hline
$r_{\mathbf{x}_1}^{(ik,j\ell)}$ & 8  & 8   &  10 &  11 &   9   &  10  \\
\hline
$r_{\mathbf{x}_2}^{(ik,j\ell)}$ & 8  &  8 &  11 &  11 & 10 &  11  \\
\hline
$r_{\mathbf{x}}^{(ik,j\ell)}$   &  12  & 12 &  17 &  18  &  15  &  17    \\
\hline
\end{tabular}
\label{table:ranks}
\end{center}
\end{table}

\begin{figure}
\subfigure[True misclassification probability]{\includegraphics[width=0.5\textwidth]{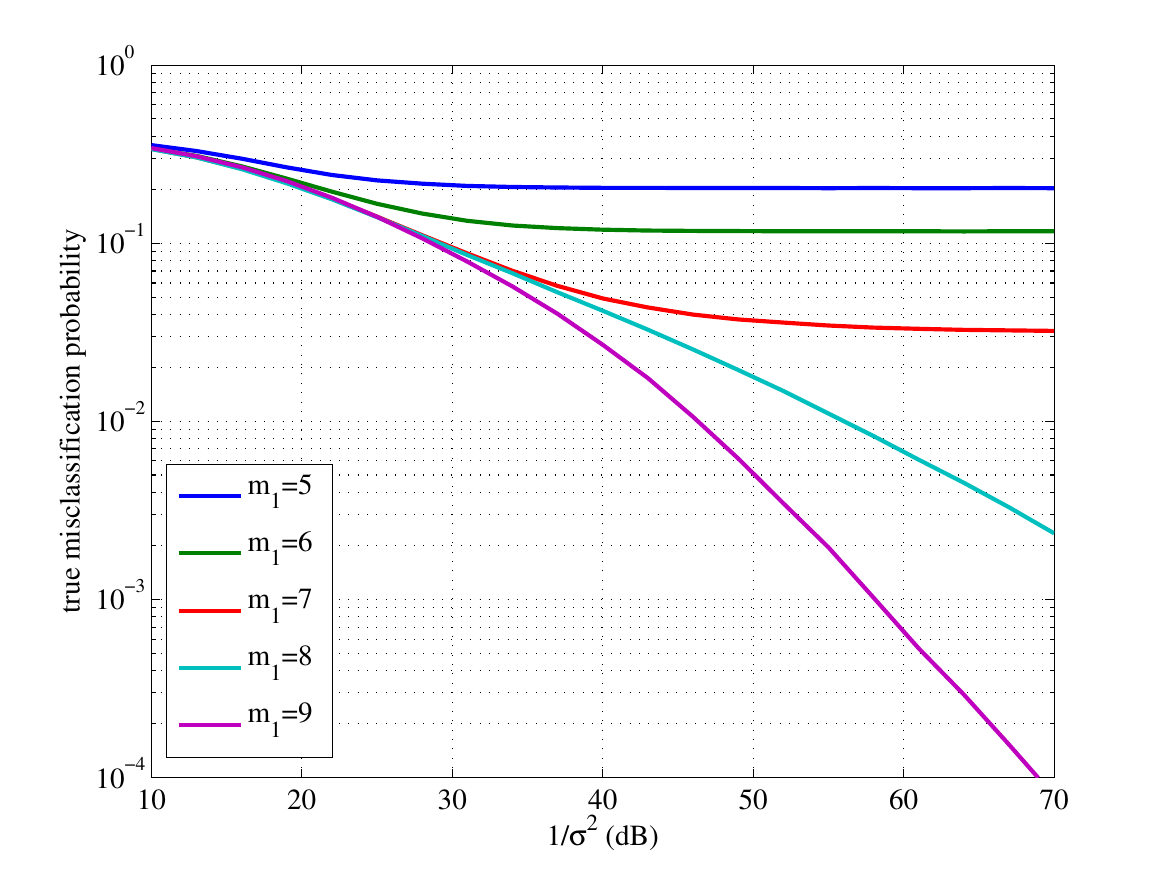}}
\subfigure[Misclassification probability upper bound in (\ref{eq:PerrUpUnion})]{\includegraphics[width=0.5\textwidth]{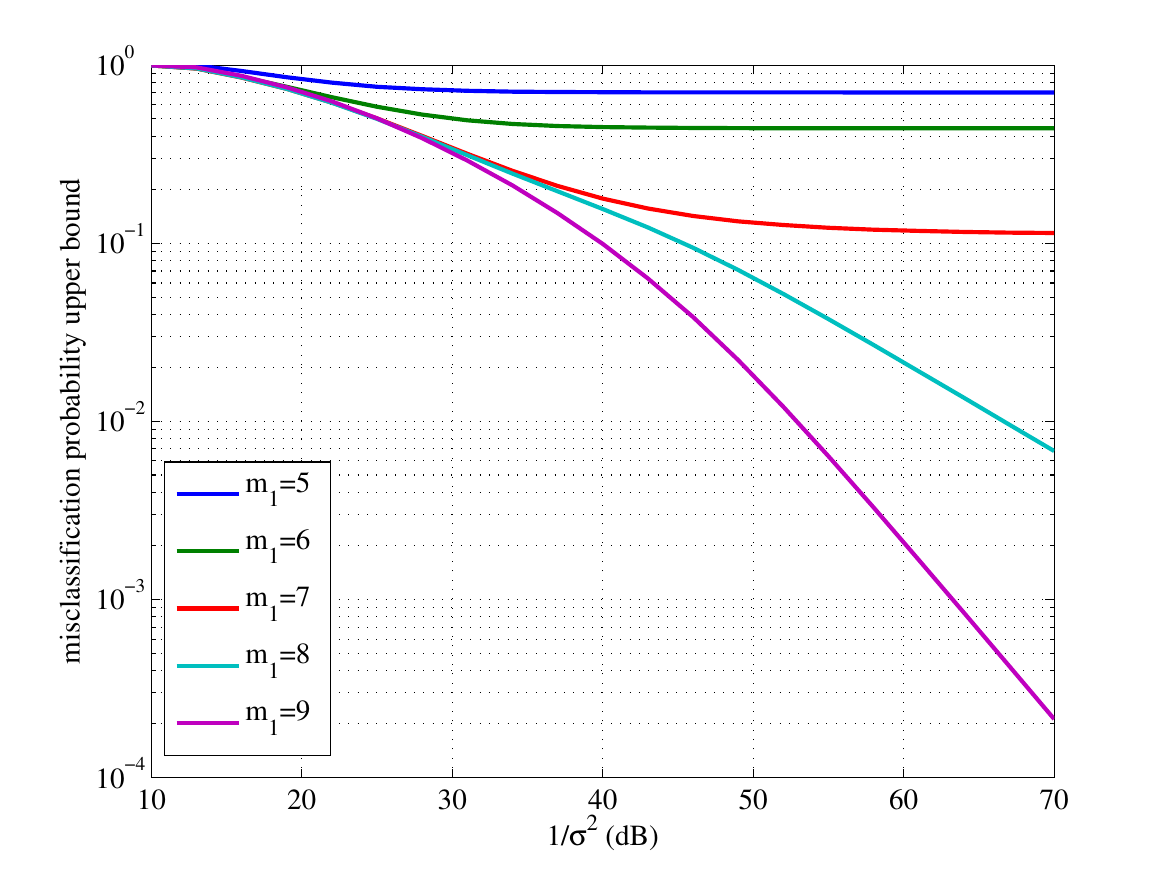}}
\caption{True misclassification probability (based on numerical experiments) and upper bound vs. $1/\sigma^2$ for classification without side information (i.e., $m_2=0$). }
\label{fig:Perr_nosi}
\end{figure}

\rev{We consider the case $\sigma_1^2 =\sigma_2^2 =\sigma^2$ and} we compare the \rev{number of features required for error free classification} and the diversity-orders yielded by the Bhattacharyya-based upper bound (\ref{eq:BhattM1union}) with the error probability obtained by numerical simulation. We report in Fig.~\ref{fig:Perr_nosi}(a) the experimental error probability and in Fig.~\ref{fig:Perr_nosi}(b) the upper bound $\bar{P}\sub{err}^{\sf U}$ in (\ref{eq:PerrUpUnion}) for the case in which no side information is available to the decoder (cf. \cite{Reboredo14}), i.e., $m_2= 0$.  In this case, the misclassification probability \rev{approaches zero as $\sigma^2 \to 0$} when $m_1 > 7$ \cite{Reboredo14}, and we note how the analysis based on the upper bound reflects well the behavior of the true error probability both in terms of \rev{number of features needed for reliable classification} and diversity-order.

\begin{figure}
\subfigure[True misclassification probability]{\includegraphics[width=0.5\textwidth]{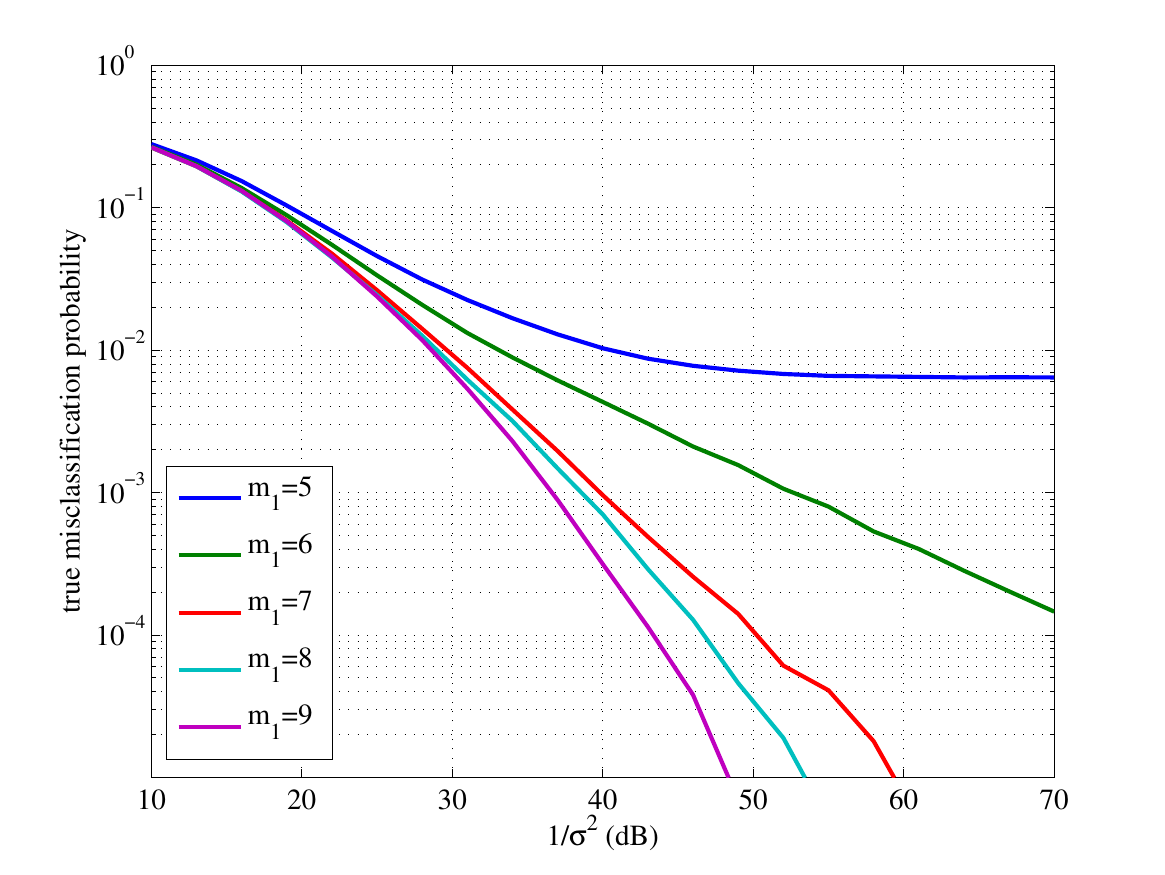}}
\subfigure[Misclassification probability upper bound in (\ref{eq:PerrUpUnion})]{\includegraphics[width=0.5\textwidth]{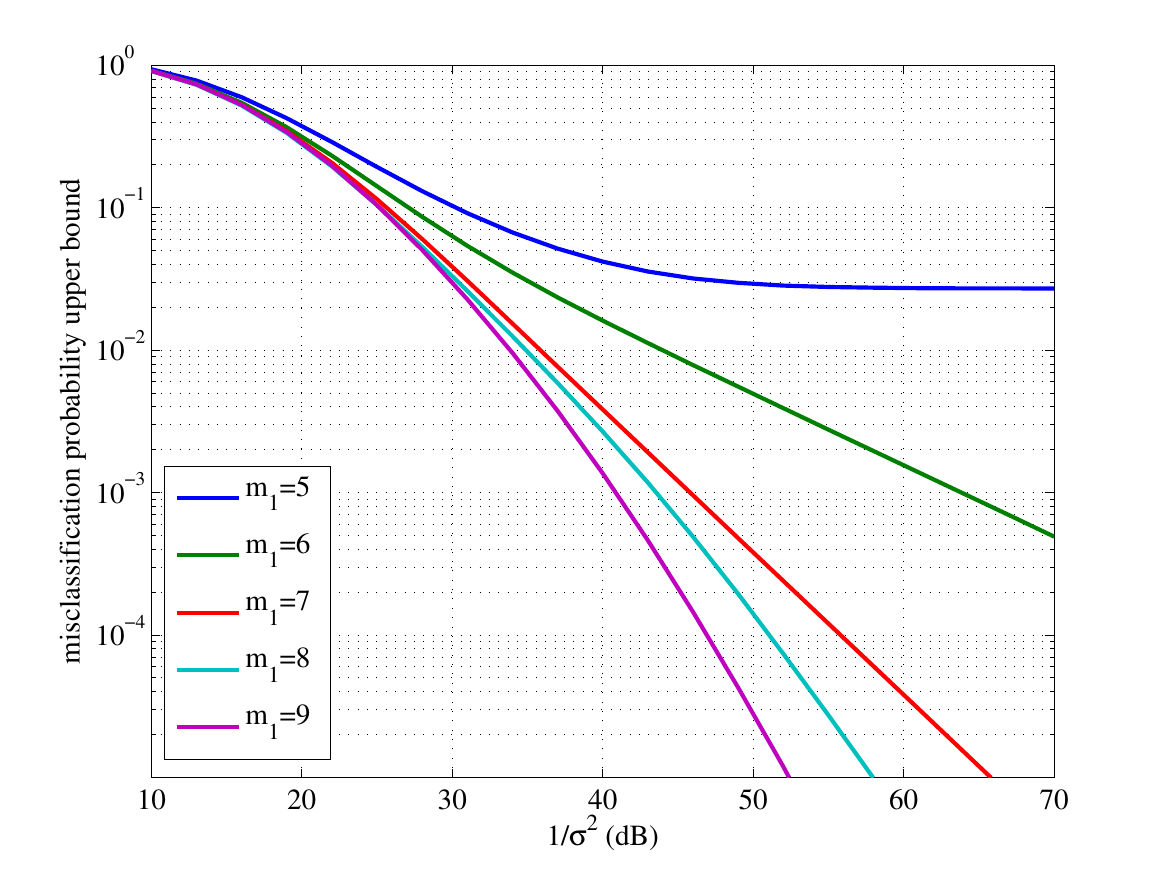}}
\caption{True misclassification probability (based on numerical experiments) and upper bound vs. $1/\sigma^2$ for classification with side information with $m_2=4$.}
\label{fig:Perr_si_M2_4}
\end{figure}

We now evaluate the impact of the side information $\mathbf{y}_2$ in the classification of the input signal $\mathbf{x}_1$. We consider the case in which the number of features representing the side information is $m_2=4$ and for different values of $m_1$. In Fig.~\ref{fig:Perr_si_M2_4}(a) we show the experimental error probability and in Fig.~\ref{fig:Perr_si_M2_4}(b) the upper bound $\bar{P}\sub{err}^{\sf U}$ in (\ref{eq:PerrUpUnion}). We observe how the presence of side information can be leveraged in order to obtain \rev{error free classification} with only $m_1 >5$ features on the input signal. In fact, when $m_1+ m_2>9$, the linear spaces spanned collectively by the projections of signals $\bar{\mathbf{x}}_1$ and $\bar{\mathbf{x}}_2$ drawn from different Gaussian components are not completely overlapping, since they are 9-dimensional spaces in $\mathbb{R}^{m_1 + m_2}$. Moreover, increasing the number of linear features extracted above 4 leads to increased diversity-order values. Also in this case, we note how the behavior analytically predicted from the characterization of the Bhattacharyya-based upper bound matches well the true behavior of the actual error probability both in terms of \rev{number of features required for reliable classification} and diversity-order.

\subsection{\rev{Synthetic Data: Reconstruction, Low-Rank Model}}

We now aim to show how numerical results for reconstruction of synthetic signals also align well with the analysis reported in Section~\ref{par:reconstruction}, in particular for what regards the characterization of the number of features needed to drive the \ac{MMSE} to zero when \rev{$\sigma_1^2, \sigma_2^2 \to 0$}. We start by considering the case in which $\mathbf{x}_1$ and $\mathbf{x}_2$ are described by a single Gaussian joint distribution. In particular, we set the signal sizes to $n_1=5$ and $n_2=4$, and we build the joint input covariance matrix using the common/innovation component representation in (\ref{eq:x1comp}) and (\ref{eq:x2comp}), where $\mathbf{P}\sub{c_1}\in \mathbb{R}^{5 \times 2},\mathbf{P}\sub{c_2}\in \mathbb{R}^{4 \times 2},\mathbf{P}_1\in \mathbb{R}^{5 \times 1}$ and $\mathbf{P}_2\in \mathbb{R}^{4 \times 1}$ have \ac{i.i.d.}, zero-mean, unit-variance Gaussian entries, thus obtaining $r_{\mathbf{x}_1}=3, r_{\mathbf{x}_2}=3$ and $r_{\mathbf{x}}=4$. We also assume that the projection kernels $\mathbf{\Phi}_1$ and $\mathbf{\Phi}_2$ have \ac{i.i.d.}, zero-mean, Gaussian entries with fixed variance, \rev{and we modify them so that $\mathbf{\Phi}_1\mathbf{\Phi}_1\tra=\mathbf{I}$ and $\mathbf{\Phi}_2\mathbf{\Phi}_2\tra=\mathbf{I}$.}

\begin{figure}
\begin{center}
\includegraphics[width=0.5\textwidth]{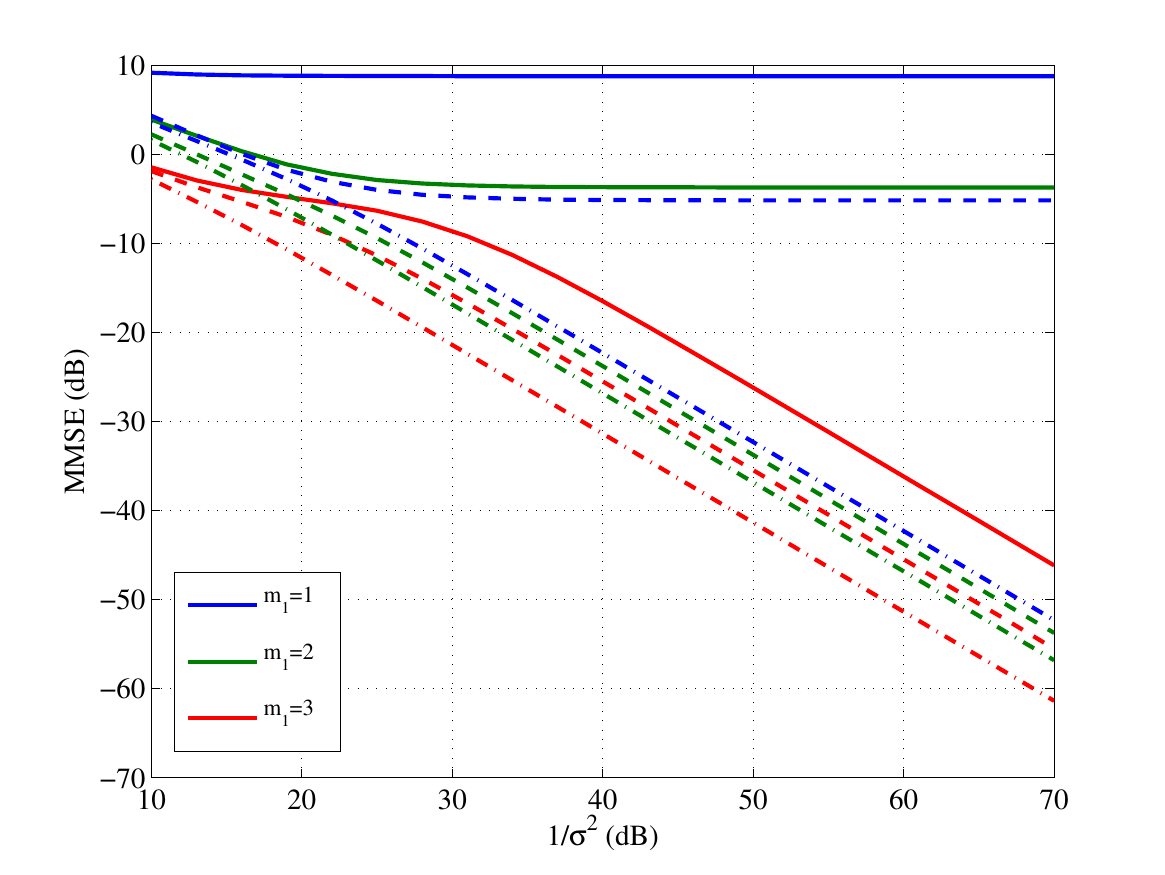}
\end{center}
\caption{Results of numerical experiments, depicting MMSE vs. $1/\sigma^2$ for Gaussian signal reconstruction with side information. $m_1=1,2,3$. $m_2=1$ (solid lines), $m_2=2$ (dashed lines) and $m_2=3$ (dashed-dotted lines).}
\label{fig:MMSEGauss}
\end{figure}

Fig.~\ref{fig:MMSEGauss} shows the values of the reconstruction \ac{MMSE} for Gaussian inputs in (\ref{eq:MMSEGsi}), for different values of the number of features $m_1$ and $m_2$. We observe that the necessary and sufficient conditions in (\ref{eq:recG}) are verified by the numerical results: in particular, when $m_2=1$, the \ac{MMSE} \rev{approaches zero in the low-rank regime} only when $m_1 \geq 3$, when $m_2=2$, \rev{the \ac{MMSE} approaches zero} with $m_1 \geq 2$ and, finally, when $m_2=3$, a single feature extracted from the input signal is sufficient to guarantee \rev{reliable reconstruction in the low-rank regime}. 

We now consider signal reconstruction for \ac{GMM} inputs. In particular, we assume that the vectors $\mathbf{x}_1$ and $\mathbf{x}_2$ are drawn from the joint \ac{GMM} prior described in Section~\ref{par:NumResClass} for the case of signal classification, and we assume again to use projection kernels with \ac{i.i.d.}, zero-mean, Gaussian entries, \rev{which are then modified in order to have orthonormal rows}. Reconstruction is performed via the conditional mean estimator, that is now given by
\begin{IEEEeqnarray}{rCl}
\hat{\mathbf{x}}_1(\mathbf{y}) &=& \E{\mathbf{x}_1 | \mathbf{y}}  = \int \mathbf{x}_1  p(\mathbf{x}_1|\mathbf{y}) d\mathbf{x}_1 \\
 & = & \int   \mathbf{x}_1   \left(    \int        \frac{  p(\mathbf{x}_1, \mathbf{x}_2) p(\mathbf{y} |  \mathbf{x}_1 , \mathbf{x}_1) }{\int   p(\mathbf{x})  p(\mathbf{y}|\mathbf{x})   d\mathbf{x}}       d\mathbf{x}_2   \right)          d\mathbf{x}_1 \\
 & = & \int \mathbf{x}_1   \left(     \int    \sum_{i=1}^{K_1}\sum_{k=1}^{K_2}     \tilde{p}_{C_1,C_2}(i,k) \mathcal{N} (\mathbf{x}_1,\mathbf{x}_2; \tilde{\boldsymbol{\mu}}_{\mathbf{x}}^{(ik)}, \tilde{{\mathbf{\Sigma}}}_{\mathbf{x}}^{(ik)})     d\mathbf{x}_2   \right)    d\mathbf{x}_1
\end{IEEEeqnarray}
where~\cite{Chen10}
\begin{IEEEeqnarray}{rCl}
\tilde{p}_{C_1,C_2}(i,k) & = & \frac{  p_{C_1,C_2}(i,k)  \mathcal{N}(\mathbf{y}; \mathbf{\Phi} \boldsymbol{\mu}_{\mathbf{x}}^{(ik)},   \mathbf{\Phi} {\mathbf{\Sigma}}_{\mathbf{x}}^{(ik)} \mathbf{\Phi}\tra  )     }{   \sum_{i=1}^{K_1}\sum_{k=1}^{K_2} p_{C_1,C_2}(i,k)  \mathcal{N}(\mathbf{y}; \mathbf{\Phi} \boldsymbol{\mu}_{\mathbf{x}}^{(ik)},   \mathbf{\Phi} {\mathbf{\Sigma}}_{\mathbf{x}}^{(ik)} \mathbf{\Phi}\tra  )   }  \\
\tilde{\boldsymbol{\mu}}_{\mathbf{x}}^{(ik)} & = &  \boldsymbol{\mu}_{\mathbf{x}}^{(ik)} + {\mathbf{\Sigma}}_{\mathbf{x}}^{(ik)} \mathbf{\Phi}\tra    ( \mathbf{\Phi} {\mathbf{\Sigma}}_{\mathbf{x}}^{(ik)} \mathbf{\Phi}\tra  )^{-1} (\mathbf{y}-\mathbf{\Phi} \boldsymbol{\mu}_{\mathbf{x}}^{(ik)}
)\\
\tilde{{\mathbf{\Sigma}}}_{\mathbf{x}}^{(ik)} & = &  {\mathbf{\Sigma}}_{\mathbf{x}}^{(ik)} - {\mathbf{\Sigma}}_{\mathbf{x}}^{(ik)} \mathbf{\Phi}\tra    ( \mathbf{\Phi} {\mathbf{\Sigma}}_{\mathbf{x}}^{(ik)} \mathbf{\Phi}\tra  )^{-1} \mathbf{\Phi} {\mathbf{\Sigma}}_{\mathbf{x}}^{(ik)},
\end{IEEEeqnarray}
and we have used the notation $\mathcal{N}(\mathbf{x};  \boldsymbol{\mu},{\mathbf{\Sigma}})$ to express explicitly the argument of the Gaussian distribution. Then, on marginalizing out $\mathbf{x}_2$, we obtain
\begin{equation}
\hat{\mathbf{x}}_1(\mathbf{y}) = \sum_{i=1}^{K_1}\sum_{k=1}^{K_2}     \tilde{p}_{C_1,C_2}(i,k) \left( \boldsymbol{\mu}_{\mathbf{x}_1}^{(ik)} + [{\mathbf{\Sigma}}_{\mathbf{x}_1}^{(ik)} \  {\mathbf{\Sigma}}_{\mathbf{x}_{12}}^{(ik)} ] \mathbf{\Phi}\tra    (  \mathbf{\Phi} {\mathbf{\Sigma}}_{\mathbf{x}}^{(ik)} \mathbf{\Phi}\tra  )^{-1} (\mathbf{y}-\mathbf{\Phi} \boldsymbol{\mu}_{\mathbf{x}}^{(ik)}
)\right)
\label{eq:cond_mean_est_GMM}
\end{equation}
and, as expected from the properties of the \ac{MMSE} estimator~\cite{Hassibi}, $\hat{\mathbf{x}}_1(\mathbf{y})$ can be also obtained by retaining the first $n_1$ entries of the joint conditional mean estimator $\hat{\mathbf{x}}(\mathbf{y}) =\E{\mathbf{x}|\mathbf{y}}$ \cite{Chen10}.

\begin{figure}
\subfigure[Without side information]{\includegraphics[width=0.5\textwidth]{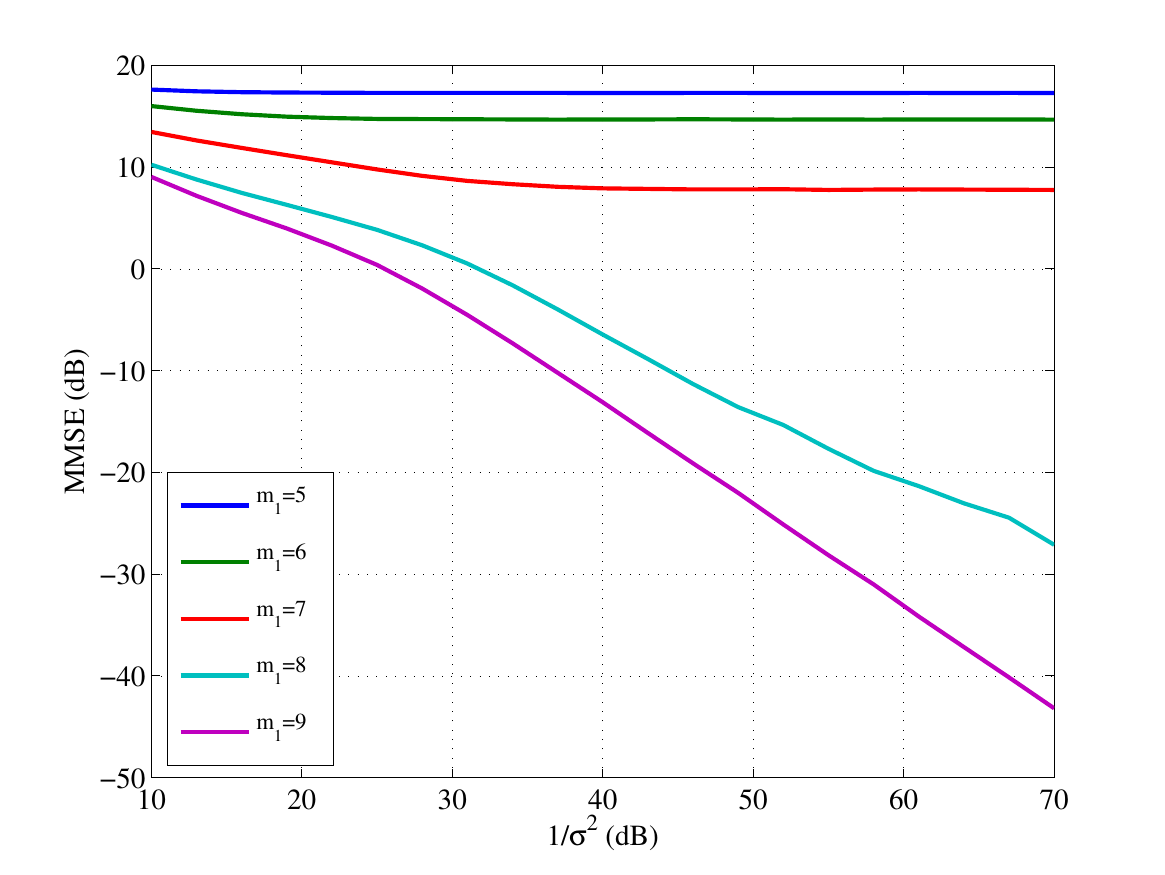}}
\subfigure[With side information]{\includegraphics[width=0.5\textwidth]{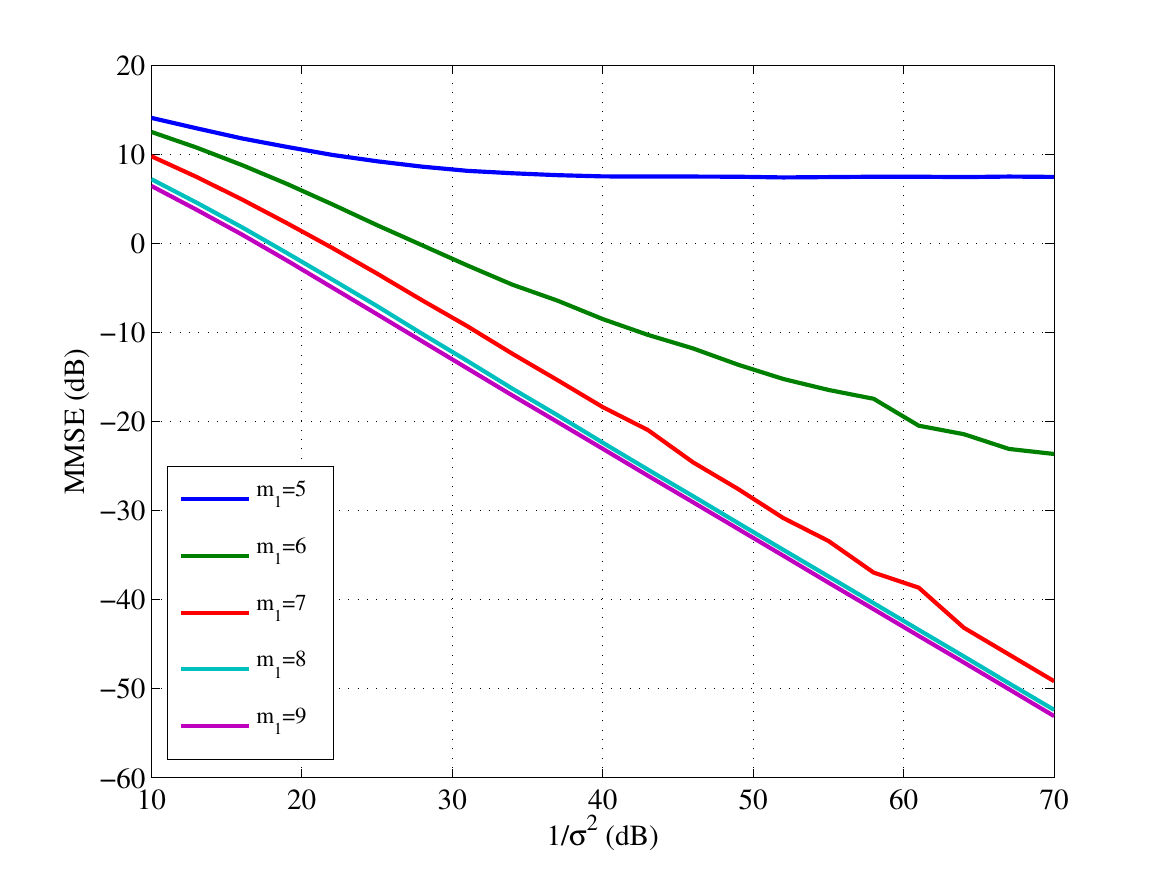}}
\caption{Results of numerical experiments, depicting MMSE vs. $1/\sigma^2$ for \ac{GMM} signal reconstruction without and with side information, i.e., $m_2=0$ and $m_2=4$, respectively.}
\label{fig:MMSE_GMM}
\end{figure}

Fig.~\ref{fig:MMSE_GMM} shows the \ac{MMSE} values for reconstruction both with side information (subfigure (a)), where we set $m_2 = 4$, and without side information (subfigure (b)), where we set $m_2=0$. When side information is not available at the decoder, \rev{reliable reconstruction} is obtained when $m_1>\max_{(i,k)}r_{\mathbf{x}_1}^{(ik)}=7$~\cite{RecJournal}. On the other hand, as predicted by the sufficient conditions in (\ref{eq:recGMM}), the presence of side information allows to guarantee \rev{reliable reconstruction} with only $m_1>5$ features. Notice also in this case how the theoretical analysis matches well the behavior shown by the numerical results.

\rev{
\subsection{Synthetic Data: Reconstruction, Approximately Low-Rank Model}

We now consider the reconstruction \ac{MMSE} obtained when the signal and the side information are drawn from an approximately low-rank model, i.e., when the values of $\sigma_1^2, \sigma_2^2 $ in Section \ref{par:SigModel} are not negligible. We set signal sizes $n_1=12$ and $n_2=8$ and we set $K_1=K_2=2$. For each class pair $(i,k)$, the corresponding joint input covariance matrix is built by using the common innovation component representation in (\ref{eq:x1comp}) and (\ref{eq:x2comp}) and by generating $\mathbf{P}\sub{c_1}^{(ik)}\in \mathbb{R}^{12 \times 2},\mathbf{P}\sub{c_2}^{(ik)}\in \mathbb{R}^{8 \times 2},\mathbf{P}_1^{(ik)}\in \mathbb{R}^{12 \times 2}$ and $\mathbf{P}_2^{(ik)}\in \mathbb{R}^{8 \times 2}$ with \ac{i.i.d.}, zero-mean, unit-variance Gaussian entries, thus obtaining $r_{\mathbf{x}_1}^{(ik)}=4, r_{\mathbf{x}_2}^{(ik)}=4$ and $r_{\mathbf{x}}^{(ik)}=6$, $\forall (i,k) \in  \mathcal{S}$. We also assume that the projection kernels $\mathbf{\Phi}_1$ and $\mathbf{\Phi}_2$ have \ac{i.i.d.}, zero-mean, Gaussian entries with fixed variance, and we modify them so that $\mathbf{\Phi}_1\mathbf{\Phi}_1\tra=\mathbf{I}$ and $\mathbf{\Phi}_2\mathbf{\Phi}_2\tra=\mathbf{I}$. Reconstruction is performed with the conditional mean estimator \eqref{eq:cond_mean_est_GMM}.

\begin{figure}
\subfigure[Without side information]{\includegraphics[width=0.5\textwidth]{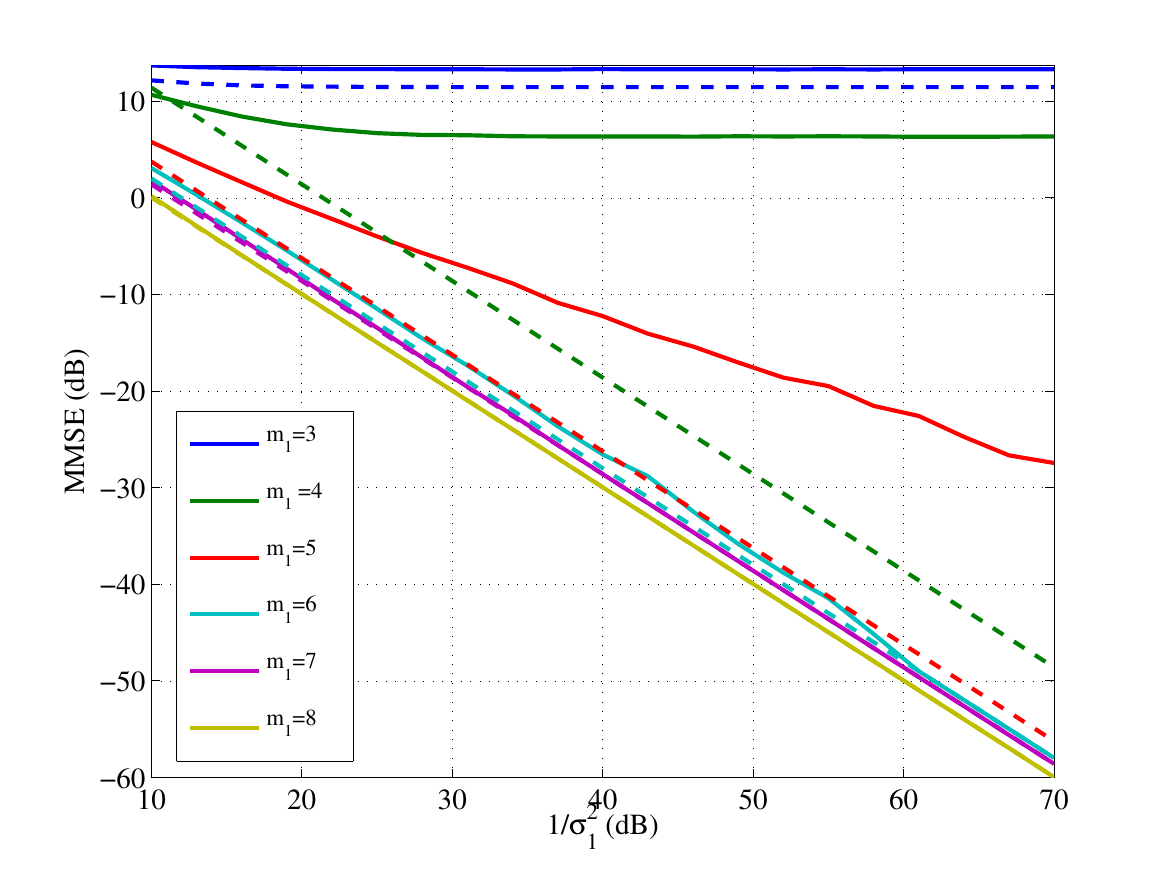}}
\subfigure[With side information]{\includegraphics[width=0.5\textwidth]{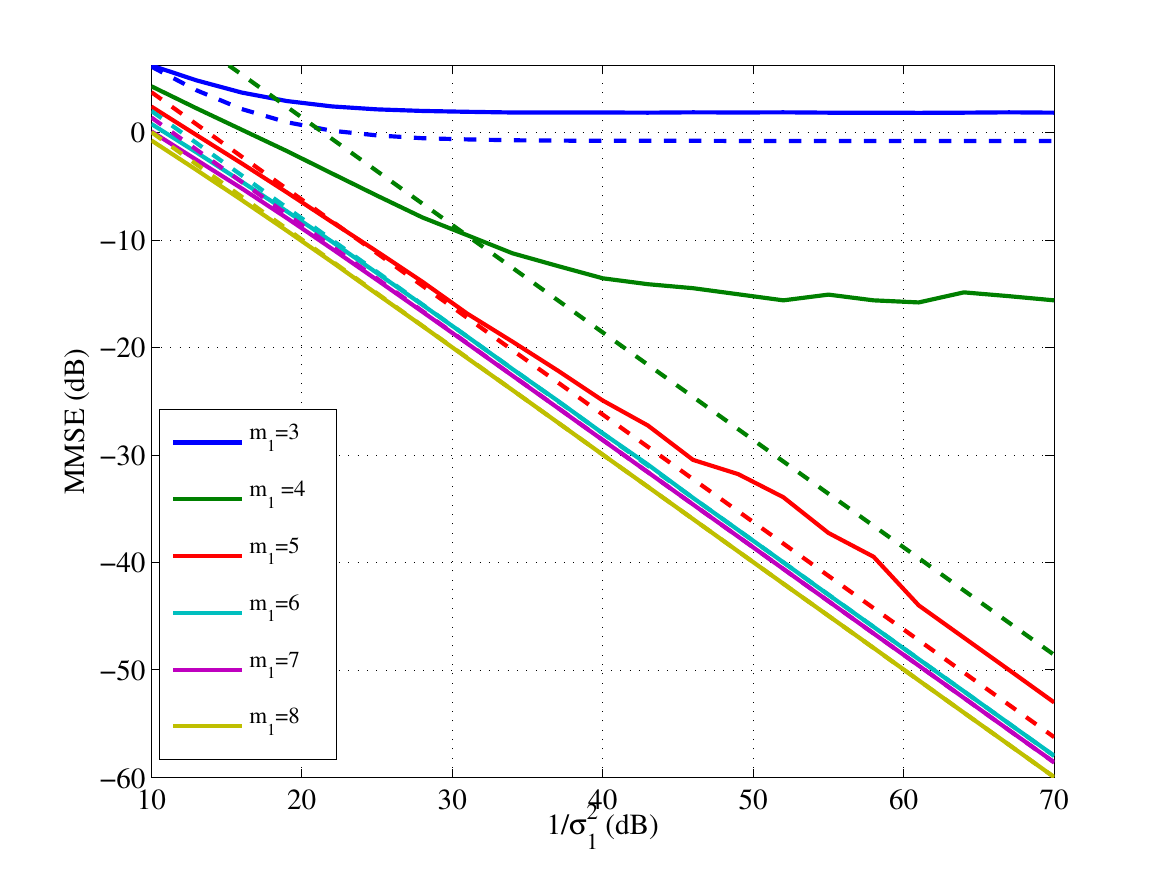}}
\caption{Results of numerical experiments, depicting \ac{MMSE} vs. $1/\sigma_1^2$ for GMM signal reconstruction without and with side information, i.e., $m_2=0$ and $m_2 = 4$. We set $\sigma_2^2 = 10^{-2}$. We report the numerical \ac{MMSE} (solid lines) and the asymptotic expansions \eqref{eq:upper_bound_exp_nosi} and \eqref{eq:upper_bound_exp} (dashed lines).}
\label{fig:approx_lowrank}
\end{figure}

Fig.~\ref{fig:approx_lowrank} reports the values of the reconstruction \ac{MMSE} for both cases  without side information and with side information ($m_2=2$, $\sigma_2^2=10^{-2}$) versus $\sigma_1^2$. We also report the values of the expansions \eqref{eq:upper_bound_exp_nosi} and \eqref{eq:upper_bound_exp}.

We note that, for $m_1 \leq 6$, side information guarantees lower MMSE values. In particular, when $m_1=3$, both the numerical \ac{MMSE} and the lower bound expansions present error floors, for both cases with and without side information. In this case, as predicted by Theorem \ref{theo:impact_si}, the presence of side information allows a lowering of the values of the error floor associated to the \ac{MMSE} lower bound, and the same behavior is also observed for the numerically evaluated \ac{MMSE}.

 On the other hand, when the conditions in Lemmas \ref{lem:upper_bound_nosi} and \ref{lem:upper_bound} are verified, i.e., when $m_1 > 6$, the expansions \eqref{eq:upper_bound_exp_nosi} and \eqref{eq:upper_bound_exp} predict accurately the behavior of the actual \ac{MMSE}. Moreover, in this case, the impact of side information on reconstruction performance is negligible, as predicted by the analysis carried out in Section~\ref{par:GMM_approx}.
}

\subsection{Experimental Results: Compressive Hyperspectral Imaging}
\label{CASSI}

Finally, we present an example to showcase how the proposed framework also offers a principled approach to design systems able to leverage effectively side information in reconstruction tasks. In this case, we do not reveal \rev{conditions on the number of features to drive exactly to zero the} reconstruction error. However, we can notice how side information can be used in order to improve reconstruction performance. 

We consider a compressive hyperspectral imaging example, in which hyperspectral images of a subject are recovered from compressive measurements in the presence of side information. In particular, we consider measurements collected by the \ac{CASSI} apparatus described in~\cite{Rajwade13}. Side information is represented in this case by an RGB snapshot of the same scene, which can be easily obtained without requiring expensive hyperspectral imaging devices. The information contained in the RGB image is expected to improve the reconstruction quality of the input signal, also due to the fact that, in contrast to the measurements taken by the \ac{CASSI} camera, the RGB image is not affected by coded aperture modulation.

In this case the vector $\mathbf{x}_1$ represents patches extracted from the hyperspectral image, whereas $\mathbf{x}_2$ represents patches extracted from the corresponding RGB image  (see \cite{Rajwade13} for details on how data from this system are analyzed). The vectors $\mathbf{x}_1$ and $\mathbf{x}_2$ are assumed to be modeled by the joint \ac{GMM} described in Section~\ref{par:SigModel} with $K_1=K_2=20$. The parameters of the joint \ac{GMM} are learned from the hyperspectral image dataset used in~\cite{Foster06JOSA}\footnote{\url{http://personalpages.manchester.ac.uk/staff/d.h.foster/Hyperspectral_images_of_natural_scenes_04.html}}, again via the \ac{EM} algorithm. Note that the images in the training dataset are associated to wavelength values that do not match perfectly those characterizing the \ac{CASSI} camera. Therefore, the training algorithm is run by selecting each time wavelengths that are closest to the nominal values of the \ac{CASSI} camera.

%
   \begin{figure}[ht!]
       \centering
       \includegraphics[scale = 0.9]{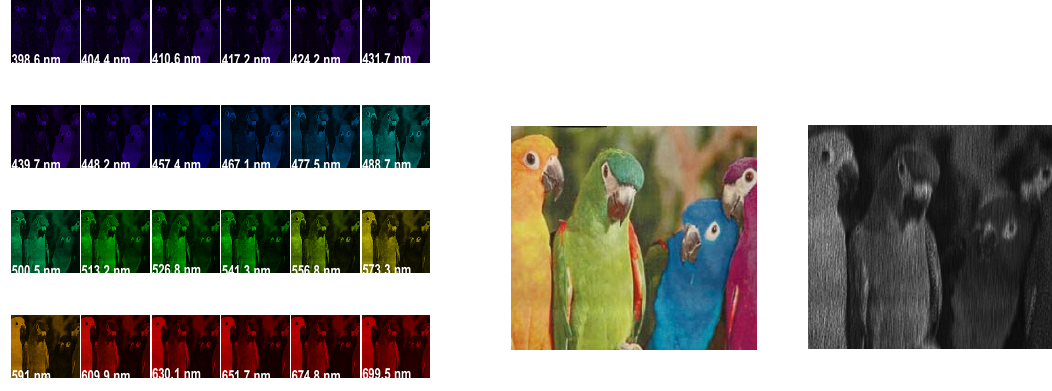}\\
       \caption{Left: hyperspectral image (reference). Middle: RGB image. Right: CASSI measurement.}
       \label{Fig:Real_Mea}
   \end{figure} 
   \begin{figure}[ht!]
       \centering
       \includegraphics[scale = 0.6]{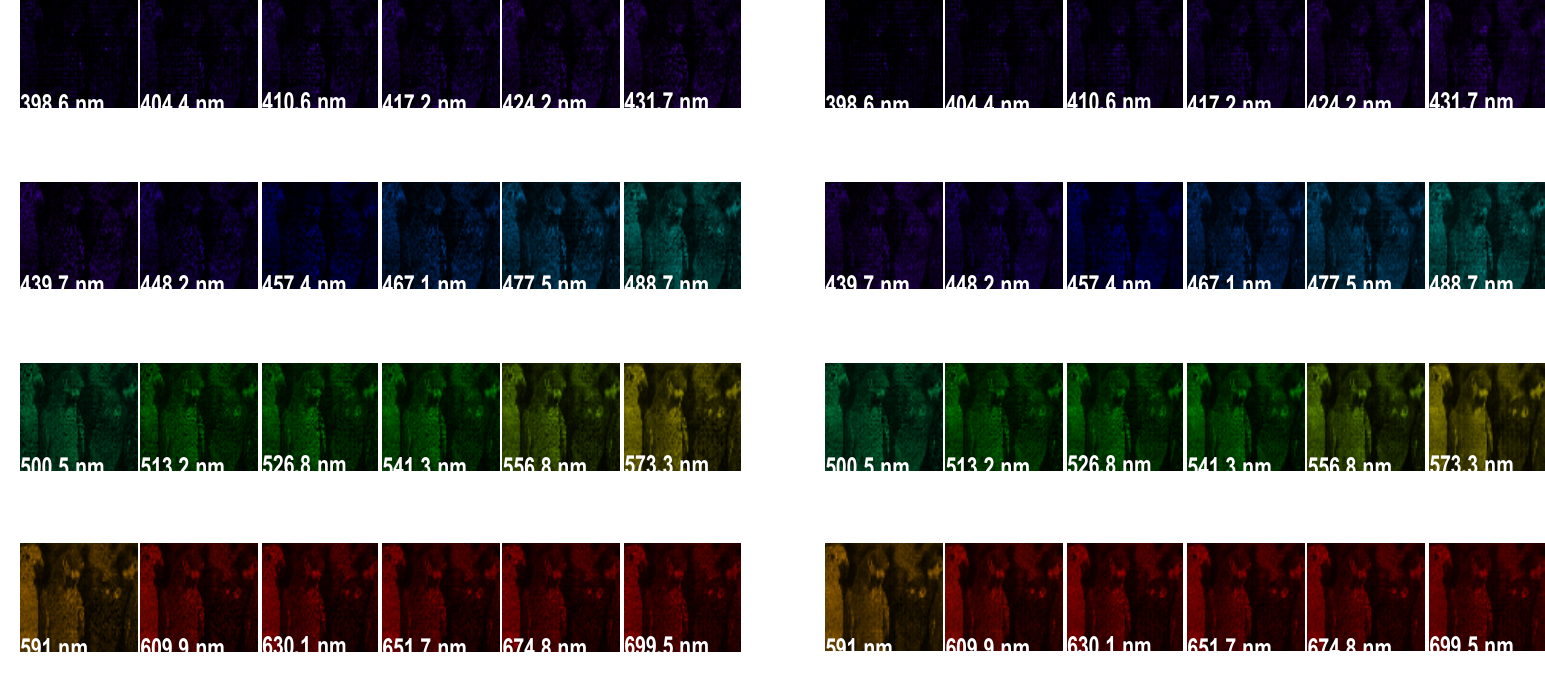}\\
       \caption{Left: reconstruction without side information. Right: reconstruction with side information.}
       \label{Fig:Real_result}
   \end{figure} 
   \begin{figure}[ht!]
       \centering
       \includegraphics[scale = 0.6]{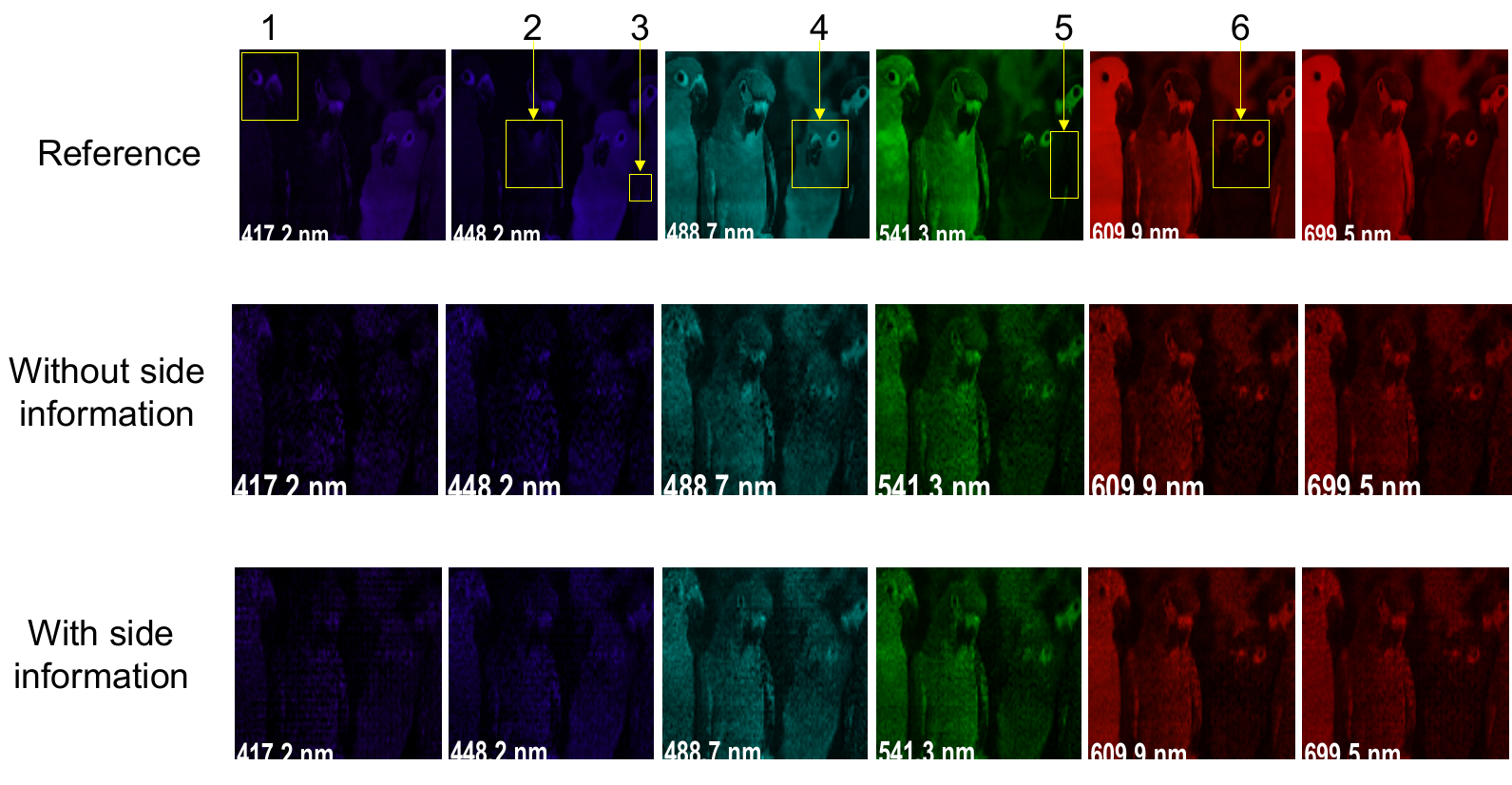}\\
       \caption{Zoom-in of reconstructed images for six different channels. The corresponding PSNRs for reconstruction with and without side information are reported in Table~\ref{Table:HSI_PSNR}.}
       \label{Fig:Real_result_zoom}
   \end{figure} 
   \begin{table}[htbp]
   \caption{Reconstruction PSNR (in dB) for the six selected regions in Fig.~\ref{Fig:Real_result_zoom}, with and without side information.}
   \centering
   \begin{tabular}{|c|c|c|c|c|c|c|}
   \hline Region & 1 & 2 & 3 & 4 & 5 & 6\\
   \hline \hline
    without side information & 12.50 & 16.42 & 12.71 & 15.49  &  16.20 & 18.86 \\
   \hline with side information & 13.81  & 18.66 & 17.24 & 17.80 &  19.21 & 20.07  \\
   \hline
   \end{tabular}
   \label{Table:HSI_PSNR}
   \end{table}

We consider real data captured by the CASSI camera, so that the entries of the projection kernel $\mathbf{\Phi}_1$ reflect the physical implementation of the compressive imaging system~\cite{Kittle10CASSI}, and they are constrained to belong to the interval $[0,1]$. On the other hand, the side information RGB image is not compressed, so that we have $\mathbf{\Phi}_2=\mathbf{I}$. The bird dataset is used and a single measurement is used, thus meaning that $24$ images of size $1021 \times 703$ corresponding to $24$ different wavelengths from $398.6$\,nm to $699.5$\,nm are compressed into a single snapshot of the same size. In order to evaluate the reconstruction accuracy, reference images are acquired using a different (and
non-compressive) hyperspectral imaging setup. Therefore, the reference images and the side information image are not perfectly aligned with the CASSI measurement shown in the right part of Fig.~\ref{Fig:Real_Mea}.
The reconstructed hyperspectral images without and with side information are shown in Fig.~\ref{Fig:Real_result}. It can be seen clearly that the reconstruction with side information has better quality. Furthermore, though the reference is not aligned well with the CASSI measurement, we can still compare the reconstruction \ac{PSNR} in correspondence of some selected blocks in the image. Fig.~\ref{Fig:Real_result_zoom} shows the reconstruction of six channels and the corresponding \ac{PSNR} values are reported in Table~\ref{Table:HSI_PSNR}. It  can be noticed that the \ac{PSNR} improvement due to side information is significant.

\section{Conclusions}
\label{par:conclusions}

We have developed a principled framework that can be used not only to study fundamental limits in the classification and reconstruction of high-dimensional signals from low-dimensional signal features in the presence of side information, but also to obtain state-of-the-art results in imaging problems.

In particular, we have considered a linear feature-extraction model, where a decoder has access to linear features of both the signal of interest and the side information signal, in order to carry out either classification or reconstruction. We have also considered a model where the joint distribution of the signal of interest and the side information, conditioned on some underlying class labels is a multivariate Gaussian, which embodies the correlation between these signals. The marginal distribution of the signal conditioned on a class label is a Gaussian mixture, and likewise the marginal distribution of the side information conditioned on the class label is also a Gaussian mixture.

This modeling approach, which can be used to encapsulate a wide range of distributions, has then offered the opportunity to capitalize on tractable bounds to the misclassification probability and the reconstruction error, to construct an asymptotic characterization of the behavior of these quantities in the low-rank regime. In addition, this modeling approach has also led to a characterization of sharp sufficient conditions for \rev{reliable classification in the low-rank regime} and necessary and sufficient conditions for \rev{reliable reconstruction in the low-rank regime}, as a function of the geometry of the sources, the geometry of the linear feature extraction process and their interplay, reminiscent of the Slepian-Wolf and the Wyner-Ziv conditions. \rev{Moreover, we have provided expansions that characterize the effect of deviations from exactly low-rank models on the reconstruction error. By capitalizing on the analysis of such expansions, we have also defined the operational regime when side information has a more significant impact on the reconstruction performance.}

It has been shown that our theory is well aligned with practice via a range of numerical results associated with low-rank \rev{and approximately low-rank} data models. Of particular relevance, it has also been shown that our framework offers a principled mechanism to integrate side information in data classification and reconstruction problems in the context of compressive hyperspectral imaging in the presence of side information.

This work also points to various possible future directions:
\begin{itemize}
\item It is of interest to extend the results from consideration of only one side information source to settings where there are multiple sources of side information. It is possible to generalize the models immediately, but the analysis is considerably more complex (as pointed out in Appendix~\ref{app:B}). 
\item There is interest in generalization of the results from the scenario where the linear features are extracted randomly to scenarios where the linear features are designed \cite{Carson12,Chen12,RecJournal} (or indeed nonlinear features are designed \cite{WangICML14}) is relevant. This could lead to additional gains in \rev{the number of features required for reliable classification or reconstruction}.
\item The generalization of the results from scenarios where only the decoder has access to the side information to scenarios where both the encoder and the decoder have access to the side information is also relevant. This may also lead to additional gains both in the presence of random linear features or designed ones.
\item Finally, it is believed that the framework, which applies to settings where both the signal of interest and the side information signal follow correlated Gaussian mixture models, can also be generalized to other data models -- this can then translate into applications of the framework to scenarios where signals conform to different modalities. 
\end{itemize}
\appendices

\section{Proof of Theorem~\ref{theo:zeromean}}
\label{app:B}

We start by considering the case, $K_1=2$. We recall that the Batthacharyya upper bound for the misclassification probability of $C_1$ is given by
%
\begin{IEEEeqnarray}{rCl}
\bar{P}\sub{err} & = & \sqrt{p_{C_1}(1) p_{C_1} (2)} \int \sqrt{p(\mathbf{y} | C_1=1)p(\mathbf{y} | C_1=2)}d\mathbf{y} \\
&=& \sqrt{p_{C_1}(1) p_{C_1} (2)} \int  \sqrt{    \sum_{k.\ell=1}^{K_2}  p_{C_2|C_1}(k|1) p_{C_2|C_1}(\ell |2) p(\mathbf{y} | C_1=1, C_2=k)  p(\mathbf{y} | C_1=2, C_2=\ell)  }d\mathbf{y}. \IEEEeqnarraynumspace  
\label{eq:BhattM1app}
\end{IEEEeqnarray}

An upper and a lower bound to the expression in (\ref{eq:BhattM1app}) are simply obtained by considering the following fact. Given $n$ non-negative numbers $a_1, \dots, a_n \geq 0$, it holds
\begin{equation}
\frac{1}{\sqrt{n}}  \sum_{i=1}^n  \sqrt{a_i}  \leq \sqrt{\sum_{i=1}^n a_i}  \leq  \sum_{i=1}^n  \sqrt{a_i},
\label{eq:sqrt_inequality}
\end{equation}
where the first inequality derives from the concavity of the function $f(x)= \sqrt{x}$ and the second inequality can be simply proved by induction starting from $n=2$.

Then, an upper bound to $\bar{P}\sub{err}$ is obtained as 
\begin{equation}
\bar{P}\sub{err}^{\sf U} = \sqrt{p_{C_1}(1) p_{C_1} (2)}     \sum_{k.\ell=1}^{K_2}  \sqrt{ p_{C_2|C_1}(k|1) p_{C_2|C_1}(\ell |2)} \int  \sqrt{ p(\mathbf{y} | C_1=1, C_2=k)  p(\mathbf{y} | C_1=2, C_2=\ell)  } d\mathbf{y},
\end{equation}
and, similarly, a lower bound is given by $\bar{P}\sub{err}^{\sf L}=  \bar{P}\sub{err}^{\sf U}/K_2$. The generalization of this result to the case $K_1 >2$, is based on the evaluation of the union bound (\ref{eq:BhattM1union}), which, together with (\ref{eq:sqrt_inequality}), yields the upper bound 
\begin{equation}
\bar{P}\sub{err}^{\sf U} = \sum_{i=1}^{K_1}\sum_{\substack{j=1\\ j\neq i}}^{K_1}  {\color{black}\sqrt{p_{C_1}(i) p_{C_1} (j)}}    \sum_{k.\ell=1}^{K_2}  \sqrt{ p_{C_2|C_1}(k|i) p_{C_2|C_1}(\ell |j)} \int  \sqrt{ p(\mathbf{y} | C_1=i, C_2=k)  p(\mathbf{y} | C_1=j, C_2=\ell)  } d\mathbf{y},
\label{eq:PerrUpUnion}
\end{equation}
and the corresponding lower bound $\bar{P}\sub{err}^{\sf L}=  \bar{P}\sub{err}^{\sf U}/K_2$. 

Note that the lower and the upper bounds differ only by the multiplicative constant $K_2$. Therefore, they are tight bounds in terms of the diversity-order, and it is possible to derive the diversity-order associated to $\bar{P}\sub{err}$ from the analysis of such bounds. 

We now observe that the integral in (\ref{eq:PerrUpUnion}) also appears in the analysis of the upper bound to the misclassification probability associated to the classification between two Gaussian distributions without side information as described in \cite{Reboredo14}. In particular, \rev{on assuming $\sigma_1^2=\sigma_2^2=\sigma^2$}, such integral can be expressed as
\begin{equation}
 \int  \sqrt{ p(\mathbf{y} | C_1=i, C_2=k)  p(\mathbf{y} | C_1=j, C_2=\ell)  } d\mathbf{y}= e^{-K(ik,j\ell)},
 \label{eq:intexp}
\end{equation}
where
\begin{IEEEeqnarray}{rCl}
\nonumber
K(ik,j\ell) & = & \frac{1}{8}   (\boldsymbol{\mu}_{\mathbf{x}}^{(ik)}   - \boldsymbol{\mu}_{\mathbf{x}}^{(j\ell)})\tra \mathbf{\Phi}\tra  \left[     \frac{       \mathbf{\Phi}   ( \bar{\mathbf{\Sigma}}_{\mathbf{x}}^{(ik)}    +   \bar{\mathbf{\Sigma}}_{\mathbf{x}}^{(j\ell)}) \mathbf{\Phi}\tra   + 2 \sigma^2 \mathbf{I}       }{2}    \right]^{-1} \mathbf{\Phi}  (\boldsymbol{\mu}_{\mathbf{x}}^{(ik)}   - \boldsymbol{\mu}_{\mathbf{x}}^{(j\ell)}) \\
\label{eq:Kikjell}
& & + \frac{1}{2} \log \frac{  \mathrm{det}\left(  \frac{       \mathbf{\Phi}   ( \bar{\mathbf{\Sigma}}_{\mathbf{x}}^{(ik)}    +   \bar{\mathbf{\Sigma}}_{\mathbf{x}}^{(j\ell)}) \mathbf{\Phi}\tra   + 2 \sigma^2 \mathbf{I}       }{2}     \right)  }{    \sqrt{\mathrm{det} ( \mathbf{\Phi}    \bar{\mathbf{\Sigma}}_{\mathbf{x}}^{(ik)}     \mathbf{\Phi}\tra   +  \sigma^2 \mathbf{I}     )}    \sqrt{\mathrm{det} ( \mathbf{\Phi}    \bar{\mathbf{\Sigma}}_{\mathbf{x}}^{(j\ell)}     \mathbf{\Phi}\tra   +  \sigma^2 \mathbf{I}     )}    }.
\end{IEEEeqnarray}
For the case of zero-mean classes, i.e., assuming $\boldsymbol{\mu}_{\mathbf{x}}^{(ik)}=\boldsymbol{\mu}_{\mathbf{x}}^{(j\ell)}=\mathbf{0}$, a low-rank expansion for the integral in (\ref{eq:PerrUpUnion}) is given by~\cite[Theorem 1]{Reboredo14}
\begin{equation}
e^{-K(ik,j\ell)} = A^{(ik,j\ell)} \cdot (\sigma^2)^{d(ik,j\ell)} + o\left((\sigma^2)^{d(ik,j\ell)}\right),
\label{eq:expansionexp}
\end{equation}
for a fixed constant $A^{(ik,j\ell)}>0$, and with $d(ik,j\ell)$ given by
\begin{equation}
d(ik, j\ell) = \frac{1}{2} \left( r^{(ik,j\ell)}- \frac{r^{(ik)} + r^{(j\ell)}}{2}    \right),
\label{dikjl}
\end{equation}
where
\begin{equation}
r^{(ik)}=\rank\left(   \mathbf{\Phi}  \bar{\mathbf{\Sigma}}_{\mathbf{x}}^{(ik)} \mathbf{\Phi}\tra \right) 
\qv
r^{(j\ell)}=\rank\left(   \mathbf{\Phi}  \bar{\mathbf{\Sigma}}_{\mathbf{x}}^{(j\ell)} \mathbf{\Phi}\tra \right) 
\qv
r^{(ik,j\ell)}=\rank\left(   \mathbf{\Phi} ( \bar{\mathbf{\Sigma}}_{\mathbf{x}}^{(ik)} + \bar{\mathbf{\Sigma}}_{\mathbf{x}}^{(j\ell)}) \mathbf{\Phi}\tra  \right).
\label{eq:ranks}
\end{equation}
Therefore, we can conclude that a low-rank expansion for the upper bound of the misclassification probability (\ref{eq:BhattM1union}) is given by 
\begin{equation}
\bar{P}\sub{err}(\sigma^2) = A \cdot  (\sigma^2)^{d} + o\left((\sigma^2)^{d}\right),
\end{equation}
where $A>0$ is a fixed constant and
\begin{equation}
d= \min_{i\neq j \in \{1,\dots, K_1\}} \min_{k,\ell \in \{1,\ldots,K_2 \}}  d(ik, j\ell) = \min_{i,k,j,\ell, i\neq j} d(ik, j\ell)
\end{equation}
is the worst-case diversity-order associated to the misclassification of pairs of Gaussian distributions identified by the index pairs $(i,k)$ and $(j,\ell)$, such that $i \neq j$.

It is then clear that the computation of the expansion of $\bar{P}\sub{err}$ for classification with side information requires the computation of the diversity-order terms (\ref{dikjl}), and, therefore, the computation of the ranks $r^{(ik)}=\rank(\mathbf{\Gamma}^{(ik)})$ and $r^{(ik,j\ell)}=\rank(\mathbf{\Gamma}^{(ik,j\ell)})$, with $\mathbf{\Gamma}^{(ik)}=\mathbf{\Phi} \mathbf{P}^{(ik)}$ and $\mathbf{\Gamma}^{(ik,j\ell)}=\mathbf{\Phi} \mathbf{P}^{(ik,j\ell)}$, where
\begin{IEEEeqnarray}{rCl}
\mathbf{P}^{(ik)} & = &\left[
\begin{array}{ccc}
 \mathbf{P}^{(ik)}\sub{c_1} &  \mathbf{P}^{(ik)}_1& \mathbf{0} \\
 \mathbf{P}^{(ik)}\sub{c_2} & \mathbf{0} & \mathbf{P}^{(ik)}_2
\end{array}
\right] \\
\mathbf{P}^{(ik,j\ell)}& = &
\left[
\begin{array}{ccc}
\mathbf{P}\sub{c_1}^{(ik,j\ell)} & \mathbf{P}_1^{(ik,j\ell)} & \mathbf{0} \\
\mathbf{P}\sub{c_2}^{(ik,j\ell)} & \mathbf{0} & \mathbf{P}_2^{(ik,j\ell)}
\end{array}
\right]
\label{eq:P}
\end{IEEEeqnarray}
and
\begin{equation}
\mathbf{\Phi} = \left[
\begin{array}{ccc}
\mathbf{\Phi}_1 & \mathbf{0} \\
\mathbf{0}  & \mathbf{\Phi}_2
\end{array}
\right].
\label{eq:Phi2}
\end{equation}

Therefore, in the following we will provide a characterization of such ranks as a function of the numbers of features $m_1$ and $m_2$. For ease of a compact notation, we drop superscripts when results hold for all possible choices of index pairs $(i,k)$ or quadruples $(i,k,j,\ell)$. 
For the ease of notation, we will assume in the following $n_1 \geq m_1$ and $n_2  \geq m_2$. However, the extension to the case where $n_1 < m_1$ or $n_2  < m_2$ is straightforward.
 
\begin{lemma}
\label{theo:1bis}
Let $\mathbf{P}\sub{c_1} \in \mathbb{R}^{n_1 \times s\sub c},\mathbf{P}\sub{c_2} \in \mathbb{R}^{n_2 \times s\sub{c}}, \mathbf{P}_1 \in \mathbb{R}^{n_1 \times s_1}, \mathbf{P}_2 \in \mathbb{R}^{n_2 \times s_2}$. Let $\mathbf{\Phi} \in \mathbb{R}^{(m_1 + m_2) \times (n_1 + n_2)}$ as in (\ref{eq:Phi2}), such that the row spaces associated to $\mathbf{\Phi}_1$ and $\mathbf{\Phi}_2$ are $m_1-$ and $m_2-$dimensional spaces, isotropically distributed at random in $\mathbb{R}^{n_1}$ and $\mathbb{R}^{n_2}$, respectively, and let $\mathbf{P}$ as
\begin{equation}
\mathbf{P} =\left[
\begin{array}{ccc}
 \mathbf{P}\sub{c_1} &  \mathbf{P}_1& \mathbf{0} \\
 \mathbf{P}\sub{c_2} & \mathbf{0} & \mathbf{P}_2
\end{array}
\right].
\end{equation}
Then, with probability 1, the rank of the matrix $\mathbf{\Gamma}=\mathbf{\Phi} \mathbf{P}$ is given by
\begin{equation}
r =\mathrm{rank}(\mathbf{\Gamma}) =  \min \left\{  r_{\mathbf{x}},   \min\{  m_1, r_{\mathbf{x}_1}  \}  + \min\{  m_2, r_{\mathbf{x}_2}  \}   \right\},
\label{eq:theo1bis}
\end{equation}
where $r_{\mathbf{x}_1}=\rank[\mathbf{P}\sub{c_1} \, \mathbf{P}_1], r_{\mathbf{x}_2}=\rank[\mathbf{P}\sub{c_2} \, \mathbf{P}_2]$ and $r_{\mathbf{x}}=\rank (\mathbf{P})$.
\end{lemma}

\begin{IEEEproof}
It is easy to observe that the expression in (\ref{eq:theo1bis}) represents an upper bound to the rank $r=\rank(\mathbf{\Phi} \mathbf{P})$ as $\rank ( \mathbf{\Phi} \mathbf{P}) \leq \rank(\mathbf{P})$ and $\rank(\mathbf{\Gamma})$ is always less than or equal to the sum of the ranks of the matrices obtained by considering separately its first $m_1$ and the remaining $m_2$ rows, i.e., $\rank(\mathbf{\Phi}_1[ \mathbf{P}\sub{c_1} \,  \mathbf{P}_1 \, \mathbf
{0}])$ and $\rank(\mathbf{\Phi}_2[ \mathbf{P}\sub{c_2} \, \mathbf{0} \,  \mathbf{P}_2])$. Therefore, in the rest of the proof, we will aim at showing that such upper bound is actually tight, by proving that we can find at least $r$ linear independent columns in $\mathbf{\Gamma}$.

We start by considering the special case in which we impose $\mathbf{\Phi}_2 =\mathbf{I}_{n_2}$, and we show that in this case it holds
\begin{equation}
r_{I_2}= \rank \left(\left[
\begin{array}{ccc}
\mathbf{\Phi}_1 & \mathbf{0} \\
\mathbf{0}  & \mathbf{I}
\end{array}
\right]  \left[
\begin{array}{ccc}
 \mathbf{P}\sub{c_1} &  \mathbf{P}_1& \mathbf{0} \\
 \mathbf{P}\sub{c_2} & \mathbf{0} & \mathbf{P}_2
\end{array}
\right] \right) = \min \{ r_{\mathbf{x}} , \min\{m_1, r_{\mathbf{x}_1} \} + r_{\mathbf{x}_2}  \}.
\end{equation}
On recalling Sylvester's rank theorem \cite{Meyer00}, which states
\begin{equation}
\rank( \mathbf{A} \mathbf{B}) = \rank(\mathbf{B}) - \dim(\mathrm{Im}(\mathbf{B})   \cap \mathrm{Null}(\mathbf{A})),
\end{equation}
we can write
\begin{equation}
r_{I_2} = \rank (\mathbf{P}) -\dim \left( \mathrm{Im}(\mathbf{P}) \cap  \mathrm{Null}  \left[
\begin{array}{ccc}
\mathbf{\Phi}_1 & \mathbf{0} \\
\mathbf{0}  & \mathbf{I}
\end{array}
\right]   \right).
\end{equation}
Then, we consider the matrix $\mathbf{\Psi}_1 \in \mathbb{R}^{n_1 \times (n_1 -m_1)}$, whose columns form a basis for the null space $\mathrm{Null}(\mathbf{\Phi}_1)$, which is isotropically distributed among the $(n_1-m_1)$-dimensional spaces in $\mathbb{R}^{n_1}$. It is then straightforward to show that the columns of the matrix $[\mathbf{\Psi}_1\tra \, \mathbf{0}_{n_2 \times (n_1-m_1)}\tra]\tra$ span the null space 
\begin{equation}
\mathrm{Null}  \left[
\begin{array}{ccc}
\mathbf{\Phi}_1 & \mathbf{0} \\
\mathbf{0}  & \mathbf{I}
\end{array}
\right] 
\end{equation}
and we can write 
\begin{IEEEeqnarray}{rCl}
r_{I_2} &=& r_{\mathbf{x}} - \dim \left( \mathrm{Im}(\mathbf{P}) \cap   \mathrm{Im}  \left[
\begin{array}{cc}
\mathbf{\Psi}_1 \\
\mathbf{0}
\end{array}
\right]   \right) \\
&= &\rank \left[
\begin{array}{cccc}
 \mathbf{P}\sub{c_1} &  \mathbf{P}_1& \mathbf{0} & \mathbf{\Psi}_1 \\
 \mathbf{P}\sub{c_2} & \mathbf{0} & \mathbf{P}_2 & \mathbf{0}
\end{array}
\right] - (n_1 - m_1),
\end{IEEEeqnarray}
in which we have leveraged the rank equality for block matrices~\cite{Tian06},
\begin{equation}
\rank\left[\mathbf{A} \,  \mathbf{B} \right] = \rank(\mathbf{A}) + \rank(\mathbf{B})   - \dim(\mathrm{Im}(\mathbf{A}) \cap \mathrm{Im}(\mathbf{B}) ),
\label{eq:rankAB1}
\end{equation}
and the fact that $\rank(\mathbf{\Psi}_1) = n_1 - m_1$ and $\rank(\mathbf{P}) = r_{\mathbf{x}}$. Consider now the computation of the rank
\begin{equation}
r_{\mathbf{\Psi}_1} = \rank \left[
\begin{array}{cccc}
 \mathbf{P}\sub{c_1} &  \mathbf{P}_1& \mathbf{0} & \mathbf{\Psi}_1 \\
 \mathbf{P}\sub{c_2} & \mathbf{0} & \mathbf{P}_2 & \mathbf{0}
\end{array}
\right] .
\end{equation}
In order to compute such rank, we will leverage the \ac{GSVD} as described in \cite{Paige81}. In particular, consider two matrices $\mathbf{A}\in \mathbb{R}^{n \times p}$ and $\mathbf{B} \in \mathbb{R}^{m \times p}$, with the same number of columns, and with $r_{\mathbf{A}}=\rank(\mathbf{A}), r_{\mathbf{B}}=\rank(\mathbf{B}),r_{\mathbf{A}\mathbf{B}}=\rank[\mathbf{A}\tra \mathbf{B}\tra]\tra$ and $s_{\mathbf{AB}}= r_{\mathbf{A}}+ r_{\mathbf{B}}-r_{\mathbf{AB}}$. Then, there exist two orthogonal matrices $\mathbf{U}\in \mathbb{R}^{n \times n}, \mathbf{V}\in \mathbb{R}^{m \times m}$ and a non-singular matrix $\mathbf{X}\in \mathbb{R}^{p \times p}$ such that
\begin{equation}
\mathbf{U}\tra \mathbf{A} \mathbf{X} = [\mathbf{\Lambda}_{\mathbf{A}} \, \mathbf{0}_{n \times (p-r_{\mathbf{A}\mathbf{B}})}]  \qv  \mathbf{V}\tra \mathbf{B} \mathbf{X} = [\mathbf{\Lambda}_{\mathbf{B}} \, \mathbf{0}_{m \times (p-r_{\mathbf{A}\mathbf{B}})}] ,
\label{eq:GSVD1}
\end{equation}
where
\begin{equation}
\mathbf{\Lambda}_{\mathbf{A}}=
\kbordermatrix{~    & r_{\mathbf{AB}}-r_{\mathbf{B}} & s_{\mathbf{AB}} & r_{\mathbf{AB}}-r_{\mathbf{A}} \cr
		      r_{\mathbf{AB}}-r_{\mathbf{B}}  & \mathbf{I} &  & \cr
		      s_{\mathbf{AB}}    &    & \mathbf{D}_\mathbf{A} &  \cr
		      n-r_{\mathbf{A}}  &   &  & \mathbf{0}
		      }
\qv  
\mathbf{\Lambda}_{\mathbf{B}}=
\kbordermatrix{~    & r_{\mathbf{AB}}-r_{\mathbf{B}} & s_{\mathbf{AB}} & r_{\mathbf{AB}}-r_{\mathbf{A}} \cr
		      m-r_{\mathbf{B}} & \mathbf{0} &  &  \cr
		      s_{\mathbf{AB}}    &    & \mathbf{D}_{\mathbf{B}} &  \cr
		      r_{\mathbf{AB}}-r_{\mathbf{A}}  &    & & \mathbf{I}
		      }.
		 \label{eq:GSVD2}     
\end{equation}
and $\mathbf{D}_{\mathbf{A}} =\mathrm{diag}(\alpha_1, \ldots, \alpha_{s_{\mathbf{AB}}})$, $\mathbf{D}_{\mathbf{B}} =\mathrm{diag}(\beta_1, \ldots, \beta_{s_{\mathbf{AB}}})$, such that $1<\alpha_1 \leq \cdots \leq \alpha_{s_{\mathbf{AB}}}<0$ and $0<\beta_1 \leq \cdots \leq \beta_{s_{\mathbf{AB}}}<1$, and $\alpha_i^2+\beta_i^2=1$, for $i=1,\ldots,s_{\mathbf{AB}}$.

Therefore, on applying the \ac{GSVD} to the two matrices $[\mathbf{P}\sub{c_1} \, \mathbf{P}_1 \, \mathbf{0}]$ and $[\mathbf{P}\sub{c_2} \, \mathbf{0} \, \mathbf{P}_2]$, we can write
\begin{IEEEeqnarray}{rCl}
r_{\mathbf{\Psi}_1} &=& \rank\left(
\left[
\begin{array}{ccc}
\mathbf{U}\tra & \mathbf{0} \\
\mathbf{0} & \mathbf{V}\tra
\end{array}
\right]
 \left[
\begin{array}{cccc}
 \mathbf{P}\sub{c_1} &  \mathbf{P}_1& \mathbf{0} & \mathbf{\Psi}_1 \\
 \mathbf{P}\sub{c_2} & \mathbf{0} & \mathbf{P}_2 & \mathbf{0}
\end{array}
\right]
\left[
\begin{array}{ccc}
\mathbf{X} & \mathbf{0} \\
\mathbf{0} & \mathbf{I}_{n_1-m_1}
\end{array}
\right]
\right)\\
\label{eq:matrank}
& = & \rank \left[
\begin{array}{cccc}
\mathbf{\Lambda}_1 & \mathbf{0} & \mathbf{\Psi}_1' \\
\mathbf{\Lambda}_2 & \mathbf{0} & \mathbf{0}
\end{array}
\right]
\end{IEEEeqnarray}
where
\begin{equation}
\mathbf{\Lambda}_{1}=
\kbordermatrix{~    & r_{\mathbf{x}}-r_{\mathbf{x}_2} & r_{\mathbf{x}_1}+r_{\mathbf{x}_2}-r_{\mathbf{x}} & r_{\mathbf{x}}-r_{\mathbf{x}_1} \cr
		      r_{\mathbf{x}}-r_{\mathbf{x}_2}  & \mathbf{I} &  & \cr
		      r_{\mathbf{x}_1}+r_{\mathbf{x}_2}-r_{\mathbf{x}}     &    & \mathbf{D}_{1} &  \cr
		      n_1-r_{\mathbf{x}_1}  &   &  & \mathbf{0}
		      }
\qv  
\mathbf{\Lambda}_{2}=
\kbordermatrix{~    & r_{\mathbf{x}}-r_{\mathbf{x}_2} & r_{\mathbf{x}_1}+r_{\mathbf{x}_2}-r_{\mathbf{x}} & r_{\mathbf{x}}-r_{\mathbf{x}_1} \cr
		      n_2-r_{\mathbf{x}_2} & \mathbf{0} &  &  \cr
		      r_{\mathbf{x}_1}+r_{\mathbf{x}_2}-r_{\mathbf{x}}       &    & \mathbf{D}_{2} &  \cr
		      r_{\mathbf{x}}-r_{\mathbf{x}_1}  &    & & \mathbf{I}
		      }
\end{equation}
and where $\mathbf{\Psi}_1'=\mathbf{U}\tra \mathbf{\Psi}_1$ is a matrix whose column space is still isotropically distributed at random among the $(n_1-m_1)-$dimensional subspaces of $\mathbb{R}^{n_1}$. Now, by considering the first $r_{\mathbf{x}} -r_{\mathbf{x}_2}$ columns of the matrix in (\ref{eq:matrank}) together with its last $n_1 -m_1$ columns, given the fact the columns in $\mathbf{\Psi}'_1$ form a random space in $\mathbb{R}^{n_1}$, we can conclude that, with probability 1, we can pick from such columns $\min\{ r_{\mathbf{x}} -r_{\mathbf{x}_2}+n_1-m_1,n_1\}$ independent columns, which are also independent from the remaining $(r_{\mathbf{x}_1}+r_{\mathbf{x}_2}-r_{\mathbf{x}})+( r_{\mathbf{x}}-r_{\mathbf{x}_1})=r_{\mathbf{x}_2}$ non-zero columns of the same matrix. 
Therefore, we have 
\begin{equation}
r_{\mathbf{\Psi}_1}=\min \{ r_{\mathbf{x}} -r_{\mathbf{x}_2}+n_1-m_1,n_1\}+r_{\mathbf{x}_2}
\end{equation}
and then
\begin{IEEEeqnarray}{rCl}
r_{I_2} & =& \min \{ r_{\mathbf{x}} -r_{\mathbf{x}_2}+n_1-m_1,n_1\}+r_{\mathbf{x}_2} -(n_1-m_1)\\
& = & \min \{r_{\mathbf{x}} , m_1 + r_{\mathbf{x}_2}  \} \\
& = &\min \{r_{\mathbf{x}} , \min \{m_1, r_{\mathbf{x}_1} \} + r_{\mathbf{x}_2}  \} ,
\label{eq:rI2}
\end{IEEEeqnarray}
where the last equality is obtained by observing that $r_{\mathbf{x}}\leq r_{\mathbf{x}_1} + r_{\mathbf{x}_2}$.

Consider now the general case, in which $\mathbf{\Phi}_2$ is not forced to be equal to the identity matrix. In this case, by leveraging (\ref{eq:rI2}), we can write
\begin{IEEEeqnarray}{rCl}
r & = &  \rank \left(\left[
\begin{array}{ccc}
\mathbf{\Phi}_1 & \mathbf{0} \\
\mathbf{0}  & \mathbf{\Phi}_2
\end{array}
\right]  \left[
\begin{array}{ccc}
 \mathbf{P}\sub{c_1} &  \mathbf{P}_1& \mathbf{0} \\
 \mathbf{P}\sub{c_2} & \mathbf{0} & \mathbf{P}_2
\end{array}
\right] \right) \\
& = &  \rank \left(\left[
\begin{array}{ccc}
\mathbf{\Phi}_1 & \mathbf{0} \\
\mathbf{0}  & \mathbf{I}
\end{array}
\right]  \left[
\begin{array}{ccc}
 \mathbf{P}\sub{c_1} &  \mathbf{P}_1& \mathbf{0} \\
\mathbf{\Phi}_2 \mathbf{P}\sub{c_2} & \mathbf{0} &\mathbf{\Phi}_2 \mathbf{P}_2
\end{array}
\right] \right)\\
& = & \min \{   r_{I_1}, \min \{ m_1, r_{\mathbf{x}_1} \}  +  \min \{ m_2, r_{\mathbf{x}_2} \}    \},
\end{IEEEeqnarray}
in which we have introduced the symbol
\begin{IEEEeqnarray}{rCl}
r_{I_1} &=& \rank  \left[
\begin{array}{ccc}
 \mathbf{P}\sub{c_1} &  \mathbf{P}_1& \mathbf{0} \\
\mathbf{\Phi}_2 \mathbf{P}\sub{c_2} & \mathbf{0} &\mathbf{\Phi}_2 \mathbf{P}_2
\end{array}
\right]\\
&=&
\rank \left(\left[
\begin{array}{ccc}
\mathbf{I} & \mathbf{0} \\
\mathbf{0}  & \mathbf{\Phi}_2
\end{array}
\right]  \left[
\begin{array}{ccc}
 \mathbf{P}\sub{c_1} &  \mathbf{P}_1& \mathbf{0} \\
 \mathbf{P}\sub{c_2} & \mathbf{0} & \mathbf{P}_2
\end{array}
\right] \right),
\end{IEEEeqnarray}
and where we have used the fact that $\rank [\mathbf{\Phi}_2 \mathbf{P}\sub{c_2} \, \mathbf{0} \, \mathbf{\Phi}_2 \mathbf{P}_2]=\min \{ m_2, r_{\mathbf{x}_2} \}$. Then, with a procedure similar to that used to compute $r_{I_2}$, it is possible to show that $r_{I_1} = \min \{r_{\mathbf{x}}, r_{\mathbf{x}_1}+ \min \{m_2, r_{\mathbf{x}_2} \} \}$, thus leading to
\begin{IEEEeqnarray}{rCl}
r &=& \min \left\{     \min \{r_{\mathbf{x}}, r_{\mathbf{x}_1}+ \min \{m_2, r_{\mathbf{x}_2} \} \}, \min\{ m_1,r_{\mathbf{x}_1}  \}  + \min\{ m_2,r_{\mathbf{x}_2}  \}     \right\}   \\
 & = &  \min \{r_{\mathbf{x}}, \min\{ m_1,r_{\mathbf{x}_1}  \}  + \min\{ m_2,r_{\mathbf{x}_2}  \}    \} .
\end{IEEEeqnarray}

\end{IEEEproof}

Finally, note that Lemma~\ref{theo:1bis} can be immediately applied to compute $r^{(ik)},r^{(j\ell)}$ and $r^{(ik,j\ell)}$, thus concluding the proof of Theorem~\ref{theo:zeromean}.

We also note in passing that the generalization of Lemma~\ref{theo:1bis} to the case of multiple side information sources, $\mathbf{x}_2,\mathbf{x}_3, \ldots, \mathbf{x}_L$, seems to be considerably more complex, due to the absence of a transform akin to the GSVD in (\ref{eq:GSVD1}) and (\ref{eq:GSVD2}) that jointly diagonalizes more than two matrices.

\section{Proof of Corollary \ref{cor:phasetrans}}
\label{app:C}

On the basis of the low-rank regime expansion for the upper bound to the misclassification probability (\ref{eq:BhattM1union}) contained in Theorem~\ref{theo:zeromean}, we can state that \rev{condition} (\ref{eq:phasetrans}) \rev{is verified} if and only if $d > 0$, which is equivalent to $d(ik,j\ell)> 0$, $\forall (i,k,j,\ell) \in \mathcal{S}\sub{SIC}$. Moreover, on observing that the matrices $ \mathbf{\Phi}  \bar{\mathbf{\Sigma}}_{\mathbf{x}}^{(ik)} \mathbf{\Phi}\tra$ and $ \mathbf{\Phi}  \bar{\mathbf{\Sigma}}_{\mathbf{x}}^{(j\ell)} \mathbf{\Phi}\tra$ are positive semidefinite, we can immediately state that $d(ik,j\ell)=0$ if and only if $r^{(ik,j\ell)}=r^{(ik)}=r^{(j\ell)}$, which is verified if only if~\cite[Lemma 2]{RecJournal}
\begin{equation}
\mathrm{Im}( \mathbf{\Phi}  \bar{\mathbf{\Sigma}}_{\mathbf{x}}^{(ik)} \mathbf{\Phi}\tra) = \mathrm{Im} ( \mathbf{\Phi}  \bar{\mathbf{\Sigma}}_{\mathbf{x}}^{(j\ell)}\mathbf{\Phi}\tra).
\label{eq:sameIm}
\end{equation}
Then, $r_{\mathbf{x}}^{(ik,j\ell)}=r_{\mathbf{x}}^{(ik)}=r_{\mathbf{x}}^{j\ell)}$ implies that $\mathrm{Im} (  \bar{\mathbf{\Sigma}}_{\mathbf{x}}^{(ik)} )  = \mathrm{Im}   (\bar{\mathbf{\Sigma}}_{\mathbf{x}}^{(j\ell)})$ and, therefore, (\ref{eq:sameIm}) holds regardless of the expression of the projection kernel $\mathbf{\Phi}$, thus leading to $d(ik,j\ell)=0$.

Assume now $r_{\mathbf{x}}^{(ik,j\ell)}>r_{\mathbf{x}}^{(ik)},r_{\mathbf{x}}^{j\ell)}$. We can then use the rank expression (\ref{eq:theo1bis}) and consider separately the following cases:
\begin{enumerate}
\item $r_{\mathbf{x}_1}^{(ik,j\ell)}>r_{\mathbf{x}_1}^{(ik)},r_{\mathbf{x}_1}^{(j\ell)}$ and $r_{\mathbf{x}_2}^{(ik,j\ell)}>r_{\mathbf{x}_2}^{(ik)},r_{\mathbf{x}_2}^{(j\ell)}$: in this case, if $m_1>\min \{ r_{\mathbf{x}_1}^{(ik)}, r_{\mathbf{x}_1}^{(j\ell)}  \}$ or $m_2>\min \{ r_{\mathbf{x}_2}^{(ik)}, r_{\mathbf{x}_2}^{(j\ell)}  \}$, we can immediately conclude that $d(ik,j\ell)>0$, by simply considering the classification from the observation of $\mathbf{y}_1$ or $\mathbf{y}_2$ alone, respectively, and by leveraging the results in~\cite[Theorem 2]{Reboredo14}. On the other hand, if we assume $m_1\leq\min \{ r_{\mathbf{x}_1}^{(ik)}, r_{\mathbf{x}_1}^{(j\ell)}  \}, m_2 \leq\min \{ r_{\mathbf{x}_2}^{(ik)}, r_{\mathbf{x}_2}^{(j\ell)}  \}$ and $m_1+m_2>\min\{ r_{\mathbf{x}}^{(ik)},r_{\mathbf{x}}^{(j\ell)}  \}$, then we have $r^{(ik)}=\min \{ r_{\mathbf{x}}^{(ik)}, m_1+m_2 \}$, $r^{(j\ell)}=\min \{ r_{\mathbf{x}}^{(j\ell)}, m_1+m_2 \}$ and $r^{(ik,j\ell)}=\min \{ r_{\mathbf{x}}^{(ik,j\ell)}, m_1+m_2 \}$. Then, since $m_1+m_2>\min\{ r_{\mathbf{x}}^{(ik)},r_{\mathbf{x}}^{(j\ell)}  \}$, we have immediately that $r^{(ik,j\ell)}>\min\{r^{(ik)},r^{(j\ell)} \}$, and thus $d(ik,j\ell)>0$. Such sufficient conditions on the minimum number of measurements $m_1,m_2$ needed to guarantee $d(ik,j\ell)>0$ are also necessary. In fact, if $m_1\leq\min \{ r_{\mathbf{x}_1}^{(ik)}, r_{\mathbf{x}_1}^{(j\ell)}  \}, m_2 \leq\min \{ r_{\mathbf{x}_2}^{(ik)}, r_{\mathbf{x}_2}^{(j\ell)}  \}$ and $m_1+m_2 \leq \min\{ r_{\mathbf{x}}^{(ik)},r_{\mathbf{x}}^{(j\ell)}  \}$, then $r^{(ik,j\ell)}=r^{(ik)}=r^{(j\ell)}=m_1+m_2$.
\item $r_{\mathbf{x}_1}^{(ik,j\ell)}=r_{\mathbf{x}_1}^{(ik)}=r_{\mathbf{x}_1}^{(j\ell)}$ and $r_{\mathbf{x}_2}^{(ik,j\ell)}=r_{\mathbf{x}_2}^{(ik)}=r_{\mathbf{x}_2}^{(j\ell)}$: in this case, we note that $\min\{ m_1, r_{\mathbf{x}_1}^{(ik)} \}+ \min\{ m_2, r_{\mathbf{x}_2}^{(ik)}\}=\min\{ m_1, r_{\mathbf{x}_1}^{(j\ell)}\} + \min\{ m_2, r_{\mathbf{x}_2}^{(j\ell)}\}=\min\{ m_1, r_{\mathbf{x}_1}^{(ik,j\ell)} \}+ \min\{ m_2, r_{\mathbf{x}_2}^{(ik,j\ell)}\}=D$, and then $d(ik,j\ell)>0$ if and only if $D>\min \{  r_{\mathbf{x}}^{(ik)} ,  r_{\mathbf{x}}^{(j\ell)}  \}$. Then, we can split the analysis in further subcases as follows:
\begin{itemize}
\item if $m_1 \leq r_{\mathbf{x}_1}^{(ik)}$ and $m_2 \leq r_{\mathbf{x}_2}^{(ik)}$, then $d(ik,j\ell)>0$ if and only if $m_1+ m_2> \min \{  r_{\mathbf{x}}^{(ik)} ,  r_{\mathbf{x}}^{(j\ell)}  \} $;
\item if $m_1 > r_{\mathbf{x}_1}^{(ik)}$ and $m_2 \leq r_{\mathbf{x}_2}^{(ik)}$, then $d(ik,j\ell)>0$ if and only if $m_2> \min \{  r_{\mathbf{x}}^{(ik)}- r_{\mathbf{x}_1}^{(ik)} ,  r_{\mathbf{x}}^{(j\ell)} -r_{\mathbf{x}_1}^{(j\ell)}  \} $;
\item if $m_1 \leq r_{\mathbf{x}_1}^{(ik)}$ and $m_2 > r_{\mathbf{x}_2}^{(ik)}$, then $d(ik,j\ell)>0$ if and only if $m_1> \min \{  r_{\mathbf{x}}^{(ik)}- r_{\mathbf{x}_2}^{(ik)} ,  r_{\mathbf{x}}^{(j\ell)} -r_{\mathbf{x}_2}^{(j\ell)}  \} $;
\item if $m_1 > r_{\mathbf{x}_1}^{(ik)}$ and $m_2 > r_{\mathbf{x}_2}^{(ik)}$, then $r_{\mathbf{x}_1}^{(ik)}+r_{\mathbf{x}_2}^{(ik)}=r_{\mathbf{x}_1}^{(j\ell)}+r_{\mathbf{x}_2}^{(j\ell)}> \min \{  r_{\mathbf{x}}^{(ik)} ,  r_{\mathbf{x}}^{(j\ell)}  \}$ and therefore $d(ik,j\ell)>0$.
\end{itemize}
Finally, we can combine the previous expressions to obtain necessary and sufficient conditions to guarantee $d(ik,j\ell)>0$ as 
\begin{equation}
\left\{
\begin{array}{lll}
m_1> \min \{  r_{\mathbf{x}}^{(ik)} -r_{\mathbf{x}_2}^{(ik)},  r_{\mathbf{x}}^{(j\ell)} -r_{\mathbf{x}_2}^{(j\ell)}   \} \\
m_2> \min \{  r_{\mathbf{x}}^{(ik)} -r_{\mathbf{x}_1}^{(ik)},  r_{\mathbf{x}}^{(j\ell)} -r_{\mathbf{x}_1}^{(j\ell)}   \} \\
m_1+ m_2> \min \{  r_{\mathbf{x}}^{(ik)} ,  r_{\mathbf{x}}^{(j\ell)}  \} \\
\end{array}
\right. .
\end{equation}
\item $r_{\mathbf{x}_1}^{(ik,j\ell)}>r_{\mathbf{x}_1}^{(ik)},r_{\mathbf{x}_1}^{(j\ell)}$ and $r_{\mathbf{x}_2}^{(ik,j\ell)}=r_{\mathbf{x}_2}^{(ik)}=r_{\mathbf{x}_2}^{(j\ell)}$: we can prove immediately that, in this case, if $m_1>\min \{ r_{\mathbf{x}_1}^{(ik)},r_{\mathbf{x}_1}^{(j\ell)} \}$, then $d(ik,j\ell)>0$, simply by considering classification based on the observation of only $\mathbf{y}_1$. Assume now $m_1 \leq \min \{ r_{\mathbf{x}_1}^{(ik)},r_{\mathbf{x}_1}^{(j\ell)} \}$. Then, it holds $r^{(ik)} = \min \{ r_{\mathbf{x}}^{(ik)}, m_1 + \min\{ m_2,r_{\mathbf{x}_2}^{(ik)} \}  \} $, $r^{(j\ell)} = \min \{ r_{\mathbf{x}}^{(j\ell)}, m_1 + \min\{ m_2,r_{\mathbf{x}_2}^{(j\ell)} \}  \} $ and $r^{(ik,j\ell)} = \min \{ r_{\mathbf{x}}^{(ikj\ell)}, m_1 + \min\{ m_2,r_{\mathbf{x}_2}^{(ik,j\ell)} \}  \} $, and, on observing that $\min\{ m_2,r_{\mathbf{x}_2}^{(ik)} \}=\min\{ m_2,r_{\mathbf{x}_2}^{(j\ell)} \}=\min\{ m_2,r_{\mathbf{x}_2}^{(ik,j\ell)} \}$, we have immediately that $d(ik,j\ell)> 0$ if and only if $\min \{ r_{\mathbf{x}}^{(ik)}, r_{\mathbf{x}}^{(j\ell)} \} <  m_1 + \min\{ m_2,r_{\mathbf{x}_2}^{(ik)} \}$. In particular, if $m_2 \leq r_{\mathbf{x}_2}^{(ik)}$, then, $d(ik,j\ell)>0$ if and only if $m_1+ m_2 > \min\{ r_{\mathbf{x}}^{(ik)},r_{\mathbf{x}}^{(j\ell)}  \}$, whereas if $m_2>r_{\mathbf{x}_2}^{(ik)}$, then $d(ik, j\ell)>0$ if and only if $m_1>\min\{ r_{\mathbf{x}}^{(ik)} - r_{\mathbf{x}_2}^{(ik)} , r_{\mathbf{x}}^{(j\ell)} - r_{\mathbf{x}_2}^{(j\ell)}  \}$. Finally, on combining these expressions, we can write necessary and sufficient conditions for $d(ik,j\ell)>0$ as
\begin{equation}
m_1 > \min \{ r_{\mathbf{x}_1}^{(ik)},r_{\mathbf{x}_1}^{(j\ell)}  \}  \quad \mathrm{or}  \quad 
\left\{
\begin{array}{lll}
m_1> \min \{  r_{\mathbf{x}}^{(ik)} -r_{\mathbf{x}_2}^{(ik)},  r_{\mathbf{x}}^{(j\ell)} -r_{\mathbf{x}_2}^{(j\ell)}   \} \\
m_1+ m_2> \min \{  r_{\mathbf{x}}^{(ik)} ,  r_{\mathbf{x}}^{(j\ell)}  \} \\
\end{array}
\right. .
\end{equation}
\item $r_{\mathbf{x}_1}^{(ik,j\ell)}=r_{\mathbf{x}_1}^{(ik)}=r_{\mathbf{x}_1}^{(j\ell)}$ and $r_{\mathbf{x}_2}^{(ik,j\ell)}>r_{\mathbf{x}_2}^{(ik)},r_{\mathbf{x}_2}^{(j\ell)}$: the proof for this case follows steps similar to the case $r_{\mathbf{x}_1}^{(ik,j\ell)}>r_{\mathbf{x}_1}^{(ik)},r_{\mathbf{x}_1}^{(j\ell)}$ and $r_{\mathbf{x}_2}^{(ik,j\ell)}=r_{\mathbf{x}_2}^{(ik)}=r_{\mathbf{x}_2}^{(j\ell)}$.
\end{enumerate}

\section{Proof of Theorem \ref{theo:nonzero_SI}}
\label{app:D}

The characterization of the low-rank expansion of the upper bound to the misclassification probability in (\ref{eq:BhattM1union}) for the case of nonzero-mean classes starts from the analysis of its lower and upper bounds presented in Appendix~\ref{app:B}. We focus on the expressions in (\ref{eq:PerrUpUnion}), (\ref{eq:intexp}) and (\ref{eq:Kikjell}), and we leverage the low-rank expansion of the integral in (\ref{eq:intexp}) presented in~\cite[Theorem 3]{Reboredo14} for the case of two nonzero-mean Gaussian classes. Namely, we recall that 
\begin{equation}
e^{-K(ik,j\ell)} = B^{(ik,j\ell)} \cdot e^{-C^{(ik,j\ell)}/\sigma^2} + o \left(  e^{-C^{(ik,j\ell)}/\sigma^2}  \right),
\end{equation}
for fixed constants $B^{(ik,j\ell)},C^{(ik,j\ell)} > 0$ if and only if
\begin{equation}
\mathbf{\Phi}(\boldsymbol{\mu}_{\mathbf{x}}^{(ik)}-\boldsymbol{\mu}_{\mathbf{x}}^{(j\ell)}) \notin \mathrm{Im} \left(   \mathbf{\Phi} ( \bar{\mathbf{\Sigma}}_{\mathbf{x}}^{(ik)} + \bar{\mathbf{\Sigma}}_{\mathbf{x}}^{(j\ell)}) \mathbf{\Phi}\tra  \right).
\label{eq:nonzero_hugo}
\end{equation}
Otherwise, the integral in (\ref{eq:intexp}) can be expanded as in (\ref{eq:expansionexp}). Therefore, if condition (\ref{eq:nonzero_hugo}) is verified for all the index quadruples $(i,k,j,\ell) \in \mathcal{S}\sub{SIC}$, then we can expand the upper bound to the misclassification probability in (\ref{eq:BhattM1union}) as 
\begin{equation}
\bar{P}\sub{err}(\sigma^2) = B \cdot e^{-C/\sigma^2} + o \left(  e^{-C^/\sigma^2}  \right),
\end{equation}
for fixed constants $B,C >0$. Otherwise, the upper bound of the misclassification probability is expanded as
\begin{equation}
\bar{P}\sub{err}(\sigma^2) = A \cdot (\sigma^2)^d + o \left(  (\sigma^2)^d  \right),
\end{equation}
for a fixed $A>0$ and where 
\begin{equation}
d=\min_{(i,k,j,\ell) \in \mathcal{S}'} d(ik,j\ell),
\end{equation}
where $\mathcal{S}'$ is the set of the index quadruples $(i,k,j,\ell) \in \mathcal{S}\sub{SIC}$ for which (\ref{eq:nonzero_hugo}) is not verified and $d(ik,j\ell)$ is as in~(\ref{dikjl}).

We can now provide necessary and sufficient conditions on $m_1$ and $m_2$ such that (\ref{eq:nonzero_hugo}) is verified. We observe that (\ref{eq:nonzero_hugo}) holds if and only if $r_{\boldsymbol{\mu}}^{(ik,j\ell)}>r^{(ik,j\ell)}$, where we have defined
\begin{equation}
r_{\boldsymbol{\mu}}^{(ik,j\ell)} =  \rank \left(\left[
\begin{array}{ccc}
\mathbf{\Phi}_1 & \mathbf{0} \\
\mathbf{0}  & \mathbf{\Phi}_2
\end{array}
\right]  \left[
\begin{array}{cccc}
\boldsymbol{\mu}_{\mathbf{x}_1 }^{(ik)}-\boldsymbol{\mu}_{\mathbf{x}_1 }^{(j\ell)}& \mathbf{P}\sub{c_1}^{(ik,j\ell)} &  \mathbf{P}_1^{(ik,j\ell)}& \mathbf{0} \\
\boldsymbol{\mu}_{ \mathbf{x}_2}^{(ik)}-\boldsymbol{\mu}_{\mathbf{x}_2}^{(j\ell)}& \mathbf{P}\sub{c_2}^{(ik,j\ell)} & \mathbf{0} & \mathbf{P}_2^{(ik,j\ell)}
\end{array}
\right] \right).
\end{equation}
Assume first that $\boldsymbol{\mu}_{\mathbf{x}}^{(ik)}-\boldsymbol{\mu}_{\mathbf{x}}^{(j\ell)} \in \mathrm{Im} \left(  \bar{\mathbf{\Sigma}}_{\mathbf{x}}^{(ik)} + \bar{\mathbf{\Sigma}}_{\mathbf{x}}^{(j\ell)}\right)$. Then $r_{\boldsymbol{\mu}}^{(ik,j\ell)}=r^{(ik,j\ell)}$, and, therefore (\ref{eq:nonzero_hugo}) does not hold, irrespectively of the exact matrix $\mathbf{\Phi}$. 

Assume now $\boldsymbol{\mu}_{\mathbf{x}}^{(ik)}-\boldsymbol{\mu}_{\mathbf{x}}^{(j\ell)} \notin \mathrm{Im} \left(  \bar{\mathbf{\Sigma}}_{\mathbf{x}}^{(ik)} + \bar{\mathbf{\Sigma}}_{\mathbf{x}}^{(j\ell)}  \right)$. We can use the rank expression (\ref{eq:theo1bis}) and similar steps to those in the proof of Corollary \ref{cor:phasetrans} in order to consider separately the following cases:
\begin{enumerate}
\item $\boldsymbol{\mu}_{\mathbf{x}_1}^{(ik)}-\boldsymbol{\mu}_{\mathbf{x}_1}^{(j\ell)}  \notin \mathrm{Im} (\bar{\mathbf{\Sigma}}_{\mathbf{x}_1}^{(ik)}+\bar{\mathbf{\Sigma}}_{\mathbf{x}_1}^{(j\ell)})$ and $\boldsymbol{\mu}_{\mathbf{x}_2}^{(ik)}-\boldsymbol{\mu}_{\mathbf{x}_2}^{(j\ell)}  \notin \mathrm{Im} (\bar{\mathbf{\Sigma}}_{\mathbf{x}_2}^{(ik)}+\bar{\mathbf{\Sigma}}_{\mathbf{x}_2}^{(j\ell)})$: on leveraging \cite[Lemma 3]{RecJournal}, we can observe that if $m_1 > r_{\mathbf{x}_1}^{(ik,j\ell)}$ or $m_2 > r_{\mathbf{x}_2}^{(ik,j\ell)}$, then (\ref{eq:nonzero_hugo}) holds, as the upper bound to the error probability obtained by classification based on the observation of $\mathbf{y}_1$ or $\mathbf{y}_2$ alone, respectively, decreases exponentially with $1/\sigma^2$. On the other hand, if $m_1 \leq r_{\mathbf{x}_1}^{(ik,j\ell)}, m_2 \leq r_{\mathbf{x}_2}^{(ik,j\ell)}$ and $m_1+m_2>r_{\mathbf{x}}^{(ik,j\ell)}$, then it holds $r_{\boldsymbol{\mu}}^{(ik,j\ell)}=\min \{ r_{\mathbf{x}}^{(ik,j\ell)}+1, m_1 + m_2  \} > r^{(ik,j\ell)} $ and thus (\ref{eq:nonzero_hugo}) is verified. The previous conditions are also shown to be necessary by noting that, if $m_1 \leq r_{\mathbf{x}_1}^{(ik,j\ell)}, m_2 \leq r_{\mathbf{x}_2}^{(ik,j\ell)}$ and $m_1+m_2 \leq r_{\mathbf{x}}^{(ik,j\ell)}$, then $r_{\boldsymbol{\mu}}^{(ik,j\ell)}=r^{(ik,j\ell)}=m_1+m_2$.
\item $\boldsymbol{\mu}_{\mathbf{x}_1}^{(ik)}-\boldsymbol{\mu}_{\mathbf{x}_1}^{(j\ell)}  \in \mathrm{Im} (\bar{\mathbf{\Sigma}}_{\mathbf{x}_1}^{(ik)}+\bar{\mathbf{\Sigma}}_{\mathbf{x}_1}^{(j\ell)})$ and $\boldsymbol{\mu}_{\mathbf{x}_2}^{(ik)}-\boldsymbol{\mu}_{\mathbf{x}_2}^{(j\ell)}  \in \mathrm{Im} (\bar{\mathbf{\Sigma}}_{\mathbf{x}_2}^{(ik)}+\bar{\mathbf{\Sigma}}_{\mathbf{x}_2}^{(j\ell)})$: in this case, $r_{\boldsymbol{\mu}}^{(ik,j\ell)}=\min\{ r_{\mathbf{x}_1,\mathbf{x}_2}^{(ik,j\ell)} + 1, \min \{ m_1, r_{\mathbf{x}_1}^{(ik,j\ell)}\} + \min \{ m_2, r_{\mathbf{x}_2}^{(ik,j\ell)}\} \}$, so that (\ref{eq:nonzero_hugo}) holds if and only if $\min \{ m_1, r_{\mathbf{x}_1}^{(ik,j\ell)}\} + \min \{ m_2, r_{\mathbf{x}_2}^{(ik,j\ell)}\}> r_{\mathbf{x}}^{(ik,j\ell)}$. Then, we can split the analysis in the following subcases:
\begin{itemize}
\item if $m_1 \leq r_{\mathbf{x}_1}^{(ik,j\ell)}$ and $m_2 \leq r_{\mathbf{x}}^{(ik,j\ell)}$, then (\ref{eq:nonzero_hugo}) is verified if and only if $m_1 + m_2 > r_{\mathbf{x}}^{(ik,j\ell)}$;
\item if $m_1 > r_{\mathbf{x}_1}^{(ik,j\ell)}$ and $m_2 \leq r_{\mathbf{x}}^{(ik,j\ell)}$, then (\ref{eq:nonzero_hugo}) is verified if and only if $m_2> r_{\mathbf{x}}^{(ik,j\ell)} -r_{\mathbf{x}_1}^{(ik,j\ell)}$;
\item if $m_1 \leq r_{\mathbf{x}_1}^{(ik,j\ell)}$ and $m_2 > r_{\mathbf{x}}^{(ik,j\ell)}$, then (\ref{eq:nonzero_hugo}) is verified if and only if $m_1> r_{\mathbf{x}}^{(ik,j\ell)} -r_{\mathbf{x}_2}^{(ik,j\ell)}$;
\item  if $m_1 > r_{\mathbf{x}_1}^{(ik,j\ell)}$ and $m_2 > r_{\mathbf{x}}^{(ik,j\ell)}$ then (\ref{eq:nonzero_hugo}) is verified, since
\begin{equation}
r_{\mathbf{x}_1}^{(ik,j\ell)} + r_{\mathbf{x}_2}^{(ik,j\ell)} = \rank [\boldsymbol{\mu}_{\mathbf{x}_1 }^{(ik)}-\boldsymbol{\mu}_{\mathbf{x}_1 }^{(j\ell)} \, \mathbf{P}\sub{c_1}^{(ik,j\ell)} \,  \mathbf{P}_1^{(ik,j\ell)} ]  + \rank [\boldsymbol{\mu}_{\mathbf{x}_2 }^{(ik)}-\boldsymbol{\mu}_{\mathbf{x}_2 }^{(j\ell)} \, \mathbf{P}\sub{c_2}^{(ik,j\ell)} \,  \mathbf{P}_2^{(ik,j\ell)} ]  \geq r_{\mathbf{x}}^{(ik,j\ell)}+1.
\end{equation}
\end{itemize}
Finally, we can combine the previous expressions and write necessary and sufficient conditions to guarantee (\ref{eq:nonzero_hugo}) as
\begin{equation}
\left\{
\begin{array}{lll}
m_1>   r_{\mathbf{x}}^{(ik,j\ell)} -r_{\mathbf{x}_2}^{(ik,j\ell)} \\
m_2>  r_{\mathbf{x}}^{(ik,j\ell)} -r_{\mathbf{x}_1}^{(ik,j\ell)} \\
m_1+ m_2> r_{\mathbf{x}}^{(ik,j\ell)} \\
\end{array}
\right. .
\end{equation}
\item $\boldsymbol{\mu}_{\mathbf{x}_1}^{(ik)}-\boldsymbol{\mu}_{\mathbf{x}_1}^{(j\ell)}  \notin \mathrm{Im} (\bar{\mathbf{\Sigma}}_{\mathbf{x}_1}^{(ik)}+\bar{\mathbf{\Sigma}}_{\mathbf{x}_1}^{(j\ell)})$ and $\boldsymbol{\mu}_{\mathbf{x}_2}^{(ik)}-\boldsymbol{\mu}_{\mathbf{x}_2}^{(j\ell)}  \in \mathrm{Im} (\bar{\mathbf{\Sigma}}_{\mathbf{x}_2}^{(ik)}+\bar{\mathbf{\Sigma}}_{\mathbf{x}_2}^{(j\ell)})$: in this case, if $m_1>r_{\mathbf{x}}^{(ik,j\ell)}$, then we can state that (\ref{eq:nonzero_hugo}) is true by considering simply classification on the basis of the observation of $\mathbf{y}_1$ alone. Therefore, assume now that $m_1 \leq r_{\mathbf{x}}^{(ik,j\ell)}$. In this case $r_{\boldsymbol{\mu}}^{(ik,j\ell)}=\min \{  r_{\mathbf{x}}^{(ik,j\ell)}+1, m_1+\min \{ m_2,r_{\mathbf{x}_2}^{(ik,j\ell)}  \}  \}$. Therefore, if $m_2 \leq r_{\mathbf{x}_2}^{(ik,j\ell)}$, then (\ref{eq:nonzero_hugo}) holds if and only if $m_1+m_2>r_{\mathbf{x}}^{(ik,j\ell)}$. On the other hand, if $m_2 > r_{\mathbf{x}_2}^{(ik,j\ell)}$, then (\ref{eq:nonzero_hugo}) holds if and only if $m_1>r_{\mathbf{x}}^{(ik,j\ell)}-r_{\mathbf{x}_2}^{(ik,j\ell)}$. We can combine the previous expressions and write necessary and sufficient conditions to guarantee (\ref{eq:nonzero_hugo}) in this case as
\begin{equation}
m_1 > r_{\mathbf{x}_1}^{(ik,j\ell)}  \quad \mathrm{or}  \quad 
\left\{
\begin{array}{lll}
m_1> r_{\mathbf{x}}^{(ik,j\ell)} -r_{\mathbf{x}_2}^{(ik,j\ell)} \\
m_1+ m_2>   r_{\mathbf{x}}^{(ik,j\ell)}  \\
\end{array}
\right. .
\end{equation}
\item $\boldsymbol{\mu}_{\mathbf{x}_1}^{(ik)}-\boldsymbol{\mu}_{\mathbf{x}_1}^{(j\ell)}  \in \mathrm{Im} (\bar{\mathbf{\Sigma}}_{\mathbf{x}_1}^{(ik)}+\bar{\mathbf{\Sigma}}_{\mathbf{x}_1}^{(j\ell)})$ and $\boldsymbol{\mu}_{\mathbf{x}_2}^{(ik)}-\boldsymbol{\mu}_{\mathbf{x}_2}^{(j\ell)}  \notin \mathrm{Im} (\bar{\mathbf{\Sigma}}_{\mathbf{x}_2}^{(ik)}+\bar{\mathbf{\Sigma}}_{\mathbf{x}_2}^{(j\ell)})$: the proof for this case follows steps similar to the case $\boldsymbol{\mu}_{\mathbf{x}_1}^{(ik)}-\boldsymbol{\mu}_{\mathbf{x}_1}^{(j\ell)}  \notin \mathrm{Im} (\bar{\mathbf{\Sigma}}_{\mathbf{x}_1}^{(ik)}+\bar{\mathbf{\Sigma}}_{\mathbf{x}_1}^{(j\ell)})$ and $\boldsymbol{\mu}_{\mathbf{x}_2}^{(ik)}-\boldsymbol{\mu}_{\mathbf{x}_2}^{(j\ell)}  \in \mathrm{Im} (\bar{\mathbf{\Sigma}}_{\mathbf{x}_2}^{(ik)}+\bar{\mathbf{\Sigma}}_{\mathbf{x}_2}^{(j\ell)})$.
\end{enumerate}

\section{Proof of Theorem \ref{theo:recG}}
\label{app:E}

We start by proving that conditions (\ref{eq:recG}) are sufficient in order to drive the \ac{MMSE} to zero in the low-rank regime. 
\rev{The first condition in (\ref{eq:recG}) reflects the fact that it is possible to drive the reconstruction \ac{MMSE} to zero in the low-rank regime from the observation of $\mathbf{y}_1$ alone, provided that $m_1\geq r_{\mathbf{x}_1}$. This is obtained by considering a slight modification of the result in~\cite[Theorem 1]{RecJournal}. The modification is required since the framework adopted in \cite{RecJournal} assumes that the signal $\mathbf{x}_1$ is drawn from an exactly low-rank model, and the linear features $\mathbf{y}_1$ are noisy. On the other hand, in this work we consider noiseless linear features, but we assume that $\mathbf{x}_1$ is described via the approximately low-rank model presented in Section \ref{par:SigModel}.

Consider the \ac{MMSE} associated to the recovery of $\mathbf{x}_1$ from $\mathbf{y}_1$ and assume that $\sigma_1^2=\sigma_2^=\sigma^2$. We can write \begin{IEEEeqnarray}{rCl}
\MMSE_{1|1,2}^{\sf G}(\sigma^2) &\leq& \MMSE_{1|1}^{\sf G}(\sigma^2)\\ 
 & = & \tr \left(  {\mathbf{\Sigma}}_{\mathbf{x}_1}  -  {\mathbf{\Sigma}}_{\mathbf{x}_1} \mathbf{\Phi}_1\tra \left(  \mathbf{\Phi}_1 {\mathbf{\Sigma}}_{\mathbf{x}_1} \mathbf{\Phi}_1\tra   \right)^{-1} \mathbf{\Phi}_1 {\mathbf{\Sigma}}_{\mathbf{x}_1}   \right)  \\
 &= & \tr \left( ( \bar{\mathbf{\Sigma}}_{\mathbf{x}_1}   +  \sigma^2\mathbf{I})  -  (\bar{\mathbf{\Sigma}}_{\mathbf{x}_1}+\sigma^2\mathbf{I}) \mathbf{\Phi}_1\tra \left( \sigma^2 \mathbf{I} +  \mathbf{\Phi}_1 \bar{\mathbf{\Sigma}}_{\mathbf{x}_1} \mathbf{\Phi}_1\tra   \right)^{-1} \mathbf{\Phi}_1 (\bar{\mathbf{\Sigma}}_{\mathbf{x}_1}  +\sigma^2\mathbf{I}) \right)\\
 \nonumber
& = & \tr \left(  \bar{\mathbf{\Sigma}}_{\mathbf{x}_1}  -  \bar{\mathbf{\Sigma}}_{\mathbf{x}_1} \mathbf{\Phi}_1\tra \left( \sigma^2 \mathbf{I} +  \mathbf{\Phi}_1 \bar{\mathbf{\Sigma}}_{\mathbf{x}_1} \mathbf{\Phi}_1\tra   \right)^{-1} \mathbf{\Phi}_1 \bar{\mathbf{\Sigma}}_{\mathbf{x}_1}   \right) + n_1 \sigma^2 \\
\nonumber
&& -2 \sigma^2 \tr \left(  \mathbf{\Phi}_1 \bar{\mathbf{\Sigma}}_{\mathbf{x}_1} \mathbf{\Phi}_1\tra \left( \sigma^2 \mathbf{I} +  \mathbf{\Phi}_1 \bar{\mathbf{\Sigma}}_{\mathbf{x}_1} \mathbf{\Phi}_1\tra   \right)^{-1}\right)\\
&& - \sigma^4 \tr \left( \left( \sigma^2 \mathbf{I} +  \mathbf{\Phi}_1 \bar{\mathbf{\Sigma}}_{\mathbf{x}_1} \mathbf{\Phi}_1\tra   \right)^{-1}    \right).
\label{eq:MMSE_G_new}
\end{IEEEeqnarray}
The first term in \eqref{eq:MMSE_G_new} represents the \ac{MMSE} studied in \cite[Appendix B]{RecJournal} and it converges to zero when $\sigma^2 \to 0$ if and only if 
\begin{equation}
\rank \left(  \mathbf{\Phi}_1 \bar{\mathbf{\Sigma}}_{\mathbf{x}_1} \mathbf{\Phi}_1\tra  \right) = \rank \left( \bar{\mathbf{\Sigma}}_{\mathbf{x}_1}  \right),
\label{eq:rank_rel_new}
\end{equation}
which is verified if and only if $m_1 \geq r_{\mathbf{x}_1}$. In fact, we can introduce the eigenvalue decomposition 
\begin{equation}
 \mathbf{\Xi}= \bar{\mathbf{\Sigma}}_{\mathbf{x}_1}^{\frac{1}{2}} \mathbf{\Phi}_1\tra \mathbf{\Phi}_1 \bar{\mathbf{\Sigma}}_{\mathbf{x}_1}^{\frac{1}{2}} = \mathbf{U}_{\mathbf{\Xi}} \mathbf{\Lambda}_{\mathbf{\Xi}} \mathbf{U}_{\mathbf{\Xi}}\tra,
\end{equation}
where $\mathbf{\Lambda}_{\mathbf{\Xi}}  = \diag (  \lambda_{\mathbf{\Xi},1}, \ldots, \lambda_{\mathbf{\Xi},r_{\mathbf{\Xi}}}, 0, \ldots, 0    )$ and $r_{\mathbf{\Xi}} = \rank (\mathbf{\Xi})=\min\{r_{\mathbf{x}_1},m_1  \}$, 
and by using the inversion Lemma~\cite[\S 0.7.4]{Horn} we can write
\begin{IEEEeqnarray}{rCl}
\tr \left(  \bar{\mathbf{\Sigma}}_{\mathbf{x}_1}  -  \bar{\mathbf{\Sigma}}_{\mathbf{x}_1} \mathbf{\Phi}_1\tra \left( \sigma^2 \mathbf{I} +  \mathbf{\Phi}_1 \bar{\mathbf{\Sigma}}_{\mathbf{x}_1} \mathbf{\Phi}_1\tra   \right)^{-1} \mathbf{\Phi}_1 \bar{\mathbf{\Sigma}}_{\mathbf{x}_1}   \right)  & = & \tr \left(  \bar{\mathbf{\Sigma}}_{\mathbf{x}_1} \left( \mathbf{I}  + 1/\sigma^2 \bar{\mathbf{\Sigma}}_{\mathbf{x}_1 }^{\frac{1}{2}} \mathbf{\Phi}_1\tra \mathbf{\Phi}_1 \bar{\mathbf{\Sigma}}_{\mathbf{x}_1 }^{\frac{1}{2}}   \right)^{-1}   \right) \\
 & = &  \tr \left(   \bar{\mathbf{\Sigma}}_{\mathbf{x}_1} \mathbf{U}_{\mathbf{\Xi}}  \left(  \mathbf{I}  + 1/\sigma^2 \mathbf{\Lambda}_{\mathbf{\Xi}}  \right)^{-1} \mathbf{U}_{\mathbf{\Xi}}\tra  \right) \\
 & = & \tr \left(  \bar{\mathbf{\Sigma}}_{\mathbf{x}_1 }  \mathbf{U}_{\mathbf{\Xi}}  \tilde{\mathbf{\Lambda}}_{\mathbf{\Xi}} \mathbf{U}_{\mathbf{\Xi}}\tra  \right),
\end{IEEEeqnarray}
where $\tilde{\mathbf{\Lambda}}_{\mathbf{\Xi}} = \mathrm{diag}  \left(    \frac{1}{1+ \lambda_{\mathbf{\Xi},1}/\sigma^2}, \ldots, \frac{1}{1+ \lambda_{\mathbf{\Xi},r_{\mathbf{\Xi}}}/\sigma^2}, 1, \ldots, 1   \right)$. It is then clear that the first term in \eqref{eq:MMSE_G_new} approaches zero, when $\sigma^2 \to 0$, if and only if
\begin{equation}
\mathrm{Null} \left(  \mathbf{\Xi}  \right)  \subseteq \mathrm{Null}  \left(  \bar{\mathbf{\Sigma}}_{\mathbf{x}_1 }  \right).
\label{eq:NullEq_new}
\end{equation}
Moreover, on noting that $\mathrm{Null}  \left(  \bar{\mathbf{\Sigma}}_{\mathbf{x}_1 }  \right) \subseteq\mathrm{Null} \left(  \mathbf{\Xi}   \right)$, we immediately conclude that (\ref{eq:NullEq_new}) is equivalent to (\ref{eq:rank_rel_new}).

We also need to show that, for any value of $m_1$, the remaining terms in \eqref{eq:MMSE_G_new} approach zero, when $\sigma^2 \to 0$. This is done by considering the eigenvalue decomposition of $\mathbf{\Phi}_1 \bar{\mathbf{\Sigma}}_{\mathbf{x}_1} \mathbf{\Phi}_1\tra$. In fact, we can note that the positive eigenvalues of $\mathbf{\Phi}_1 \bar{\mathbf{\Sigma}}_{\mathbf{x}_1} \mathbf{\Phi}_1\tra$ are the same of $\mathbf{\Xi}$, and therefore, we can write,
\begin{IEEEeqnarray}{rCl}
\sigma^2 \tr \left(  \mathbf{\Phi}_1 \bar{\mathbf{\Sigma}}_{\mathbf{x}_1} \mathbf{\Phi}_1\tra \left( \sigma^2 \mathbf{I} +  \mathbf{\Phi}_1 \bar{\mathbf{\Sigma}}_{\mathbf{x}_1} \mathbf{\Phi}_1\tra   \right)^{-1}\right) & = & \sigma^2 \sum_{t=1}^{r_{\mathbf{\Xi}}}   \frac{\lambda_{\mathbf{\Xi},t}}{\lambda_{\mathbf{\Xi},t}+\sigma^2}  = r_{\mathbf{\Xi}} \cdot \sigma^2 + o(\sigma^2) \\
 \sigma^4 \tr \left( \left( \sigma^2 \mathbf{I} +  \mathbf{\Phi}_1 \bar{\mathbf{\Sigma}}_{\mathbf{x}_1} \mathbf{\Phi}_1\tra   \right)^{-1}    \right) & = &  \sigma^4 \sum_{t=1}^{r_{\mathbf{\Xi}}}  \frac{1}{ \lambda_{\mathbf{\Xi},t}   + \sigma^2  }   +  (m_1- r_{\mathbf{\Xi}}) \sigma^2 ,
\end{IEEEeqnarray}
thus noting immediately that such terms converge to zero when $\sigma^2\to0$.
}

Consider now the upper bound associated to the distributed reconstruction problem, i.e., the \ac{MMSE} incurred in recovering \emph{both} $\mathbf{x}_1$ and $\mathbf{x}_2$ from $\mathbf{y}_1$ and $\mathbf{y}_2$ (or, equivalently, $\mathbf{x}$ from $\mathbf{y}$). 
Then, we can write
\rev{
\begin{IEEEeqnarray}{rCl}
\MMSE_{1|1,2}^{\sf G}(\sigma^2) & \leq & \MMSE_{1,2|1,2}^{\sf G}(\sigma^2)\\
 & = & \tr \left(  {\mathbf{\Sigma}}_{\mathbf{x}}  -  {\mathbf{\Sigma}}_{\mathbf{x}} \mathbf{\Phi}\tra \left( \mathbf{\Phi} {\mathbf{\Sigma}}_{\mathbf{x}} \mathbf{\Phi}\tra   \right)^{-1} \mathbf{\Phi} {\mathbf{\Sigma}}_{\mathbf{x}}   \right)\\
& = & \tr \left(  (\bar{\mathbf{\Sigma}}_{\mathbf{x}} + \sigma^2 \mathbf{I} ) - ( \bar{\mathbf{\Sigma}}_{\mathbf{x}}+ \sigma^2 \mathbf{I} ) \mathbf{\Phi}\tra \left( \sigma^2 \mathbf{I} +  \mathbf{\Phi} \bar{\mathbf{\Sigma}}_{\mathbf{x}} \mathbf{\Phi}\tra   \right)^{-1} \mathbf{\Phi} (\bar{\mathbf{\Sigma}}_{\mathbf{x}} + \sigma^2 \mathbf{I} )  \right)\\
\nonumber
& = & \tr \left(  \bar{\mathbf{\Sigma}}_{\mathbf{x}}  -  \bar{\mathbf{\Sigma}}_{\mathbf{x}} \mathbf{\Phi}\tra \left( \sigma^2 \mathbf{I} +  \mathbf{\Phi} \bar{\mathbf{\Sigma}}_{\mathbf{x}} \mathbf{\Phi}\tra   \right)^{-1} \mathbf{\Phi} \bar{\mathbf{\Sigma}}_{\mathbf{x}}   \right) + (n_1 + n_2 ) \sigma^2 \\
\nonumber
&& -2 \sigma^2 \tr \left(  \mathbf{\Phi} \bar{\mathbf{\Sigma}}_{\mathbf{x}} \mathbf{\Phi}\tra \left( \sigma^2 \mathbf{I} +  \mathbf{\Phi} \bar{\mathbf{\Sigma}}_{\mathbf{x}} \mathbf{\Phi}\tra   \right)^{-1}\right)\\
&& - \sigma^4 \tr \left( \left( \sigma^2 \mathbf{I} +  \mathbf{\Phi} \bar{\mathbf{\Sigma}}_{\mathbf{x}} \mathbf{\Phi}\tra   \right)^{-1}    \right).
\label{eq:MMSE_G12_new}
\end{IEEEeqnarray}
By using similar steps to those considered for $\MMSE_{1|1}^{\sf G}(\sigma^2)$ we can show that $\MMSE_{1,2|1,2}^{\sf G}(\sigma^2)$ approaches zero when $\sigma^2\to 0$ if and only if
}
\begin{equation}
\rank \left(  \mathbf{\Phi} \bar{\mathbf{\Sigma}}_{\mathbf{x}} \mathbf{\Phi}\tra  \right) = \rank \left( \bar{\mathbf{\Sigma}}_{\mathbf{x}}  \right).
\label{eq:rank_rel}
\end{equation}
We can now determine conditions on the number of features $m_1$ and $m_2$ needed in order to verify (\ref{eq:rank_rel}) by leveraging the rank expression (\ref{eq:theo1bis}). In particular, note that (\ref{eq:rank_rel}) holds if and only if
\begin{equation}
\min \{ m_1, r_{\mathbf{x}_1}  \} + \min \{ m_2, r_{\mathbf{x}_2}  \} \geq r_{\mathbf{x}}.
\end{equation}
We can consider separately four different cases, and observe that, if $m_1 \leq r_{\mathbf{x}_1}$ and $m_2 \leq r_{\mathbf{x}_2}$, then (\ref{eq:rank_rel}) is verified if and only if $m_1+m_2 \geq r_{\mathbf{x}}$. If $m_1 \leq r_{\mathbf{x}_1}$ and $m_2 > r_{\mathbf{x}_2}$, then (\ref{eq:rank_rel}) holds if and only if $m_1 \geq r_{\mathbf{x}}-r_{\mathbf{x}_2}$ and symmetrically,  $m_1 > r_{\mathbf{x}_1}$ and $m_2 \leq r_{\mathbf{x}_2}$, then (\ref{eq:rank_rel}) holds if and only if $m_2 \geq r_{\mathbf{x}}-r_{\mathbf{x}_1}$. Finally, $m_1 > r_{\mathbf{x}_1}$ and $m_2 > r_{\mathbf{x}_2}$, then (\ref{eq:rank_rel}) always holds since $r_{\mathbf{x}_1} + r_{\mathbf{x}_2} \geq r_{\mathbf{x}}$. Then, the four previous cases can be summarized by stating that (\ref{eq:rank_rel}) is true if and only if $m_1$ and $m_2$ verify the conditions
\begin{equation}
\left\{
\begin{array}{lll}
m_1 \geq   r_{\mathbf{x}} -r_{\mathbf{x}_2} \\
m_2 \geq r_{\mathbf{x}} -r_{\mathbf{x}_1} \\
m_1+ m_2  \geq  r_{\mathbf{x}} \\
\end{array}
\right. .
\label{eq:DSregion}
\end{equation}  
Then, the proof of sufficiency is concluded by simply considering the union of the set of values $(m_1,m_2)$ which verify (\ref{eq:DSregion}) with the set $m_1 \geq r_{\mathbf{x}_1}$.

We now prove that conditions (\ref{eq:recG}) are also necessary to guarantee that the \ac{MMSE} approaches zero when $\sigma^2 \to 0$. In the following, we will denote the \ac{MMSE} associated to the estimation of the random vector $\mathbf{u}$ from the observation vector $\mathbf{v}$ by 
\begin{equation}
\MMSE(\mathbf{u} | \mathbf{v}) = \E{\|  \mathbf{u} - \E{\mathbf{u} |   \mathbf{v}}   \|^2 },
\end{equation}
where the expectation is taken with respect to the joint distribution of $(\mathbf{u},\mathbf{v})$. Then, we obtain a lower bound to $\MMSE_{1|1,2}^{\sf G}(\sigma^2)$ by observing that, for all $\sigma^2 > 0$, we have
\rev{
\begin{IEEEeqnarray}{rCl}
\MMSE_{1|1,2}^{\sf G}(\sigma^2) & = & \MMSE (\mathbf{x}_1 | \mathbf{y}_1,\mathbf{y}_2)  \geq \MMSE (\mathbf{x}_1 | \mathbf{y}_1,\mathbf{y}_2, \mathbf{w}_1, \mathbf{w}_2) = \MMSE (\bar{\mathbf{x}}_1 |\mathbf{\Phi}_1 \bar{\mathbf{x}}_1, \mathbf{\Phi}_2 \bar{\mathbf{x}}_2). 
\label{eq:MMSEphi12}
\end{IEEEeqnarray}
}
On the other hand, by observing that the \ac{MMSE} does not depend on the value of the mean of the input signal to estimate, and by taking the expectation in the right hand side of (\ref{eq:MMSEphi12}) with respect to the random variables \rev{$\bar{\mathbf{x}}_1 | \mathbf{\Phi}_2 \bar{\mathbf{x}}_2$} and \rev{$\mathbf{\Phi}_2 \bar{\mathbf{x}}_2$}, separately, it is possible to show that
\rev{
\begin{equation}
\MMSE (\bar{\mathbf{x}}_1 |\mathbf{\Phi}_1 \bar{\mathbf{x}}_1, \mathbf{\Phi}_2 \bar{\mathbf{x}}_2)= \MMSE (\bar{\mathbf{z}}| \mathbf{\Phi}_1 \bar{\mathbf{z}}),
\end{equation}}
where $\bar{\mathbf{z}} \in \mathbb{R}^{n_1}$ is a Gaussian vector with covariance matrix equal to the conditional covariance of $\bar{\mathbf{x}}_1$ given $\mathbf{\Phi}_2 \bar{\mathbf{x}}_2$, i.e., $\bar{\mathbf{z}} \sim \mathcal{N}(\mathbf{0}, \bar{\mathbf{\Sigma}}_{\mathbf{z}})$, where 
\begin{equation}
\bar{\mathbf{\Sigma}}_{\mathbf{z}} = \mathrm{Cov}(\mathbf{x}_1| \mathbf{\Phi}_2 \mathbf{x}_2) = \bar{\mathbf{\Sigma}}_{\mathbf{x}_1} - \bar{\mathbf{\Sigma}}_{\mathbf{x}_{12}} \mathbf{\Phi}_2\tra  (  \mathbf{\Phi}_2 \bar{\mathbf{\Sigma}}_{\mathbf{x}_2}  \mathbf{\Phi}_2\tra  )^{\dag} \mathbf{\Phi}_2 \bar{\mathbf{\Sigma}}_{\mathbf{x}_{21}}.
\end{equation}
Then, by leveraging the result in \cite[Theorem 1]{RecJournal}, or by simply considering the set of linear equations corresponding to the rows of the matrix $ \mathbf{\Phi}_1 \bar{\mathbf{\Sigma}}_{\mathbf{z}}^{1/2}$, a necessary condition for $\MMSE (\bar{\mathbf{z}}| \mathbf{\Phi}_1 \bar{\mathbf{z}})=0$, and therefore, a necessary condition for $ \lim_{\sigma^2 \to 0}\MMSE_{1|1,2}^{\sf G}(\sigma^2)= 0 $, is given by
\begin{equation}
m_1 \geq r_{\mathbf{z}} = \rank(\bar{\mathbf{\Sigma}}_{\mathbf{z}}). 
\label{eq:m1rz}
\end{equation}
We complete the proof by computing the rank $r_{\mathbf{z}}$ using a result on the generalized Schur complement of a positive semidefinite matrix \cite{Wang99}. Namely, $\bar{\mathbf{\Sigma}}_{\mathbf{z}}$ can be viewed as the generalized Schur complement of the block $\mathbf{\Phi}_2 \bar{\mathbf{\Sigma}}_{\mathbf{x}_2} \mathbf{\Phi}_2\tra$ of the positive semidefinite matrix
\begin{equation}
\bar{\mathbf{\Sigma}}_{\mathbf{x}_1 \mathbf{\Phi}_2 \mathbf{x}_2} =  \E{\left[ 
\begin{array}{cc}
\bar{\mathbf{x}}_1\\
\mathbf{\Phi}_2 \bar{\mathbf{x}}_2
\end{array}
\right]
[\bar{\mathbf{x}}_1\tra  \,  (\mathbf{\Phi}_2 \bar{\mathbf{x}}_2)\tra ]  }= \left[
\begin{array}{cccc}
\bar{\mathbf{\Sigma}}_{\mathbf{x}_1}  & \bar{\mathbf{\Sigma}}_{\mathbf{x}_{12}} \mathbf{\Phi}_2\tra \\
\mathbf{\Phi}_2 \bar{\mathbf{\Sigma}}_{\mathbf{x}_{21}}   & \mathbf{\Phi}_2 \bar{\mathbf{\Sigma}}_{\mathbf{x}_2} \mathbf{\Phi}_2\tra
\end{array}
\right],
\end{equation}
and, with probability 1, we have \cite{Wang99}
\begin{equation}
\rank (\bar{\mathbf{\Sigma}}_{\mathbf{x}_1 \mathbf{\Phi}_2 \mathbf{x}_2}) = r_{\mathbf{z}} + \rank(\mathbf{\Phi}_2 \bar{\mathbf{\Sigma}}_{\mathbf{x}_2} \mathbf{\Phi}_2\tra) = r_{\mathbf{z}} + \min \{   m_2, r_{\mathbf{x}_2} \}.
\label{eq:rz1}
\end{equation}
In addition, on considering the matrix
\begin{equation}
\bar{\mathbf{\Sigma}}_{ \mathbf{\Phi}_2 \mathbf{x}_2\mathbf{x}_1} =  \E{\left[ 
\begin{array}{cc}
\mathbf{\Phi}_2 \bar{\mathbf{x}}_2\\
\bar{\mathbf{x}}_1
\end{array}
\right]
[ (\mathbf{\Phi}_2 \bar{\mathbf{x}}_2)\tra \, \bar{\mathbf{x}}_1\tra   ]  }= \left[
\begin{array}{cccc}
\mathbf{\Phi}_2 \bar{\mathbf{\Sigma}}_{\mathbf{x}_2} \mathbf{\Phi}_2\tra &   \mathbf{\Phi}_2 \bar{\mathbf{\Sigma}}_{\mathbf{x}_{21}}   \\
\bar{\mathbf{\Sigma}}_{\mathbf{x}_{12}} \mathbf{\Phi}_2\tra &  \bar{\mathbf{\Sigma}}_{\mathbf{x}_1} 
\end{array}
\right],
\end{equation}
and on applying the same rank computation, we also have
\begin{equation}
\rank (\bar{\mathbf{\Sigma}}_{ \mathbf{\Phi}_2 \mathbf{x}_2 \mathbf{x}_1} ) =\rank (\bar{\mathbf{\Sigma}}_{\mathbf{x}_1 \mathbf{\Phi}_2 \mathbf{x}_2}) = r_{\mathbf{x}_1} + \rank (\mathrm{Cov}(\mathbf{\Phi}_2 \mathbf{x}_2 | \mathbf{x}_1))=r_{\mathbf{x}_1} + \rank \left( \mathbf{\Phi}_2 (  \bar{\mathbf{\Sigma}}_{\mathbf{x}_2}  - \bar{\mathbf{\Sigma}}_{\mathbf{x}_{21}} \bar{\mathbf{\Sigma}}_{\mathbf{x}_1}^{\dag}  \bar{\mathbf{\Sigma}}_{\mathbf{x}_{12}}   ) \mathbf{\Phi}_2\tra  \right).
\label{eq:rz2}
\end{equation}
Then, on recalling that the projection kernel $\mathbf{\Phi}_2$ is rotation-invariant, and by using again the generalized Schur complement rank computation, with probability 1, we have
\begin{equation}
\rank (\mathrm{Cov}(\mathbf{\Phi}_2 \mathbf{x}_2 | \mathbf{x}_1)) = \min \{ m_2, r_{\mathbf{x}} - r_{\mathbf{x}_1} \}.
\label{eq:rz3}
\end{equation}
Finally, by substituting (\ref{eq:rz2}) and (\ref{eq:rz3}) in (\ref{eq:rz1}), we can rewrite (\ref{eq:m1rz}) as
\begin{equation}
m_1 \geq r_{\mathbf{x}_1}  - \min \{ m_2, r_{\mathbf{x}_2} \}   +   \min \{   m_2, r_{\mathbf{x}} - r_{\mathbf{x}_1} \},
\end{equation}
which can be immediately shown to be equivalent to conditions (\ref{eq:recG}), thus concluding the necessity part of the proof.

\section{Proof of Theorem~\ref{theo:recGMM}}
\label{app:F} 

This proof in based on steps similar to those in \cite[Appendix C]{RecJournal}. Nevertheless, we report here the key ideas used in the proof for completeness. 
On defining
\begin{equation}
\mathcal{W}_{\mathbf{x}_1}^{(ik)}=  \boldsymbol{\mu}_{\mathbf{x}_1}^{(ik)} + \mathbf{W}_{\mathbf{x}_1}^{(ik)}\left(
\mathbf{y}
-\mathbf{\Phi} 
\boldsymbol{\mu}_{\mathbf{x}}^{(ik)}
\right),
\end{equation}
where $ \mathbf{W}_{\mathbf{x}_1}^{(ik)}$ is as in (\ref{eq:Wiener_indexed}), and by using the law of total probability, we can write 
\begin{IEEEeqnarray}{rCl}
\nonumber
\MSE^{\mathrm{CR}}(\sigma^2)  & \leq &  \sum_{i, k}   p_{C_1,C_2}(i,k)  \E{ \| \mathbf{x}_1 -  \mathcal{W}_{\mathbf{x}_1}^{(ik)}(\mathbf{y})   \|^2  |C_1=i,C_2=k }  \\
\nonumber
&  &  +  \sum_{i,k}  p_{C_1,C_2}(i,k)   \sum_{(j,\ell) \neq (i,k)} p(\hat{C}_1=j,\hat{C}_2=\ell|C_1=i,C_2=k)\\
& & \cdot \E{ \| \mathbf{x}_1 -  \mathcal{W}_{\mathbf{x}_1}^{(j\ell)}(\mathbf{y})   \|^2  |\hat{C}_1=j,\hat{C}_2=\ell,C_1=i,C_2=k } .
\end{IEEEeqnarray}
We can observe immediately that, assuming the conditions in (\ref{eq:recGMM}) are verified, Theorem~\ref{theo:recG} guarantees that the terms $\E{ \| \mathbf{x}_1 -  \mathcal{W}_{\mathbf{x}_1}^{(ik)}(\mathbf{y})   \|^2  |C_1=i,C_2=k }$ approach zero when $\sigma^2 \to 0$. Then, we are left with proving 
\begin{equation}
\lim_{\sigma^2 \to 0} p(\hat{C}_1=j,\hat{C}_2=\ell|C_1=i,C_2=k) \E{ \| \mathbf{x}_1 -  \mathcal{W}_{\mathbf{x}_1}^{(j\ell)}(\mathbf{y})   \|^2  |\hat{C}_1=j,\hat{C}_2=\ell,C_1=i,C_2=k }  =0
\label{eq:limterm}
\end{equation}
whenever $(j,\ell) \neq (i,k)$. Given the conditions (\ref{eq:recGMM}) hold, if  $m_1 > r_{\mathbf{x}_1}^{(ik)}$ and $m_1 > r_{\mathbf{x}_1}^{(j\ell)}$, \rev{then we can leverage a result akin to that in \cite[Appendix C]{RecJournal} on the reconstruction of $\mathbf{x}_1$ from $\mathbf{y}_1$ alone to show that (\ref{eq:limterm}) holds. In particular, we consider two different cases: i) $r_{\mathbf{x}_1}^{(ik,j\ell)}> r_{\mathbf{x}_1}^{(ik)},r_{\mathbf{x}_1}^{(j\ell)}$ and i) $r_{\mathbf{x}_1}^{(ik,j\ell)} = r_{\mathbf{x}_1}^{(ik)} =r_{\mathbf{x}_1}^{(j\ell)}$. In the first case, the range spaces $\mathrm{Im}(\bar{\mathbf{\Sigma}}_{\mathbf{x}_1}^{(ik)})$ and $\mathrm{Im}(\bar{\mathbf{\Sigma}}_{\mathbf{x}_1}^{(j\ell)})$ are distinct, and we can leverage the characterization of the misclassification probability in order to prove (\ref{eq:limterm}). In particular, by following steps similar to those in the proof of Corollary \ref{cor:phasetrans}, we can show that 
\begin{equation}
\lim_{\sigma^2 \to 0}p(\hat{C}_1=j,\hat{C}_2=\ell|C_1=i,C_2=k)=0.
\end{equation}
Therefore, observe that the misclassification probability $p(\hat{C}_1=j,\hat{C}_2=\ell|C_1=i,C_2=k)$ is the measure of the set representing the decision region of the \ac{MAP} classifier associated to the classes $(j,\ell)$ with respect to the Gaussian measure induced by the Gaussian distribution of classes $(i,k)$. Then, it is also possible to show that, in the limit $\sigma^2\to 0$, the product in (\ref{eq:limterm}) is upper bounded by the integral of a measurable function over a set with measure zero, and then it converges to zero. 

In the second case, instead, we have $\mathrm{Im}(\bar{\mathbf{\Sigma}}_{\mathbf{x}_1}^{(ik)})=\mathrm{Im}(\bar{\mathbf{\Sigma}}_{\mathbf{x}_1}^{(j\ell)})$, and we can consider separately further two cases. If 
\begin{equation}
\boldsymbol{\mu}_{\mathbf{x}_1}^{(ik)}  - \boldsymbol{\mu}_{\mathbf{x}_1}^{(ik)} \notin \mathrm{Im}(\bar{\mathbf{\Sigma}}_{\mathbf{x}_1}^{(ik)}+\bar{\mathbf{\Sigma}}_{\mathbf{x}_1}^{(j\ell)})= \mathrm{Im}(\bar{\mathbf{\Sigma}}_{\mathbf{x}_1}^{(ik)}),
\end{equation} 
then Theorem~\ref{theo:nonzero_SI} states that $p(\hat{C}_1=j,\hat{C}_2=\ell|C_1=i,C_2=k)$ approaches zero in the low-rank regime, and we can prove that (\ref{eq:limterm}) holds by following a similar procedure to that used for case i). On the other hand, if 
\begin{equation}
\boldsymbol{\mu}_{\mathbf{x}_1}^{(ik)}  - \boldsymbol{\mu}_{\mathbf{x}_1}^{(ik)} \in \mathrm{Im}(\bar{\mathbf{\Sigma}}_{\mathbf{x}_1}^{(ik)}+\bar{\mathbf{\Sigma}}_{\mathbf{x}_1}^{(j\ell)}) = \mathrm{Im}(\bar{\mathbf{\Sigma}}_{\mathbf{x}_1}^{(ik)}),
\end{equation} 
then the misclassification probability associated to the estimation of $(C_1,C_2)$ from $\mathbf{y}_1$ is not guaranteed to approach zero in the low-rank regime. 
However, on using the law of total probability and the definition of \ac{MSE}, we can notice that the argument of the limit in (\ref{eq:limterm}) for the case of reconstruction of $\mathbf{x}_1$ from $\mathbf{y}_1$ alone is upper bounded by
\begin{equation}
   \E{ \| \mathbf{x}_1 -  \mathcal{W}_{\mathbf{x}_1}^{(j\ell)}(\mathbf{y}_1)   \|^2  | C_1=i,C_2=k },
  \label{eq:MMSEmis_New}
\end{equation}
where the Wiener filter associated to the reconstruction of $\mathbf{x}_1$ from $\mathbf{y}_1$ alone is given by
\begin{equation}
\mathcal{W}_{\mathbf{x}_1}^{(j\ell)}(\mathbf{y}_1)=   \boldsymbol{\mu}_{\mathbf{x}_1}^{(j\ell)} + \mathbf{W}_{\mathbf{x}_1}^{(j\ell)}\left(
\mathbf{y}_1
-\mathbf{\Phi}_1 
\boldsymbol{\mu}_{\mathbf{x}_1}^{(j\ell)}
\right)
\end{equation}
where
\begin{equation}
\mathbf{W}_{\mathbf{x}_1}^{(j\ell)} =   (\bar{\mathbf{\Sigma}}_{\mathbf{x}_1}^{(j\ell)}  + \sigma^2\mathbf{I}) \mathbf{\Phi}_1\tra \left( \sigma^2 \mathbf{I}  + \mathbf{\Phi}_1 \bar{\mathbf{\Sigma}}_{\mathbf{x}_1}^{(j\ell)} \mathbf{\Phi}_1\tra   \right)^{-1}.
\end{equation}
Then, we can show that (\ref{eq:MMSEmis_New}) approaches zero when $\sigma^2 \to 0$ by using steps similar to those in \cite[Appendix C-B]{RecJournal}. This reflects the fact that the mismatched \ac{MSE} for Gaussian sources reaches zero in the low-rank regime, provided that the estimated input covariance has the same range space than the true input covariance. 
In particular, on denoting by ${\mathbf{\Sigma}}_{\mathbf{y}_1}^{(ik)} = \sigma^2 \mathbf{I} + \mathbf{\Phi}_1  \bar{\mathbf{\Sigma}}_{\mathbf{x}_1}^{(ik)} \mathbf{\Phi}_1\tra$ the covariance matrix of $\mathbf{y}_1$ conditioned on $(C_1,C_2)=(i,k)$, and on introducing the symbol $\mathbf{M}_1^{(ik,j\ell)} = (\boldsymbol{\mu}_{\mathbf{x}_1}^{(ik)}- \boldsymbol{\mu}_{\mathbf{x}_1}^{(j\ell)}) (\boldsymbol{\mu}_{\mathbf{x}_1}^{(ik)}- \boldsymbol{\mu}_{\mathbf{x}_1}^{(j\ell)})\tra$, we can write
\begin{IEEEeqnarray}{rCl}
\MSE_{1|1}^{(ik,j\ell)} (\sigma^2) & = & \E{ \| \mathbf{x}_1 -  \mathcal{W}_{\mathbf{x}_1}^{(j\ell)}(\mathbf{y}_1)   \|^2  | C_1=i,C_2=k } \\
& = & \E{  \tr \left(    \left( \mathbf{x}_1 -\mathcal{W}_{\mathbf{x}_1}^{(j\ell)} (\mathbf{y}_1)  \right)   \left( \mathbf{x}_1 -\mathcal{W}_{\mathbf{x}_1}^{(j\ell)} (\mathbf{y}_1)   \right)\tra      \right) | C_1=i,C_2=k  }  \\
\nonumber
 & = & \tr \left(  \bar{\mathbf{\Sigma}}_{\mathbf{x}_1}^{(ik)} \right) + n_1\sigma^2  - 2 \tr \left(   \mathbf{W}_{\mathbf{x}_1}^{(ik)}  {\mathbf{\Sigma}}_{\mathbf{y}_1}^{(ik)} (\mathbf{W}_{\mathbf{x}_1}^{(j\ell)})\tra \right)  + \tr \left( \mathbf{W}_{\mathbf{x}_1}^{(j\ell)} {\mathbf{\Sigma}}_{\mathbf{y}_1}^{(ik)} (\mathbf{W}_{\mathbf{x}_1}^{(j\ell)})\tra    \right)\\ 
 &&+ \tr \left(  \mathbf{M}_1^{(ik,j\ell)} \right)  - 2 \tr \left( \mathbf{M}_1^{(ik,j\ell)} \mathbf{\Phi}_1\tra (\mathbf{W}_{\mathbf{x}_1}^{j\ell})\tra \right) + \tr \left( \mathbf{\Phi}_1 \mathbf{W}_{\mathbf{x}_1}^{(j\ell)} \mathbf{M}_1^{(ik,j\ell)} (\mathbf{W}_{\mathbf{x}_1}^{(j\ell)})\tra \right),
\end{IEEEeqnarray}
and we can prove that
\begin{equation}
 \lim_{\sigma^2 \to 0}  \tr \left(   \mathbf{W}_{\mathbf{x}_1}^{(ik)}  {\mathbf{\Sigma}}_{\mathbf{y}_1}^{(ik)} (\mathbf{W}_{\mathbf{x}_1}^{(j\ell)})\tra \right) =  \tr(\bar{\mathbf{\Sigma}}_{\mathbf{x}_1}^{(ik)});
 \label{eq:fact1_New}
 \end{equation}
\begin{equation} 
 \label{eq:fact2_New}
\lim_{\sigma^2 \to 0}  \tr \left( \mathbf{W}_{\mathbf{x}_1}^{(j\ell)} {\mathbf{\Sigma}}_{\mathbf{y}_1}^{(ik)}  (\mathbf{W}_{\mathbf{x}_1}^{(j\ell)})\tra \right)
 =  \tr(\bar{\mathbf{\Sigma}}_{\mathbf{x}_1}^{(ik)});
 \end{equation}
 \begin{equation}
  \lim_{\sigma^2 \to 0}  \tr \left( \mathbf{M}_1^{(ik,j\ell)} (\mathbf{W}_{\mathbf{x}_1}^{(j\ell)})\tra \right)
 =  \tr \left(\mathbf{M}_1^{(ik,j\ell)} \right);
 \label{eq:fact3_New}
 \end{equation}
 \begin{equation}
  \lim_{\sigma^2 \to 0}  \tr \left(  \mathbf{W}_{\mathbf{x}_1}^{(j\ell)}\mathbf{\Phi}_1 \mathbf{M}_1^{(ik,j\ell)} \mathbf{\Phi}_1\tra (\mathbf{W}_{\mathbf{x}_1}^{(j\ell)})\tra \right)
 =  \tr \left(\mathbf{M}_1^{(ik,j\ell)} \right).
 \label{eq:fact4_New}
 \end{equation}

The proof is based on the use of the inversion Lemma~\cite{Horn}
\begin{equation}
\mathbf{A} (\mathbf{I}c^{-1} + \mathbf{B}\mathbf{A} )^{-1} \mathbf{B} = \mathbf{I} -( \mathbf{I}  + c \mathbf{A}\mathbf{B})^{-1},
\end{equation}
in which we choose $\mathbf{A}=\mathbf{\Phi}_1\tra$ and $\mathbf{B}=\mathbf{\Phi}_1 \bar{\mathbf{\Sigma}}_{\mathbf{x}_1}^{(j\ell)}$ and we write
\begin{IEEEeqnarray}{rCl}
\tr \left(   \mathbf{W}_{\mathbf{x}_1}^{(ik)}  {\mathbf{\Sigma}}_{\mathbf{y}_1}^{(ik)} (\mathbf{W}_{\mathbf{x}_1}^{(j\ell)})\tra \right)  & = & \tr \left( (\bar{\mathbf{\Sigma}}_{\mathbf{x}_1}^{(ik)}  +\sigma^2\mathbf{I})   \mathbf{\Phi}_1\tra ( \mathbf{I} \sigma^2 +  \mathbf{\Phi}_1 \bar{\mathbf{\Sigma}}_{\mathbf{x}_1}^{(j\ell)}\mathbf{\Phi}_1\tra)^{-1}  \mathbf{\Phi}_1  ( \bar{\mathbf{\Sigma}}_{\mathbf{x}_1}^{(j\ell)}  +\sigma^2\mathbf{I})  \right)\\
\nonumber
& = & \tr \left( \bar{\mathbf{\Sigma}}_{\mathbf{x}_1}^{(ik)}  \right) - \tr \left( \bar{\mathbf{\Sigma}}_{\mathbf{x}_1}^{(ik)} ( \mathbf{I} +\frac{1}{ \sigma^2} \mathbf{\Phi}_1\tra \mathbf{\Phi}_1 \bar{\mathbf{\Sigma}}_{\mathbf{x}_1}^{(j\ell)} )^{-1}    \right)\IEEEeqnarraynumspace \\
\nonumber
& & + \sigma^2 \tr    \left(  \mathbf{\Phi}_1 \bar{\mathbf{\Sigma}}_{\mathbf{x}_1}^{(ik)}   \mathbf{\Phi}_1\tra ( \mathbf{I} \sigma^2 +  \mathbf{\Phi}_1 \bar{\mathbf{\Sigma}}_{\mathbf{x}_1}^{(j\ell)}\mathbf{\Phi}_1\tra)^{-1}    \right)\\
\nonumber
& & + \sigma^2 \tr    \left(  \mathbf{\Phi}_1 \bar{\mathbf{\Sigma}}_{\mathbf{x}_1}^{(j\ell)}   \mathbf{\Phi}_1\tra ( \mathbf{I} \sigma^2 +  \mathbf{\Phi}_1 \bar{\mathbf{\Sigma}}_{\mathbf{x}_1}^{(j\ell)}\mathbf{\Phi}_1\tra)^{-1}    \right)\\
& & + \sigma^4 \tr    \left(   ( \mathbf{I} \sigma^2 +  \mathbf{\Phi}_1 \bar{\mathbf{\Sigma}}_{\mathbf{x}_1}^{(j\ell)}\mathbf{\Phi}_1\tra)^{-1}    \right).
\label{eq:Wmis_New}
\end{IEEEeqnarray}
Then, on noting that the matrix $\mathbf{\Phi}_1\tra \mathbf{\Phi}_! \bar{\mathbf{\Sigma}}_{\mathbf{x}_1}^{(j\ell)}$ is diagonalizable with probability 1, and by following steps similar to those adopted in the proof of Theorem~\ref{theo:recG}, we are able to prove that the second term in \eqref{eq:Wmis_New} converges to zero when $\sigma^2 \to 0$. Moreover, on noting that $\mathrm{Null}(\mathbf{\Phi}_1 \bar{\mathbf{\Sigma}}_{\mathbf{x}_1}^{(ik)}\mathbf{\Phi}_1\tra)=\mathrm{Null}(\mathbf{\Phi}_1 \bar{\mathbf{\Sigma}}_{\mathbf{x}_1}^{(j\ell)}\mathbf{\Phi}_1\tra)$, and by following steps similar to those adopted in the proof of Theorem~\ref{theo:recG}, we are able to prove that also the third, fourth and fifth terms in \eqref{eq:Wmis_New} converge to zero when $\sigma^2 \to 0$. Finally, also (\ref{eq:fact2_New}), (\ref{eq:fact3_New}) and (\ref{eq:fact4_New}) are proved by following a similar approach.
} 

\rev{Consider now the case when $m_1 \leq r_{\mathbf{x}_1}^{(ik)}$ or $m_1 \leq r_{\mathbf{x}_1}^{(j\ell)}$, so that (\ref{eq:recGMM}) implies that $m_2 > r_{\mathbf{x}}^{(ik)} - r_{\mathbf{x}_1}^{(ik)}$ or $m_2 > r_{\mathbf{x}}^{(j\ell)} - r_{\mathbf{x}_1}^{(j\ell)}$ respectively. We can now use a similar approach to that used for the case when $m_1 > r_{\mathbf{x}_1}^{(ik)}$ and $m_1 > r_{\mathbf{x}_1}^{(j\ell)}$ in order to show that (\ref{eq:limterm}) holds. We can consider separately the two following cases: i) $r_{\mathbf{x}}^{(ik,j\ell)}> r_{\mathbf{x}}^{(ik)}, r_{\mathbf{x}}^{(j\ell)}$ and  ii) $r_{\mathbf{x}}^{(ik,j\ell)}=r_{\mathbf{x}}^{(ik)}=r_{\mathbf{x}}^{(j\ell)}$. In the first case, the range spaces $\mathrm{Im}(\bar{\mathbf{\Sigma}}_{\mathbf{x}}^{(ik)})$ and $\mathrm{Im}(\bar{\mathbf{\Sigma}}_{\mathbf{x}}^{(j\ell)})$ are distinct, and we can leverage the characterization of the misclassification probability of the distributed classification problem in order to prove (\ref{eq:limterm}). In particular, by following steps similar to those in the proof of Corollary \ref{cor:phasetrans}, we can show that 
\begin{equation}
\lim_{\sigma^2 \to 0}p(\hat{C}_1=j,\hat{C}_2=\ell|C_1=i,C_2=k)=0.
\end{equation}
Therefore, observe that the misclassification probability $p(\hat{C}_1=j,\hat{C}_2=\ell|C_1=i,C_2=k)$ is the measure of the set representing the decision region of the \ac{MAP} classifier for the distributed classification problem associated to the classes $(j,\ell)$ with respect to the Gaussian measure induced by the Gaussian distribution of classes $(i,k)$. Then, it is also possible to show that, in the limit $\sigma^2 \to 0$, the product in (\ref{eq:limterm}) is upper bounded by the integral of a measurable function over a set with measure zero, and then it converges to zero. 

In the second case, instead, we have $\mathrm{Im}(\bar{\mathbf{\Sigma}}_{\mathbf{x}}^{(ik)})=\mathrm{Im}(\bar{\mathbf{\Sigma}}_{\mathbf{x}}^{(j\ell)})$, and we can consider separately further two cases. If 
\begin{equation}
\boldsymbol{\mu}_{\mathbf{x}}^{(ik)}  - \boldsymbol{\mu}_{\mathbf{x}}^{(ik)} \notin \mathrm{Im}(\bar{\mathbf{\Sigma}}_{\mathbf{x}}^{(ik)}+\bar{\mathbf{\Sigma}}_{\mathbf{x}}^{(j\ell)})= \mathrm{Im}(\bar{\mathbf{\Sigma}}_{\mathbf{x}}^{(ik)}),
\end{equation} 
then Theorem~\ref{theo:nonzero_SI} states that $p(\hat{C}_1=j,\hat{C}_2=\ell|C_1=i,C_2=k)$ approaches zero in the low-rank regime, and we can prove that (\ref{eq:limterm}) holds by following a similar procedure to that used for case i). On the other hand, if 
\begin{equation}
\boldsymbol{\mu}_{\mathbf{x}}^{(ik)}  - \boldsymbol{\mu}_{\mathbf{x}}^{(ik)} \in \mathrm{Im}(\bar{\mathbf{\Sigma}}_{\mathbf{x}}^{(ik)}+\bar{\mathbf{\Sigma}}_{\mathbf{x}}^{(j\ell)}) = \mathrm{Im}(\bar{\mathbf{\Sigma}}_{\mathbf{x}}^{(ik)}),
\end{equation} 
then the misclassification probability is not guaranteed to approach zero in the low-rank regime. 
However, on using the law of total probability and the definition of \ac{MSE}, we can notice that the argument of the limit in (\ref{eq:limterm}) is upper bounded by $  \E{ \| \mathbf{x}_1 -  \mathcal{W}_{\mathbf{x}_1}^{(j\ell)}(\mathbf{y})   \|^2  | C_1=i,C_2=k }$ and that
\begin{equation}
  \E{ \| \mathbf{x}_1 -  \mathcal{W}_{\mathbf{x}_1}^{(j\ell)}(\mathbf{y})   \|^2  | C_1=i,C_2=k } \leq   \E{ \| \mathbf{x} -  \mathcal{W}_{\mathbf{x}}^{(j\ell)}(\mathbf{y})   \|^2  | C_1=i,C_2=k },
  \label{eq:MMSEmis}
\end{equation}
where 
\begin{equation}
\mathcal{W}_{\mathbf{x}}^{(j\ell)}(\mathbf{y})=   \boldsymbol{\mu}_{\mathbf{x}}^{(j\ell)} + \mathbf{W}_{\mathbf{x}}^{(j\ell)}\left(
\mathbf{y}
-\mathbf{\Phi} 
\boldsymbol{\mu}_{\mathbf{x}}^{(j\ell)}
\right)
\end{equation}
and\begin{equation}
\mathbf{W}_{\mathbf{x}}^{(j\ell)} =   (\bar{\mathbf{\Sigma}}_{\mathbf{x}}^{(j\ell)} + \sigma^2\mathbf{I}) \mathbf{\Phi}\tra \left( \sigma^2 \mathbf{I}  + \mathbf{\Phi} \bar{\mathbf{\Sigma}}_{\mathbf{x}}^{(j\ell)} \mathbf{\Phi}\tra   \right)^{-1}.
\end{equation}
Also in this case, we can show that the right hand side of (\ref{eq:MMSEmis}) approaches zero when $\sigma^2 \to 0$, since the mismatched \ac{MSE} for Gaussian sources reaches zero in the low-rank regime, provided that the estimated input covariance has the same range space than the true input covariance. 
In particular, on denoting by ${\mathbf{\Sigma}}_{\mathbf{y}}^{(ik)} = \sigma^2 \mathbf{I} + \mathbf{\Phi}  \bar{\mathbf{\Sigma}}_{\mathbf{x}}^{(ik)} \mathbf{\Phi}\tra$ the covariance matrix of $\mathbf{y}$ conditioned on $(C_1,C_2)=(i,k)$, and on introducing the symbol $\mathbf{M}^{(ik,j\ell)} = (\boldsymbol{\mu}_{\mathbf{x}}^{(ik)}- \boldsymbol{\mu}_{\mathbf{x}}^{(j\ell)}) (\boldsymbol{\mu}_{\mathbf{x}}^{(ik)}- \boldsymbol{\mu}_{\mathbf{x}}^{(j\ell)})\tra$, we can write
\begin{IEEEeqnarray}{rCl}
\MSE_{1,2|1,2}^{(ik,j\ell)} (\sigma^2) & = & \E{ \| \mathbf{x} -  \mathcal{W}_{\mathbf{x}}^{(j\ell)}(\mathbf{y})   \|^2  | C_1=i,C_2=k } \\
& = & \E{  \tr \left(    \left( \mathbf{x} -\mathcal{W}_{\mathbf{x}}^{(j\ell)} (\mathbf{y})  \right)   \left( \mathbf{x} -\mathcal{W}_{\mathbf{x}}^{(j\ell)} (\mathbf{y})   \right)\tra      \right) | C_1=i,C_2=k  }  \\
\nonumber
 & = & \tr \left(  \bar{\mathbf{\Sigma}}_{\mathbf{x}}^{(ik)} \right)  +(n_1+n_2)\sigma^2 - 2 \tr \left(   \mathbf{W}_{\mathbf{x}}^{(ik)}  \bar{\mathbf{\Sigma}}_{\mathbf{y}}^{(ik)} (\mathbf{W}_{\mathbf{x}}^{(j\ell)})\tra \right)  + \tr \left( \mathbf{W}_{\mathbf{x}}^{(j\ell)} \bar{\mathbf{\Sigma}}_{\mathbf{y}}^{(ik)} (\mathbf{W}_{\mathbf{x}}^{(j\ell)})\tra    \right)\\ &&
 + \tr \left(  \mathbf{M}^{(ik,j\ell)} \right)  - 2 \tr \left( \mathbf{M}^{(ik,j\ell)} \mathbf{\Phi}\tra (\mathbf{W}_{\mathbf{x}}^{j\ell})\tra \right) + \tr \left( \mathbf{\Phi} \mathbf{W}_{\mathbf{x}}^{(j\ell)} \mathbf{M}^{(ik,j\ell)} (\mathbf{W}_{\mathbf{x}}^{(j\ell)})\tra \right),
\end{IEEEeqnarray}
and we can prove that
\begin{equation}
 \lim_{\sigma^2 \to 0}  \tr \left(   \mathbf{W}_{\mathbf{x}}^{(ik)}  \bar{\mathbf{\Sigma}}_{\mathbf{y}}^{(ik)} (\mathbf{W}_{\mathbf{x}}^{(j\ell)})\tra \right) =  \tr(\bar{\mathbf{\Sigma}}_{\mathbf{x}}^{(ik)});
 \label{eq:fact1}
 \end{equation}
\begin{equation} 
 \label{eq:fact2}
\lim_{\sigma^2 \to 0}  \tr \left( \mathbf{W}_{\mathbf{x}}^{(j\ell)} \bar{\mathbf{\Sigma}}_{\mathbf{y}}^{(ik)}  (\mathbf{W}_{\mathbf{x}}^{(j\ell)})\tra \right)
 =  \tr(\bar{\mathbf{\Sigma}}_{\mathbf{x}}^{(ik)});
 \end{equation}
 \begin{equation}
  \lim_{\sigma^2 \to 0}  \tr \left( \mathbf{M}^{(ik,j\ell)} (\mathbf{W}_{\mathbf{x}}^{(j\ell)})\tra \right)
 =  \tr \left(\mathbf{M}^{(ik,j\ell)} \right);
 \label{eq:fact3}
 \end{equation}
 \begin{equation}
  \lim_{\sigma^2 \to 0}  \tr \left(  \mathbf{W}_{\mathbf{x}}^{(j\ell)}\mathbf{\Phi} \mathbf{M}^{(ik,j\ell)} \mathbf{\Phi}\tra (\mathbf{W}_{\mathbf{x}}^{(j\ell)})\tra \right)
 =  \tr \left(\mathbf{M}^{(ik,j\ell)} \right),
 \label{eq:fact4}
 \end{equation}
 by following steps similar to those adopted to prove \eqref{eq:fact1_New}-\eqref{eq:fact4_New}.

}

{\color{black}

\section{Proof of Lemma \ref{lem:lower_bound_nosi}}
\label{app:lower_bound_nosi}

The expansion of the lower bound $\MSE_{1|1}^{\mathrm{LB}}(\sigma_1^2)$ is based on an expression of the \ac{MMSE} associated to the reconstruction of Gaussian vectors in class $(i,k)$ from the linear features $\mathbf{y}_1$ akin to that reported in Appendix \ref{app:E}. In particular, we can write
\begin{IEEEeqnarray}{rCl}
\MMSE_{1|1}^{{\sf G}(i,k)}(\sigma_1^2)& = & \tr \left(  {\mathbf{\Sigma}}_{\mathbf{x}_1}^{(ik)}  -  {\mathbf{\Sigma}}_{\mathbf{x}_1}^{(ik)} \mathbf{\Phi}_1\tra \left(  \mathbf{\Phi}_1 {\mathbf{\Sigma}}_{\mathbf{x}_1}^{(ik)} \mathbf{\Phi}_1\tra   \right)^{-1} \mathbf{\Phi}_1 {\mathbf{\Sigma}}_{\mathbf{x}_1}^{(ik)}   \right)  \\
 &= & \tr \left( ( \bar{\mathbf{\Sigma}}_{\mathbf{x}_1}^{(ik)}   +  \sigma_1^2\mathbf{I})  -  (\bar{\mathbf{\Sigma}}_{\mathbf{x}_1}^{(ik)}+\sigma_1^2\mathbf{I}) \mathbf{\Phi}_1\tra \left( \sigma_1^2 \mathbf{I} +  \mathbf{\Phi}_1 \bar{\mathbf{\Sigma}}_{\mathbf{x}_1}^{(ik)} \mathbf{\Phi}_1\tra   \right)^{-1} \mathbf{\Phi}_1 (\bar{\mathbf{\Sigma}}_{\mathbf{x}_1}^{(ik)}  +\sigma_1^2\mathbf{I}) \right)\\
 \nonumber
& = & \tr \left(  \bar{\mathbf{\Sigma}}_{\mathbf{x}_1}^{(ik)}  -  \bar{\mathbf{\Sigma}}_{\mathbf{x}_1}^{(ik)} \mathbf{\Phi}_1\tra \left( \sigma_1^2 \mathbf{I} +  \mathbf{\Phi}_1 \bar{\mathbf{\Sigma}}_{\mathbf{x}_1}^{(ik)} \mathbf{\Phi}_1\tra   \right)^{-1} \mathbf{\Phi}_1 \bar{\mathbf{\Sigma}}_{\mathbf{x}_1}^{(ik)}   \right) + n_1 \sigma_1^2 \\
\nonumber
&& -2 \sigma_1^2 \tr \left(  \mathbf{\Phi}_1 \bar{\mathbf{\Sigma}}_{\mathbf{x}_1}^{(ik)} \mathbf{\Phi}_1\tra \left( \sigma_1^2 \mathbf{I} +  \mathbf{\Phi}_1 \bar{\mathbf{\Sigma}}_{\mathbf{x}_1}^{(ik)} \mathbf{\Phi}_1\tra   \right)^{-1}\right)\\
&& - \sigma_1^4 \tr \left( \left( \sigma_1^2 \mathbf{I} +  \mathbf{\Phi}_1 \bar{\mathbf{\Sigma}}_{\mathbf{x}_1}^{(ik)} \mathbf{\Phi}_1\tra   \right)^{-1}    \right).
\label{eq:MMSE_G_new_nosi}
\end{IEEEeqnarray}
Recall the definition of $\mathbf{\Xi}^{(ik)}$ in \eqref{eq:def_Xi} and its eigenvalue decomposition. Then, on following steps similar to those used in \cite[Appendix B]{RecJournal} and in Appendix \ref{app:E}, we can write 
\begin{IEEEeqnarray}{rCl}
\nonumber
\MMSE_{1|1}^{{\sf G}(i,k)}(\sigma_1^2)& = & \sum_{t=1}^{r_{\mathbf{\Xi}}^{(ik)}}    \frac{1}{1+\lambda_{\mathbf{\Xi},t}^{(ik)}/\sigma_1^2}   (\mathbf{u}_{\mathbf{\Xi},t}^{(ik)})\tra \bar{\mathbf{\Sigma}}_{\mathbf{x}_1}^{(ik)} \mathbf{u}_{\mathbf{\Xi},t}^{(ik)}
+ \sum_{t= r_{\mathbf{\Xi}}^{(ik)} +1}^{r_{\mathbf{x}_1}^{(ik)}}  (\mathbf{u}_{\mathbf{\Xi},t}^{(ik)})\tra \bar{\mathbf{\Sigma}}_{\mathbf{x}_1}^{(ik)} \mathbf{u}_{\mathbf{\Xi},t}^{(ik)} \IEEEeqnarraynumspace \\
&& + n_1 \sigma_1^2 - 2 \sigma_1^2 \sum_{t=1}^{r_{\mathbf{\Xi}}^{(ik)}}   \frac{\lambda_{\mathbf{\Xi},t}^{(ik)}}{\lambda_{\mathbf{\Xi},t}^{(ik)}+\sigma_1^2}  - \sigma_1^4 \sum_{t=1}^{r_{\mathbf{\Xi}}^{(ik)}}  \frac{1}{ \lambda_{\mathbf{\Xi},t}^{(ik)}   + \sigma_1^2  }   -  (m_1- r_{\mathbf{\Xi}}^{(ik)}) \sigma_1^2 \\
\nonumber
& = &  \sum_{t= r_{\mathbf{\Xi}}^{(ik)} +1}^{r_{\mathbf{x}_1}^{(ik)}}  (\mathbf{u}_{\mathbf{\Xi},t}^{(ik)})\tra \bar{\mathbf{\Sigma}}_{\mathbf{x}_1}^{(ik)} \mathbf{u}_{\mathbf{\Xi},t}^{(ik)}    \\
\nonumber
&&    +   \left(  n_1- m_1 - \min \{ m_1, r_{\mathbf{x}_1}^{(ik)} \}   +    \sum_{t=1}^{\min \{ m_1, r_{\mathbf{x}_1}^{(ik)} \} }    \frac{1}{1+\lambda_{\mathbf{\Xi},t}^{(ik)}/\sigma_1^2}   (\mathbf{u}_{\mathbf{\Xi},t}^{(ik)})\tra \bar{\mathbf{\Sigma}}_{\mathbf{x}_1}^{(ik)} \mathbf{u}_{\mathbf{\Xi},t}^{(ik)}  \right)   \cdot \sigma_1^2\\
&&  +    o (\sigma_1^2). \IEEEeqnarraynumspace
\end{IEEEeqnarray}

\section{Proof of Lemma \ref{lem:upper_bound_nosi}}
\label{app:upper_bound_nosi}

The expansion of $\MMSE_{1|1}(\sigma_1^2)$ is obtained by combining the result in Lemma \ref{lem:lower_bound_nosi} with the upper bound represented by the \ac{MSE} corresponding to a suboptimal classify and reconstruct approach akin to the described in Section~\ref{par:recGMM}, which we denote by $\MSE_{1|1}^{\mathrm{CR}}(\sigma_1^2)$. Note that $\MSE_{1|1}^{\mathrm{CR}}(\sigma_1^2)$ can be written as
\begin{IEEEeqnarray}{rCl}
\nonumber
\MSE_{1|1}^{\mathrm{CR}}(\sigma_1^2)  & = & \sum_{(i,k) \in \mathcal{S}} p_{C_1,C_2}(i,k)   \sum_{(j,\ell) \in \mathcal{S}}  \\
& & \cdot \int d \mathbf{x}_1 d\mathbf{y}_1 p(\mathbf{x}_1,\mathbf{y}_1|C_1=i,C_2=k)\\
\nonumber
& & \cdot  \prod_{\substack{(s,t) \in \mathcal{S} \\(s,t) \neq (j,\ell)}} u\left( \log  \frac{ p_{C_1,C_2}(j,\ell) p(\mathbf{y}_1| C_1=j,C_2=\ell)  }{p_{C_1,C_2}(s,t) p(\mathbf{y}_1| C_1=s,C_2=t)} \right) \left\|  \mathbf{x}_1   -\mathcal{W}_{\mathbf{x}_1}^{(j\ell)}(\mathbf{y}_1)  \right\|^2, \IEEEeqnarraynumspace \\
\end{IEEEeqnarray}
where $u(\cdot)$ is the unit step function and where
\begin{equation}
\mathcal{W}_{\mathbf{x}_1}^{(j\ell)}(\mathbf{y}_1)=   \boldsymbol{\mu}_{\mathbf{x}_1}^{(j\ell)} + \mathbf{W}_{\mathbf{x}_1}^{(j\ell)}\left(
\mathbf{y}_1
-\mathbf{\Phi}_1 
\boldsymbol{\mu}_{\mathbf{x}_1}^{(j\ell)}
\right)
\end{equation}
and
\begin{equation}
\mathbf{W}_{\mathbf{x}_1}^{(j\ell)} =   (\bar{\mathbf{\Sigma}}_{\mathbf{x}_1}^{(j\ell)}  + \sigma_1^2\mathbf{I}) \mathbf{\Phi}_1\tra \left( \sigma^2 \mathbf{I}  + \mathbf{\Phi}_1 \bar{\mathbf{\Sigma}}_{\mathbf{x}_1}^{(j\ell)} \mathbf{\Phi}_1\tra   \right)^{-1}.
\end{equation}
Then, on using the fact that $u(x) \leq 1, \forall x \in \mathbb{R}$, we can write the upper bound
\begin{IEEEeqnarray}{rCl}
\nonumber
\MSE_{1|1}^{\mathrm{CR}}(\sigma_1^2)  & \leq & \sum_{(i,k) \in \mathcal{S}} p_{C_1,C_2}(i,k)  \int d \mathbf{x}_1 d\mathbf{y}_1 p(\mathbf{x}_1,\mathbf{y}_1|C_1=i,C_2=k)  \left\|  \mathbf{x}_1   -\mathcal{W}_{\mathbf{x}_1}^{(ik)}(\mathbf{y}_1)  \right\|^2 \\
\nonumber
&& + \sum_{(i,k) \in \mathcal{S}} p_{C_1,C_2}(i,k)   \sum_{\substack{(j,\ell) \in \mathcal{S} \\(j,\ell) \neq (i,k)}}  \int d \mathbf{x}_1 d\mathbf{y}_1 p(\mathbf{x}_1,\mathbf{y}_1|C_1=i,C_2=k) \\
& & \cdot  u\left( \log  \frac{ p_{C_1,C_2}(j,\ell) p(\mathbf{y}_1| C_1=j,C_2=\ell)  }{p_{C_1,C_2}(i,k) p(\mathbf{y}_1| C_1=i,C_2=k)} \right) \left\|  \mathbf{x}_1   -\mathcal{W}_{\mathbf{x}_1}^{(j\ell)}(\mathbf{y}_1)  \right\|^2  \IEEEeqnarraynumspace \\
& = & \MSE_{1|1}^{\mathrm{LB}}(\sigma_1^2) \\
\nonumber
&& + \sum_{(i,k) \in \mathcal{S}} p_{C_1,C_2}(i,k)   \sum_{\substack{(j,\ell) \in \mathcal{S} \\(j,\ell) \neq (i,k)}}  \int d \mathbf{x}_1 d\mathbf{y}_1 p(\mathbf{x}_1,\mathbf{y}_1|C_1=i,C_2=k) \\
& & \cdot  u\left( \log  \frac{ p_{C_1,C_2}(j,\ell) p(\mathbf{y}_1| C_1=j,C_2=\ell)  }{p_{C_1,C_2}(i,k) p(\mathbf{y}_1| C_1=i,C_2=k)} \right) \left\|  \mathbf{x}_1   -\mathcal{W}_{\mathbf{x}_1}^{(j\ell)}(\mathbf{y}_1)  \right\|^2.  \IEEEeqnarraynumspace
\end{IEEEeqnarray}
Moreover, on using the upper bound $u(x) \leq e^{\frac{1}{2}x}, \forall x \in \mathbb{R}$, we can further upper bound $\MSE_{1|1}^{\mathrm{CR}}(\sigma_1^2)$ by
\begin{IEEEeqnarray}{rCl}
\nonumber
\MSE_{1|1}^{\mathrm{CR}}(\sigma_1^2) & \leq & \MSE_{1|1}^{\mathrm{LB}}(\sigma_1^2) \\
\nonumber
&& + \sum_{(i,k) \in \mathcal{S}}  \sum_{\substack{(j,\ell) \in \mathcal{S} \\(j,\ell) \neq (i,k)}}  \sqrt{p_{C_1,C_2}(i,k) p_{C_1,C_2}(j,\ell)      } \\
\nonumber
&& \cdot  \int  d\mathbf{y}_1 \sqrt{p(\mathbf{y}_1|C_1=i,C_2=k)  p(\mathbf{y}_1|C_1=j,C_2=\ell)   } \\
& & \int  d\mathbf{x}_1  p(\mathbf{x}_1| \mathbf{y}_1,C_1=i,C_2=k) \left\|  \mathbf{x}_1   -\mathcal{W}_{\mathbf{x}_1}^{(j\ell)}(\mathbf{y}_1)  \right\|^2.\IEEEeqnarraynumspace
\label{eq:MMSE_upper_New}
\end{IEEEeqnarray}

Then, in order to complete the proof of Lemma \ref{lem:upper_bound_nosi}, we show that the integrals in \eqref{eq:MMSE_upper_New} are $o(\sigma_1^2)$ when $m_1$ is such that $d^{\mathrm{NOSI}}(ik,j\ell)>1, \forall (i,k,j,\ell) \in \mathcal{S}\sub{DC}$, where $d^{\mathrm{NOSI}}(ik,j\ell)$ is defined as in \eqref{eq:d_nosi_ikjl}.

We first note that $p(\mathbf{x}_1|\mathbf{y}_1 , C_1=i, C_2=k) = \mathcal{N}(\tilde{\boldsymbol{\mu}}_{\mathbf{x}_1}^{(ik)}, \tilde{\mathbf{\Sigma}}_{\mathbf{x}_1}^{(ik)})$, where
\begin{IEEEeqnarray}{rCl}
\tilde{\boldsymbol{\mu}}_{\mathbf{x}_1}^{(ik)} & = & \mathbf{\Sigma}_{\mathbf{x}_1}^{(ik)} \mathbf{\Phi}_1\tra (\mathbf{\Phi}_1\mathbf{\Sigma}_{\mathbf{x}_1}^{(ik)} \mathbf{\Phi}_1\tra )^{-1}(\mathbf{y}_1 - \mathbf{\Phi}_1\boldsymbol{\mu}_{\mathbf{x}_1}^{(ik)}) +  \boldsymbol{\mu}_{\mathbf{x}_1}^{(ik)} \\
\tilde{\mathbf{\Sigma}}_{\mathbf{x}_1}^{(ik)} & = &  \mathbf{\Sigma}_{\mathbf{x}_1}^{(ik)} - \mathbf{\Sigma}_{\mathbf{x}_1}^{(ik)} \mathbf{\Phi}_1\tra (\mathbf{\Phi}_1\mathbf{\Sigma}_{\mathbf{x}_1}^{(ik)} \mathbf{\Phi}_1\tra)^{-1} \mathbf{\Phi}_1 \mathbf{\Sigma}_{\mathbf{x}_1}^{(ik)}.
\end{IEEEeqnarray}

Then, on using the triangular inequality, we can write the following upper bound:
\begin{IEEEeqnarray}{rCl}
\nonumber
\int  d\mathbf{x}_1  p(\mathbf{x}_1| \mathbf{y}_1,C_1=i,C_2=k) \left\|  \mathbf{x}_1   -\mathcal{W}_{\mathbf{x}_1}^{(j\ell)}(\mathbf{y}_1)  \right\|^2 & =& \int  d\mathbf{x}_1  p(\mathbf{x}_1| \mathbf{y}_1,C_1=i,C_2=k)\\
&& \cdot \left\|  \mathbf{x}_1 - \mathcal{W}_{\mathbf{x}_1}^{(ik)}(\mathbf{y}_1) + \mathcal{W}_{\mathbf{x}_1}^{(ik)}(\mathbf{y}_1)   -\mathcal{W}_{\mathbf{x}_1}^{(j\ell)}(\mathbf{y}_1)  \right\|^2 \IEEEeqnarraynumspace \\
& \leq & \mathrm{tr} (\tilde{\mathbf{\Sigma}}_{\mathbf{x}_1}^{(ik)} )  + \|    \mathcal{W}_{\mathbf{x}_1}^{(ik)}(\mathbf{y}_1)   -\mathcal{W}_{\mathbf{x}_1}^{(j\ell)}(\mathbf{y}_1)   \|^2\\
& \leq & \mathrm{tr} (\mathbf{\Sigma}_{\mathbf{x}_1}^{(ik)} )  + \|    \mathcal{W}_{\mathbf{x}_1}^{(ik)}(\mathbf{y}_1)   -\mathcal{W}_{\mathbf{x}_1}^{(j\ell)}(\mathbf{y}_1)   \|^2,
\end{IEEEeqnarray}
where we have leveraged the fact that the matrix $\mathbf{\Sigma}_{\mathbf{x}_1}^{(ik)} \mathbf{\Phi}_1\tra (\mathbf{\Phi}_1\mathbf{\Sigma}_{\mathbf{x}_1}^{(ik)} \mathbf{\Phi}_1\tra)^{-1} \mathbf{\Phi}_1 \mathbf{\Sigma}_{\mathbf{x}_1}^{(ik)}$ is positive semidefinite to establish the last inequality.

Consider now the integral
%
\begin{equation}
 \int  d\mathbf{y}_1 \sqrt{p(\mathbf{y}_1|C_1=i,C_2=k)  p(\mathbf{y}_1|C_1=j,C_2=\ell)   }  \cdot  \|    \mathcal{W}_{\mathbf{x}_1}^{(ik)}(\mathbf{y}_1)   -\mathcal{W}_{\mathbf{x}_1}^{(j\ell)}(\mathbf{y}_1)   \|^2.
\end{equation}
By leveraging the expression of the product of two Gaussian distributions in \cite[\S 8.1.8]{Petersen08} and on using the notation $\mathcal{N}(\mathbf{x}; \boldsymbol{\mu}, \mathbf{\Sigma})$ in order to denote explicitly the argument of the Gaussian distribution, we can write 
\begin{equation}
\sqrt{p(\mathbf{y}_1|C_1=i,C_2=k)  p(\mathbf{y}_1|C_1=j,C_2=\ell)   } = e^{-K_{1}(ik,j\ell)} \cdot \mathcal{N}(\mathbf{y}_1 ;    \boldsymbol{\mu}_1^{(ik,j\ell)}, \mathbf{\Sigma}_1^{(ik,j\ell)}),
\end{equation}
where
\begin{IEEEeqnarray}{rCl}
\nonumber
K_1(ik,j\ell) & = & \frac{1}{8}   (\boldsymbol{\mu}_{\mathbf{x}_1}^{(ik)}   - \boldsymbol{\mu}_{\mathbf{x}_1}^{(j\ell)})\tra \mathbf{\Phi}_1\tra  \left[     \frac{       \mathbf{\Phi}_1   ( \bar{\mathbf{\Sigma}}_{\mathbf{x}_1}^{(ik)}    +   \bar{\mathbf{\Sigma}}_{\mathbf{x}_1}^{(j\ell)}) \mathbf{\Phi}_1\tra   + 2 \sigma_1^2 \mathbf{I}       }{2}    \right]^{-1} \mathbf{\Phi}_1  (\boldsymbol{\mu}_{\mathbf{x}_1}^{(ik)}   - \boldsymbol{\mu}_{\mathbf{x}_1}^{(j\ell)}) \\
\label{eq:K_1ikjell}
& & + \frac{1}{2} \log \frac{  \mathrm{det}\left(  \frac{       \mathbf{\Phi}_1   ( \bar{\mathbf{\Sigma}}_{\mathbf{x}_1}^{(ik)}    +   \bar{\mathbf{\Sigma}}_{\mathbf{x}_1}^{(j\ell)}) \mathbf{\Phi}_1\tra   + 2 \sigma_1^2 \mathbf{I}       }{2}     \right)  }{    \sqrt{\mathrm{det} ( \mathbf{\Phi}_1    \bar{\mathbf{\Sigma}}_{\mathbf{x}_1}^{(ik)}     \mathbf{\Phi}_1\tra   +  \sigma_1^2 \mathbf{I}     )}    \sqrt{\mathrm{det} ( \mathbf{\Phi}_1    \bar{\mathbf{\Sigma}}_{\mathbf{x}_1}^{(j\ell)}     \mathbf{\Phi}_1\tra   +  \sigma_1^2 \mathbf{I}     )}    },
\end{IEEEeqnarray}
and where
%
%
%
%
%
\begin{IEEEeqnarray}{rCl}
 \boldsymbol{\mu}_1^{(ik,j\ell)} & = & \left(  (\mathbf{\Sigma}_{\mathbf{y}_1}^{(ik)})^{-1}  + (\mathbf{\Sigma}_{\mathbf{y}_1}^{(j\ell)})^{-1}     \right)^{-1} \left(  (\mathbf{\Sigma}_{\mathbf{y}_1}^{(ik)})^{-1} \mathbf{\Phi}_1 \boldsymbol{\mu}_{\mathbf{x}_1}^{(ik)}  + (\mathbf{\Sigma}_{\mathbf{y}_1}^{(j\ell)})^{-1} \mathbf{\Phi}_1 \boldsymbol{\mu}_{\mathbf{x}_1}^{(j\ell)}     \right) \\
 \mathbf{\Sigma}_1^{(ik,j\ell)}& = & 2 \left(  (\mathbf{\Sigma}_{\mathbf{y}_1}^{(ik)})^{-1}  + (\mathbf{\Sigma}_{\mathbf{y}_1}^{(j\ell)})^{-1} \right)^{-1},
\end{IEEEeqnarray}
where we have used the notation  $\mathbf{\Sigma}_{\mathbf{y}_1}^{(ik)}= \mathbf{\Phi}_1\bar{\mathbf{\Sigma}}_{\mathbf{x}_1}^{(ik)} \mathbf{\Phi}_1\tra + \mathbf{I} \sigma_1^2$.

Based on the analysis carried out in \cite{Reboredo14}, we can formulate the following upper bound:
\begin{equation}
e^{K_1(ik,j\ell)}  \leq A_1 (\sigma_1^2)^{d^{\mathrm{NOSI}}(ik,j\ell)} + o \left(   (\sigma_1^2)^{d^{\mathrm{NOSI}}(ik,j\ell)} \right),
\end{equation}
where $A_1$ is a positive constant and $d^{\mathrm{NOSI}}(ik,j\ell)$ is given by \eqref{eq:d_nosi_ikjl}. Therefore, our objective is to prove that the integral
\begin{equation}
\int d \mathbf{y}_1  \mathcal{N}(\mathbf{y}_1 ;    \boldsymbol{\mu}_1^{(ik,j\ell)}, \mathbf{\Sigma}_1^{(ik,j\ell)})\cdot \|    \mathcal{W}_{\mathbf{x}_1}^{(ik)}(\mathbf{y}_1)   -\mathcal{W}_{\mathbf{x}_1}^{(j\ell)}(\mathbf{y}_1)   \|^2
\label{eq:int_bounded}
\end{equation}
 is upper bounded by a constant when $\sigma_1^2  \to 0$.
%
In particular, on using the triangular inequality, we can upper bound the integral in \eqref{eq:int_bounded} as follows: 
\begin{IEEEeqnarray}{rCl}
\nonumber
 &&\int d \mathbf{y}_1  \mathcal{N}(\mathbf{y}_1 ;    \boldsymbol{\mu}_1^{(ik,j\ell)}, \mathbf{\Sigma}_1^{(ik,j\ell)})\cdot\|  \boldsymbol{\mu}_{\mathbf{x}_1}^{(ik)} - \boldsymbol{\mu}_{\mathbf{x}_1}^{(j\ell)}    + \mathbf{W}_{\mathbf{x}_1}^{(ik)} \mathbf{y}_1   - \mathbf{W}_{\mathbf{x}_1}^{(j\ell)} \mathbf{y}_1   + \mathbf{W}_{\mathbf{x}_1}^{(j\ell)} \mathbf{\Phi}_1 \boldsymbol{\mu}_{\mathbf{x}_1}^{(j\ell)}    - \mathbf{W}_{\mathbf{x}_1}^{(ik)} \mathbf{\Phi}_1 \boldsymbol{\mu}_{\mathbf{x}_1}^{(ik)}  \|^2 \\
 \nonumber
 & \leq & \|  \boldsymbol{\mu}_{\mathbf{x}_1}^{(ik)} - \boldsymbol{\mu}_{\mathbf{x}_1}^{(j\ell)} \|^2 + \|\mathbf{W}_{\mathbf{x}_1}^{(j\ell)} \mathbf{\Phi}_1 \boldsymbol{\mu}_{\mathbf{x}_1}^{(j\ell)}\|^2 + \| \mathbf{W}_{\mathbf{x}_1}^{(ik)} \mathbf{\Phi}_1 \boldsymbol{\mu}_{\mathbf{x}_1}^{(ik)}  \|^2 \\
 & & +  \int d \mathbf{y}_1  \mathcal{N}(\mathbf{y}_1 ;    \boldsymbol{\mu}_1^{(ik,j\ell)}, \mathbf{\Sigma}_1^{(ik,j\ell)}) \cdot \| \mathbf{W}_{\mathbf{x}_1}^{(ik)} \mathbf{y}_1 \|^2 +    \int d \mathbf{y}_1  \mathcal{N}(\mathbf{y}_1 ;    \boldsymbol{\mu}_1^{(ik,j\ell)}, \mathbf{\Sigma}_1^{(ik,j\ell)}) \cdot \| \mathbf{W}_{\mathbf{x}_1}^{(j\ell)} \mathbf{y}_1 \|^2  \\
 \nonumber
  & = & \|  \boldsymbol{\mu}_{\mathbf{x}_1}^{(ik)} - \boldsymbol{\mu}_{\mathbf{x}_1}^{(j\ell)} \|^2 + \|\mathbf{W}_{\mathbf{x}_1}^{(j\ell)} \mathbf{\Phi}_1 \boldsymbol{\mu}_{\mathbf{x}_1}^{(j\ell)}\|^2 + \| \mathbf{W}_{\mathbf{x}_1}^{(ik)} \mathbf{\Phi}_1 \boldsymbol{\mu}_{\mathbf{x}_1}^{(ik)}  \|^2 \\
  \nonumber
 & & + \tr  (\mathbf{W}_{\mathbf{x}_1}^{(ik)}  \mathbf{\Sigma}_1^{(ik,j\ell)} ( \mathbf{W}_{\mathbf{x}_1}^{(ik)})\tra  ) + \tr  (\mathbf{W}_{\mathbf{x}_1}^{(ik)}  \boldsymbol{\mu}_1^{(ik,j\ell)} (\boldsymbol{\mu}_1^{(ik,j\ell)})\tra ( \mathbf{W}_{\mathbf{x}_1}^{(ik)})\tra  )\\
&& + \tr  (\mathbf{W}_{\mathbf{x}_1}^{(j\ell)}  \mathbf{\Sigma}_1^{(ik,j\ell)} ( \mathbf{W}_{\mathbf{x}_1}^{(j\ell)})\tra  ) + \tr  (\mathbf{W}_{\mathbf{x}_1}^{(j\ell)}  \boldsymbol{\mu}_1^{(ik,j\ell)} (\boldsymbol{\mu}_1^{(ik,j\ell)})\tra ( \mathbf{W}_{\mathbf{x}_1}^{(j\ell)})\tra  ).
\label{eq:sum_terms_new}
\end{IEEEeqnarray}

Then, it is possible to show that all the terms in (\ref{eq:sum_terms_new}) are bounded. In particular, on leveraging the the fact that, given two positive semidefinite matrices of the same size $\mathbf{A}, \mathbf{B}$, it holds $\mathrm{tr}(\mathbf{A} \mathbf{B}) \leq \mathrm{tr}(\mathbf{A}) \mathrm{tr}(\mathbf{B})$, we can observe that $\|   \mathbf{W}_{\mathbf{x}_1}^{(ik)} \mathbf{\Phi}_1 \boldsymbol{\mu}_{\mathbf{x}_1}^{(ik)} \|^2 \leq \mathrm{tr}(\mathbf{W}_{\mathbf{x}_1}^{(ik)} (\mathbf{W}_{\mathbf{x}_1}^{(ik)})\tra) \mathrm{tr} (\mathbf{\Phi}_1 \boldsymbol{\mu}_{\mathbf{x}_1}^{(ik)} (\boldsymbol{\mu}_{\mathbf{x}_1}^{(ik)})\tra  \mathbf{\Phi}_1\tra) $ and all the terms in
\begin{IEEEeqnarray}{rCl}
\nonumber
\mathrm{tr}(\mathbf{W}_{\mathbf{x}_1}^{(ik)} (\mathbf{W}_{\mathbf{x}_1}^{(ik)})\tra) & = &  \tr \left(  \bar{\Sigma}_{\mathbf{x}_1}^{(ik)} \mathbf{\Phi}_1\tra \left(   \mathbf{\Phi}_1 \bar{\mathbf{\Sigma}}_{\mathbf{x}_1}^{(ik)}  \mathbf{\Phi}_1\tra + \sigma_1^2 \mathbf{I}   \right)^{-2}   \mathbf{\Phi}_1 \bar{\mathbf{\Sigma}}_{\mathbf{x}_1}^{(ik)}       \right) \\
\nonumber
&&+ 2 \sigma_1^2 \tr \left(  \mathbf{\Phi}_1  \bar{\Sigma}_{\mathbf{x}_1}^{(ik)} \mathbf{\Phi}_1\tra \left(   \mathbf{\Phi}_1 \bar{\mathbf{\Sigma}}_{\mathbf{x}_1}^{(ik)}  \mathbf{\Phi}_1\tra + \sigma_1^2 \mathbf{I}   \right)^{-2}        \right) \\
&&+ \sigma_1^4  \tr \left( \left(   \mathbf{\Phi}_1 \bar{\mathbf{\Sigma}}_{\mathbf{x}_1}^{(ik)}  \mathbf{\Phi}_1\tra + \sigma_1^2 \mathbf{I}   \right)^{-2}        \right)
\end{IEEEeqnarray}
are shown to be bounded by noting that $\mathrm{Null}(\mathbf{\Phi}_1  \bar{\Sigma}_{\mathbf{x}_1}^{(ik)} \mathbf{\Phi}_1\tra) = \mathrm{Null}(  \bar{\Sigma}_{\mathbf{x}_1}^{(ik)} \mathbf{\Phi}_1\tra)$ and by using steps similar to those used in Appendix \ref{app:E}. Similarly, we can write $\tr  (\mathbf{W}_{\mathbf{x}_1}^{(ik)}  \mathbf{\Sigma}_1^{(ik,j\ell)} ( \mathbf{W}_{\mathbf{x}_1}^{(ik)})\tra  )  \leq   \mathrm{tr}(\mathbf{W}_{\mathbf{x}_1}^{(ik)} (\mathbf{W}_{\mathbf{x}_1}^{(ik)})\tra) \tr ( \mathbf{\Sigma}_1^{(ik,j\ell)}     )  $ and we can note that 
\begin{IEEEeqnarray}{rCl}
\tr ( \mathbf{\Sigma}_1^{(ik,j\ell)} )   &  = &   2  \tr \left(  \mathbf{\Sigma}_{\mathbf{y}_1}^{(ik)}    \left( \mathbf{\Sigma}_{\mathbf{y}_1}^{(ik)} + \mathbf{\Sigma}_{\mathbf{y}_1}^{j\ell)}    \right)^{-1}  \mathbf{\Sigma}_{\mathbf{y}_1}^{(j\ell)}    \right)    \\
\nonumber
& = &  2 \tr   \left(   \mathbf{\Phi}_1 \bar{\mathbf{\Sigma}}_{\mathbf{x}_1}^{(ik)} \mathbf{\Phi}_1\tra   \left(  \mathbf{\Phi}_1 (\bar{\mathbf{\Sigma}}_{\mathbf{x}_1}^{(ik)}+\bar{\mathbf{\Sigma}}_{\mathbf{x}_1}^{(j\ell)}) \mathbf{\Phi}_1\tra   + 2\sigma_1^2 \mathbf{I}  \right)^{-1}         \mathbf{\Phi}_1 \bar{\mathbf{\Sigma}}_{\mathbf{x}_1}^{(j\ell)} \mathbf{\Phi}_1\tra    \right)\\
\nonumber
&& + 2 \sigma_1^2 \tr   \left(     \left(  \mathbf{\Phi}_1 (\bar{\mathbf{\Sigma}}_{\mathbf{x}_1}^{(ik)}+\bar{\mathbf{\Sigma}}_{\mathbf{x}_1}^{(j\ell)}) \mathbf{\Phi}_1\tra   + 2\sigma_1^2 \mathbf{I}  \right)^{-1}         \mathbf{\Phi}_1 \bar{\mathbf{\Sigma}}_{\mathbf{x}_1}^{(j\ell)} \mathbf{\Phi}_1\tra    \right)\\
\nonumber
&& + 2\sigma_1^2 \tr   \left(   \mathbf{\Phi}_1 \bar{\mathbf{\Sigma}}_{\mathbf{x}_1}^{(ik)} \mathbf{\Phi}_1\tra   \left(  \mathbf{\Phi}_1 (\bar{\mathbf{\Sigma}}_{\mathbf{x}_1}^{(ik)}+\bar{\mathbf{\Sigma}}_{\mathbf{x}_1}^{(j\ell)}) \mathbf{\Phi}_1\tra   + 2\sigma_1^2 \mathbf{I}  \right)^{-1}            \right)\\
&& + 2 \sigma_1^4 \tr   \left(     \left(  \mathbf{\Phi}_1 (\bar{\mathbf{\Sigma}}_{\mathbf{x}_1}^{(ik)}+\bar{\mathbf{\Sigma}}_{\mathbf{x}_1}^{(j\ell)}) \mathbf{\Phi}_1\tra   + 2\sigma_1^2 \mathbf{I}  \right)^{-1}        \right)
\label{eq:tra_12_new}
\end{IEEEeqnarray}
is also bounded, since $\mathrm{Null}(\mathbf{\Phi}_1 (\bar{\mathbf{\Sigma}}_{\mathbf{x}_1}^{(ik)}+\bar{\mathbf{\Sigma}}_{\mathbf{x}_1}^{(j\ell)}) \mathbf{\Phi}_1\tra) \subseteq  \mathrm{Null}(\mathbf{\Phi}_1 \bar{\mathbf{\Sigma}}_{\mathbf{x}_1}^{(ik)} \mathbf{\Phi}_1\tra)$ and $\mathrm{Null}(\mathbf{\Phi}_1 (\bar{\mathbf{\Sigma}}_{\mathbf{x}_1}^{(ik)}+\bar{\mathbf{\Sigma}}_{\mathbf{x}_1}^{(j\ell)}) \mathbf{\Phi}_1\tra) \subseteq  \mathrm{Null}(\mathbf{\Phi}_1 \bar{\mathbf{\Sigma}}_{\mathbf{x}_1}^{(j\ell)} \mathbf{\Phi}_1\tra)$. Finally, we can write $\tr  (\mathbf{W}_{\mathbf{x}_1}^{(ik)}  \boldsymbol{\mu}_1^{(ik,j\ell)} (\boldsymbol{\mu}_1^{(ik,j\ell)})\tra ( \mathbf{W}_{\mathbf{x}_1}^{(ik)})\tra  )    \leq    \mathrm{tr}(\mathbf{W}_{\mathbf{x}_1}^{(ik)} (\mathbf{W}_{\mathbf{x}_1}^{(ik)})\tra) \tr   ( \boldsymbol{\mu}_1^{(ik,j\ell)} (\boldsymbol{\mu}_1^{(ik,j\ell)})\tra)  $ and we can show that $\tr   ( \boldsymbol{\mu}_1^{(ik,j\ell)} (\boldsymbol{\mu}_1^{(ik,j\ell)})\tra)  $ is bounded when $\sigma_1^2 \to 0$ by noting that
\begin{IEEEeqnarray}{rCl}
\tr   ( \boldsymbol{\mu}_1^{(ik,j\ell)} (\boldsymbol{\mu}_1^{(ik,j\ell)})\tra) & = &  \left\| \mathbf{\Sigma}_1^{(ik,j\ell)}  \left((\mathbf{\Sigma}_{\mathbf{y}_1}^{(ik)})^{-1} \mathbf{\Phi}_1 \boldsymbol{\mu}_{\mathbf{x}_1}^{(ik)}  + (\mathbf{\Sigma}_{\mathbf{y}_1}^{(j\ell)})^{-1} \mathbf{\Phi}_1 \boldsymbol{\mu}_{\mathbf{x}_1}^{(j\ell)} \right)   \right\|^2/4 \\
 & \leq &  \left\| \mathbf{\Sigma}_1^{(ik,j\ell)}  (\mathbf{\Sigma}_{\mathbf{y}_1}^{(ik)})^{-1} \mathbf{\Phi}_1 \boldsymbol{\mu}_{\mathbf{x}_1}^{(ik)}    \right\|^2/4  +  \left\| \mathbf{\Sigma}_1^{(jik,j\ell)}  (\mathbf{\Sigma}_{\mathbf{y}_1}^{(j\ell)})^{-1} \mathbf{\Phi}_1 \boldsymbol{\mu}_{\mathbf{x}_1}^{(j\ell)}    \right\|^2/4\\
 & =&   \|   \mathbf{\Sigma}_{\mathbf{y}_1}^{(j\ell)}  (   \mathbf{\Sigma}_{\mathbf{y}_1}^{(ik)}  + \mathbf{\Sigma}_{\mathbf{y}_1}^{(j\ell)}     )^{-1}  \mathbf{\Phi}_1 \boldsymbol{\mu}_{\mathbf{x}_1}^{(ik)}     \|^2     + \|   \mathbf{\Sigma}_{\mathbf{y}_1}^{(ik)}  (   \mathbf{\Sigma}_{\mathbf{y}_1}^{(ik)}  + \mathbf{\Sigma}_{\mathbf{y}_1}^{(j\ell)}     )^{-1}  \mathbf{\Phi}_1 \boldsymbol{\mu}_{\mathbf{x}_1}^{(j\ell)}     \|^2     
\end{IEEEeqnarray}
and by following steps similar to those used to prove that \eqref{eq:tra_12_new} is bounded when $\sigma_1^2 \to 0$.

\section{Proof of Lemma \ref{lem:lower_bound}}
\label{app:lower_bound}
The lower bound $\MSE_{1|1,2}^{\mathrm{LB}}(\sigma_1^2)$ is defined as 
\begin{equation}
\MSE_{1|1,2}^{\mathrm{LB}}(\sigma_1^2) = \sum_{(i,k) \in \mathcal{S}}   p_{C_1,C_2}(i,k) \MMSE_{1|1,2}^{{\sf G}(i,k)}(\sigma_1^2),
\end{equation}
where $\MMSE_{1|1,2}^{{\sf G}(i,k)}(\sigma_1^2)$ is the Gaussian \ac{MMSE} associated to signals in class $(i,k)$. By following similar steps to those in Appendix \ref{app:E}, we recall that the Gaussian \ac{MMSE} does not depend on the mean, and by taking the expectation independently with respect to $\mathbf{x}_1 | \mathbf{y}_2$ and $\mathbf{y}_2$ we can write 
\begin{equation}
\MMSE_{1|1,2}^{{\sf G}(i,k)}(\sigma_1^2) = \MMSE^{{\sf G}(i,k)}(\mathbf{z}| \mathbf{\Phi}_1 \mathbf{z}),
\end{equation}
where $\mathbf{z} \sim p(\mathbf{x}_1|\mathbf{y}_2, C_1=i,C_2=k) = \mathcal{N}(\boldsymbol{\mu}_{\mathbf{z}}^{(ik)}, \mathbf{\Sigma}_{\mathbf{z}}^{(ik)})$, and
\begin{IEEEeqnarray}{rCl}
\label{eq:mu_z_new}
\boldsymbol{\mu}_{\mathbf{z}}^{(ik)} & =  & \boldsymbol{\mu}_{\mathbf{x}_1}^{(ik)}+
\bar{\mathbf{\Sigma}}_{\mathbf{x}_{12}}^{(ik)}  \mathbf{\Phi}_2\tra   (\mathbf{\Phi}_2 {\bar{\mathbf{\Sigma}}_{\mathbf{x}_2}^{(ik)}}  \mathbf{\Phi}_2\tra + \mathbf{I} \sigma_2^2  )^{-1}
(\mathbf{y}_2-  \mathbf{\Phi}_2 \boldsymbol{\mu}_{\mathbf{x}_2}^{(ik)}) \\
\mathbf{\Sigma}_{\mathbf{z}}^{(ik)}&  =  &  \bar{\mathbf{\Sigma}}_{\mathbf{z}}^{(ik)} + \sigma_1^2\mathbf{I}=  \bar{\mathbf{\Sigma}}_{\mathbf{x}_1}^{(ik)} - \bar{\mathbf{\Sigma}}_{\mathbf{x}_{12}}^{(ik)}  \mathbf{\Phi}_2\tra (  \mathbf{\Phi}_2 \bar{\mathbf{\Sigma}}_{\mathbf{x}_2}^{(ik)}   \mathbf{\Phi}_2\tra + \mathbf{I}\sigma_2^2   )^{-1}  \mathbf{\Phi}_2 \bar{\mathbf{\Sigma}}_{\mathbf{x}_{21}}^{(ik)} + \sigma_1^2\mathbf{I}.
\label{eq:sigma_z_new}
\end{IEEEeqnarray}
Then, the proof is completed by following steps similar to those in the proof of Lemma~\ref{lem:lower_bound_nosi}.

\section{Proof of Lemma \ref{lem:upper_bound}}
\label{app:upper_bound}

By taking independently the expectation with respect to $\mathbf{x}_1| \mathbf{y}_2$ and $\mathbf{y}_2$ in the definition of the \ac{MMSE} we can write
\begin{equation}
\MMSE_{1|12}(\sigma_1^2) = \E{\MMSE(\mathbf{z}|\mathbf{\Phi}_1 \mathbf{z})},
\label{eq:mmse_upper_new}
\end{equation}
where $\mathbf{z} \sim p(\mathbf{x}_1|\mathbf{y}_2)$ and where the expectation in \eqref{eq:mmse_upper_new} is taken with respect to $\mathbf{y}_2$. Then, we can note that
\begin{equation}
p(\mathbf{x}_1|\mathbf{y}_2)  = \sum_{(i,k) \in \mathcal{S}}   p(C_1=i, C_2=k|\mathbf{y}_2)  \cdot \mathcal{N}(\boldsymbol{\mu}_{\mathbf{z}}^{(ik)}, \mathbf{\Sigma}_{\mathbf{z}}^{(ik)}),
\end{equation}
where $\boldsymbol{\mu}_{\mathbf{z}}^{(ik)}$ and $\mathbf{\Sigma}_{\mathbf{z}}^{(ik)}$ are as in \eqref{eq:mu_z_new} and \eqref{eq:sigma_z_new}, and, for any value of $\mathbf{y}_2$ we can repeat the steps followed in Appendx \ref{app:upper_bound_nosi} in order to derive an upper bound to $\MMSE(\mathbf{z}|\mathbf{\Phi}_1 \mathbf{z})$ which admit the same first order expansion as $\MSE_{1|1,2}^{\mathrm{LB}}(\sigma_1^2)$. In particular, note that terms in the upper bound of $\MMSE_{1|12}(\sigma_1^2)$ which are functions of $\boldsymbol{\mu}_{\mathbf{z}}^{(ik)}$ are also bounded since 
\begin{IEEEeqnarray}{rCl}
\E{\|\boldsymbol{\mu}_{\mathbf{z}}^{(ik)}\|^2} = \|  \boldsymbol{\mu}_{\mathbf{x}_1}^{(ik)}  \|^2 +  \tr  \left(   \bar{\mathbf{\Sigma}}_{\mathbf{x}_{12}}^{(ik)}  \mathbf{\Phi}_2\tra (  \mathbf{\Phi}_2 \bar{\mathbf{\Sigma}}_{\mathbf{x}_2}^{(ik)}   \mathbf{\Phi}_2\tra + \mathbf{I}\sigma_2^2   )^{-1}  \mathbf{\Phi}_2 \bar{\mathbf{\Sigma}}_{\mathbf{x}_{21}}^{(ik)}    \right)
\end{IEEEeqnarray}
is bounded when $\sigma_1^2 \to 0$.

\section{Proof of Theorem \ref{theo:impact_si}}
\label{app:impact_si}


Note that the matrix
\begin{equation}
\bar{\mathbf{\Sigma}}_{\mathbf{z}}^{(ik)} =   \bar{\mathbf{\Sigma}}_{\mathbf{x}_1}^{(ik)} - \bar{\mathbf{\Sigma}}_{\mathbf{x}_{12}}^{(ik)}  \mathbf{\Phi}_2\tra (  \mathbf{\Phi}_2\tra {\bar{\mathbf{\Sigma}}_{\mathbf{x}_2}^{(ik)}}   \mathbf{\Phi}_2\tra + \mathbf{I}\sigma_2^2   )^{-1}  \mathbf{\Phi}_2 \bar{\mathbf{\Sigma}}_{\mathbf{x}_{21}}^{(ik)}
\end{equation}
is obtained as the Schur complement of the block $  \mathbf{\Phi}_2\tra {\bar{\mathbf{\Sigma}}_{\mathbf{x}_2}^{(ik)}}   \mathbf{\Phi}_2\tra + \mathbf{I}\sigma_2^2$ of the matrix
\begin{equation}
\left[
\begin{array}{cc}
\bar{\mathbf{\Sigma}}_{\mathbf{x}_1}^{(ik)} &  \mathbf{\Phi}_2 \bar{\mathbf{\Sigma}}_{\mathbf{x}_{12}}^{(ik)} \\
\bar{\mathbf{\Sigma}}_{\mathbf{x}_{21}}^{(ik)} \mathbf{\Phi}_2\tra &   \mathbf{\Phi}_2\tra {\bar{\mathbf{\Sigma}}_{\mathbf{x}_2}^{(ik)}}   \mathbf{\Phi}_2\tra + \mathbf{I}\sigma_2^2
\end{array}
\right].
\end{equation}
Then, on leveraging \cite[Lemma 4.1]{Ouellette81} in conjunction with \cite[Theorem 4.3]{Gallier10}, we have that
\begin{equation}
\mathrm{Im}(\bar{\mathbf{\Sigma}}_{\mathbf{z}}^{(ik)})   \subseteq  \mathrm{Im}(\bar{\mathbf{\Sigma}}_{\mathbf{x}_1}^{(ik)}).
\end{equation}
Moreover, on leveraging a rank computation akin to that in Appendix \ref{app:E}, it is possible to show that, for any $\sigma_2^2>0$, it holds $r_{\mathbf{z}}^{(ik)}= r_{\mathbf{x}_1}^{(ik)}$, and, therefore,
\begin{equation}
\mathrm{Im}(\bar{\mathbf{\Sigma}}_{\mathbf{z}}^{(ik)})   =  \mathrm{Im}(\bar{\mathbf{\Sigma}}_{\mathbf{x}_1}^{(ik)}).
\end{equation}
Then, when $m_1 < r_{\mathbf{x}_1}^{(ik)}=r_{\mathbf{z}}^{(ik)}$, we have $\mathcal{M}_{1|1}^{(i,k)}>0$ and $\mathcal{M}_{1|1,2}^{(i,k)}>0$, since $  \mathrm{Null}(\bar{\mathbf{\Sigma}}_{\mathbf{x}_1}^{(ik)}) \subset \mathrm{Null}({\mathbf{\Xi}}^{(ik)})$ and $ \mathrm{Null}(\bar{\mathbf{\Sigma}}_{\mathbf{z}}^{(ik)}) \subset \mathrm{Null}({\mathbf{\Theta}}^{(ik)})$. Moreover, $\mathcal{M}_{1|1,2}^{(i,k)} \leq \mathcal{M}_{1|1}^{(i,k)}$ follows directly from the fact $\MMSE_{1|1,2}^{{\sf G}(i,k)}(\sigma_1^2) \leq \MMSE_{1|1}^{{\sf G}(i,k)}(\sigma_1^2)$ for all $\sigma_1^2 >0$.

Consider now the case $m_1 > r_{\mathbf{x}_1}^{(ik)}$. In this case $\rank (\mathbf{\Xi}^{(ik)}) = \rank (\mathbf{\Theta}^{(ik)})= r_{\mathbf{x}_1}^{(ik)} = r_{\mathbf{z}}^{(ik)}$, therefore $  \mathrm{Null}(\bar{\mathbf{\Sigma}}_{\mathbf{x}_1}^{(ik)}) = \mathrm{Null}({\mathbf{\Xi}}^{(ik)})$ and $ \mathrm{Null}(\bar{\mathbf{\Sigma}}_{\mathbf{z}}^{(ik)}) = \mathrm{Null}({\mathbf{\Theta}}^{(ik)})$, which imply $\mathcal{M}_{1|1}^{(i,k)}=\mathcal{M}_{1|1,2}^{(i,k)}=0$. 

On the other hand, 
we can write $\mathcal{D}_{1|1,2}^{(i,k)}$ as
\begin{IEEEeqnarray}{rCl}
\mathcal{D}_{1|1,2}^{(i,k)} & = &n_1-m_1-r_{\mathbf{x}_1}^{(ik)} +  \sum_{t=1}^{r_{\mathbf{x}_1}^{(ik)}}    \frac{1}{\lambda_{\mathbf{\Theta},t}^{(ik)}}   (\mathbf{u}_{\mathbf{\Theta},t}^{(ik)})\tra  \bar{\mathbf{\Sigma}}_{\mathbf{z}}^{(ik)} \mathbf{u}_{\mathbf{\Theta},t}^{(ik)} \\
& = & n_1-m_1-r_{\mathbf{x}_1}^{(ik)}+ \mathrm{tr} \left(   \sum_{t=1}^{r_{\mathbf{x}_1}^{(ik)}}    \frac{1}{\lambda_{\mathbf{\Theta},t}^{(ik)}}  \mathbf{u}_{\mathbf{\Theta},t}^{(ik)} (\mathbf{u}_{\mathbf{\Theta},t}^{(ik)})\tra  \bar{\mathbf{\Sigma}}_{\mathbf{z}}^{(ik)}   \right).
\end{IEEEeqnarray}
Then, since $\rank(\mathbf{\Theta}_{\mathbf{z}}^{(ik)})= r_{\mathbf{z}}^{(ik)}$ we can leverage the expression of the Moore-Penrose inverse of a matrix in terms of its \ac{SVD} in order to observe that
\begin{IEEEeqnarray}{rCl}
\sum_{t=1}^{r_{\mathbf{x}_1}^{(ik)}}    \frac{1}{\lambda_{\mathbf{\Theta},t}^{(ik)}}  \mathbf{u}_{\mathbf{\Theta},t}^{(ik)} (\mathbf{u}_{\mathbf{\Theta},t}^{(ik)})\tra    & = & \left( \mathbf{\Theta}^{(ik)}  \right)^\dag \\
 & = & \left(  (\bar{\mathbf{\Sigma}}_{\mathbf{z}}^{(ik)})^{\frac{1}{2}} \mathbf{\Phi}_1\tra \mathbf{\Phi}_1 (\bar{\mathbf{\Sigma}}_{\mathbf{z}}^{(ik)})^{\frac{1}{2}}  \right)^\dag,
\end{IEEEeqnarray}
which allows us to write
\begin{IEEEeqnarray}{rCl}
\mathcal{D}_{1|1,2}^{(i,k)} & = &n_1-m_1-r_{\mathbf{x}_1}^{(ik)} + \mathrm{tr} \left( \left(  (\bar{\mathbf{\Sigma}}_{\mathbf{z}}^{(ik)})^{\frac{1}{2}} \mathbf{\Phi}_1\tra \mathbf{\Phi}_1 (\bar{\mathbf{\Sigma}}_{\mathbf{z}}^{(ik)})^{\frac{1}{2}}  \right)^\dag \bar{\mathbf{\Sigma}}_{\mathbf{z}}^{(ik)}  \right)\\
& = & n_1-m_1-r_{\mathbf{x}_1}^{(ik)} + \mathrm{tr} \left( (\bar{\mathbf{\Sigma}}_{\mathbf{z}}^{(ik)})^{\frac{1}{2}} \left(  (\bar{\mathbf{\Sigma}}_{\mathbf{z}}^{(ik)})^{\frac{1}{2}} \mathbf{\Phi}_1\tra \mathbf{\Phi}_1 (\bar{\mathbf{\Sigma}}_{\mathbf{z}}^{(ik)})^{\frac{1}{2}}  \right)^\dag (\bar{\mathbf{\Sigma}}_{\mathbf{z}}^{(ik)})^{\frac{1}{2}}  \right).
\label{eq:trace1}
\end{IEEEeqnarray}
Then, let us write the compact eigenvalue decomposition of the matrix $(\bar{\mathbf{\Sigma}}_{\mathbf{z}}^{(ik)})^{\frac{1}{2}}$ as
\begin{equation}
(\bar{\mathbf{\Sigma}}_{\mathbf{z}}^{(ik)})^{\frac{1}{2}} = \mathbf{U}_{\mathbf{z}}^{(ik)} (\mathbf{\Lambda}_{\mathbf{z}}^{(ik)})^{\frac{1}{2}} (\mathbf{U}_{\mathbf{z}}^{(ik)})\tra,
\end{equation}
where $\mathbf{U}_{\mathbf{z}}^{(ik)} \in \mathbb{R}^{n_1 \times r_{\mathbf{z}}^{(ik)}}$ has orthonormal columns and $\mathbf{\Lambda}_{\mathbf{z}}^{(ik)} \in \mathbb{R}^{r_{\mathbf{z}}^{(ik)} \times r_{\mathbf{z}}^{(ik)}}$ has positive entries. Note also that
\begin{IEEEeqnarray}{rCl}
(\mathbf{A} \mathbf{A}\tra)^\dag &  = & (\mathbf{A}\tra)^\dag \mathbf{A}^\dag ,
\end{IEEEeqnarray}
and 
\begin{equation}
(\mathbf{A} \mathbf{B} )^\dag  = \mathbf{B}^\dag \mathbf{A}^\dag,
\label{eq:prodMP}
\end{equation}
if $\mathbf{A}$ is full column rank and $\mathbf{B}$ is full row rank~\cite{Ben03}. 
Then, we can write \eqref{eq:trace1} as
\begin{IEEEeqnarray}{rCl}
\mathcal{D}_{1|1,2}^{(i,k)} & = & n_1-m_1-r_{\mathbf{x}_1}^{(ik)}+ \mathrm{tr}  \left(     (\bar{\mathbf{\Sigma}}_{\mathbf{z}}^{(ik)})^{\frac{1}{2}}(\mathbf{\Phi}_1 (\bar{\mathbf{\Sigma}}_{\mathbf{z}}^{(ik)})^{\frac{1}{2}}  )^\dag  (  (\bar{\mathbf{\Sigma}}_{\mathbf{z}}^{(ik)})^{\frac{1}{2}} \mathbf{\Phi}_1\tra )^\dag (\bar{\mathbf{\Sigma}}_{\mathbf{z}}^{(ik)})^{\frac{1}{2}}     \right) \\
\nonumber
& = & n_1-m_1-r_{\mathbf{x}_1}^{(ik)}+ \mathrm{tr}   \left(   (\mathbf{\Lambda}_{\mathbf{z}}^{(ik)})^{\frac{1}{2}}    (\mathbf{U}_{\mathbf{z}}^{(ik)})\tra  ((\mathbf{\Lambda}_{\mathbf{z}}^{(ik)})^{\frac{1}{2}}    (\mathbf{U}_{\mathbf{z}}^{(ik)})\tra)^\dag  (\mathbf{\Phi}_1 \mathbf{U}_{\mathbf{z}}^{(ik)} )^\dag   \right. \\
& & \left.   \cdot     ((\mathbf{U}_{\mathbf{z}}^{(ik)})\tra \mathbf{\Phi}_1\tra)^\dag    (\mathbf{U}_{\mathbf{z}}^{(ik)} (\mathbf{\Lambda}_{\mathbf{z}}^{(ik)})^{\frac{1}{2}})^\dag \mathbf{U}_{\mathbf{z}}^{(ik)} (\mathbf{\Lambda}_{\mathbf{z}}^{(ik)})^{\frac{1}{2}}     \right) \\
& = & n_1-m_1-r_{\mathbf{x}_1}^{(ik)} + \mathrm{tr}   \left(      \left(    (\mathbf{U}_{\mathbf{z}}^{(ik)})\tra \mathbf{\Phi}_1\tra \mathbf{\Phi}_1      \mathbf{U}_{\mathbf{z}}^{(ik)} \right)^\dag    \right),
\end{IEEEeqnarray}
where we have used the assumption $m_1 > r_{\mathbf{z}}^{(ik)}$ in order to use the property in \eqref{eq:prodMP}.
Consider now the compact eigenvalue decomposition of the matrix $(\bar{\mathbf{\Sigma}}_{\mathbf{x}_1}^{(ik)})^{\frac{1}{2}}$,
\begin{equation}
(\bar{\mathbf{\Sigma}}_{\mathbf{x}_1}^{(ik)})^{\frac{1}{2}} = \mathbf{U}_{\mathbf{x}_1}^{(ik)} (\mathbf{\Lambda}_{\mathbf{x}_1}^{(ik)})^{\frac{1}{2}} (\mathbf{U}_{\mathbf{x}_1}^{(ik)})\tra,
\end{equation}
where $\mathbf{U}_{\mathbf{x}_1}^{(ik)} \in \mathbb{R}^{n_1 \times r_{\mathbf{x}_1}^{(ik)}}$ has orthonormal columns and $\mathbf{\Lambda}_{\mathbf{x}_1}^{(ik)} \in \mathbb{R}^{r_{\mathbf{x}_1}^{(ik)} \times r_{\mathbf{x}_1}^{(ik)}}$ has positive entries. Then, on following steps similar to those used to express $\mathcal{D}_{1|1,2}^{(i,k)}$, we can also write
\begin{equation}
\mathcal{D}_{1|1}^{(i,k)}= n_1 -m_1 -r_{\mathbf{x}_1}^{(ik)}+    \mathrm{tr}   \left(      \left(    (\mathbf{U}_{\mathbf{x}_1}^{(ik)})\tra \mathbf{\Phi}_1\tra \mathbf{\Phi}_1      \mathbf{U}_{\mathbf{x}_1}^{(ik)} \right)^\dag    \right).
\end{equation}
Finally on recalling that $\mathrm{Im}(\bar{\mathbf{\Sigma}}_{\mathbf{x}_1}^{(ik)}) = \mathrm{Im}(\bar{\mathbf{\Sigma}}_{\mathbf{z}}^{(ik)})$, we observe that
\begin{equation}
\mathbf{U}_{\mathbf{x}_1}^{(i,k)} = \mathbf{U}_{\mathbf{z}}^{(i,k)} \mathbf{R},
\end{equation}
where $\mathbf{R}$ is an $r_{\mathbf{x}_1}^{(ik)} \times r_{\mathbf{x}_1}^{(ik)}$ orthogonal matrix, from which we can immediately conclude $\mathcal{D}_{1|1,2}^{(i,k)}=\mathcal{D}_{1|1}^{(i,k)}$.

}

\section*{Acknowledgment}

The work of F. Renna was supported by Project I-CITY - ICT for Future Health/Faculdade de Engenharia da Universidade do Porto, NORTE-07-0124-FEDER-000068, funded by the Fundo Europeu de Desenvolvimento Regional (FEDER) through the Programa Operacional do Norte (ON2) and by national funds, through FCT/MEC (PIDDAC). This work was also supported by the Royal Society International Exchanges Scheme IE120996, and the Duke components of the research were supported in part by the following agencies: AFOSR, ARO, DARPA, DOE, NGA and ONR.

\bibliographystyle{IEEEtran}
\bibliography{references_rec}

\end{document}